\documentclass{jpp}
\usepackage{epstopdf, epsfig}

 \usepackage{amssymb}
 \usepackage{bigints}
 \usepackage{amsmath}
 \usepackage{empheq}
 \usepackage{framed}
 \usepackage{color}
 \usepackage{enumerate}
 \usepackage{ulem}
 \usepackage{bm}
 \usepackage{appendix}
 \usepackage{multicol}
 \numberwithin{equation}{section}
 \usepackage{upgreek}
 \usepackage{textgreek}
\usepackage{stmaryrd} 
\usepackage{xfrac}    

\newcommand{\comment}[1]{}

\newcommand{\be}{\begin{equation}}
\newcommand{\ee}{\end{equation}}
\newcommand{\ba}{\[\begin{aligned}}
\newcommand{\ea}{\end{aligned}\]}
\newcommand{\bea}{\begin{eqnarray}}
\newcommand{\eea}{\end{eqnarray}}
\newcommand{\beann}{\begin{eqnarray*}}
\newcommand{\eeann}{\end{eqnarray*}}
\newcommand{\bs}{\begin{split}}
\newcommand{\es}{\end{split}}

\newcommand*{\cA}{\mathcal{A}} 

\newcommand*{\cF}{\mathcal{F}}
\newcommand*{\cG}{\mathcal{G}}

\newcommand*{\cI}{\mathcal{I}}

\newcommand*{\cL}{\mathcal{L}}

\newcommand*{\cN}{\mathcal{N}}

\newcommand*{\cP}{\mathcal{P}}

\newcommand*{\cT}{\mathcal{T}}

\newcommand*{\cW}{\mathcal{W}}
\newcommand*{\cX}{\mathcal{X}}
\newcommand*{\cY}{\mathcal{Y}}
\newcommand*{\cZ}{\mathcal{Z}}

\newcommand*{\ep}{\epsilon}
\newcommand*{\vep}{\varepsilon}

\newcommand*{\B}{\bm{B}}

\newcommand*{\J}{\bm{J}}

\newcommand*{\JxB}{\bm{J}\times \bm{B}}

\newcommand*{\jpl}{j_{||}}

\newcommand*{\dl}{\bm{\nabla}}
\newcommand*{\del}{\partial}
\newcommand*{\BD}{\bm{B}\cdot\bm{\nabla}}

\newcommand*{\dlts}{\Delta_*}

\newcommand*{\lh}{\bm{\hat{\ell}}}

\newcommand*{\rhoh}{\bm{\hat{\rho}}}

\newcommand*{\wh}{\bm{\hat{\omega}}}

\newcommand*{\lbr}{\left(}
\newcommand*{\rbr}{\right)}

\newcommand*{\zb}{\overline{z}}

\newcommand*{\cZb}{\overline{\cZ}}

\newcommand*{\Gb}{\overline{G}}
\newcommand*{\Kb}{\overline{K}}

\newcommand*{\alphaO}{\alpha^{(0)}}
\newcommand*{\alphaOcw}{\alpha_{,\omega}^{(0)}}
\newcommand*{\alphaOcz}{\alpha_{,\ell}^{(0)}}

\newcommand*{\PhiO}{\Phi^{(0)}}

\newcommand*{\GO}{G^{(0)}}

\newcommand*{\alphaOne}{\alpha^{(1)}}

\newcommand*{\PhiOne}{\Phi^{(1)}}

\newcommand*{\cAOne}{\mathcal{A}^{(1)}}

\newcommand*{\FOne}{F^{(1)}}
\newcommand*{\cFOne}{\mathcal{F}^{(1)}}
\newcommand*{\GOne}{G^{(1)}}

\newcommand*{\GammaOne}{\Gamma^{(1)}}

\newcommand*{\gOne}{g^{(1)}}

\newcommand*{\cZOne}{\cZ^{(1)}}

\newcommand*{\alphaTwo}{\alpha^{(2)}}
\newcommand*{\pTwo}{p^{(2)}}
\newcommand*{\psiTwo}{\psi^{(2)}}

\newcommand*{\PhiTwo}{\Phi^{(2)}}
\newcommand*{\phiTwo}{\phi^{(2)}}

\newcommand*{\fTwocTwo}{f^{(2)}_{c2}}
\newcommand*{\fTwosTwo}{f^{(2)}_{s2}}
\newcommand*{\GTwo}{G^{(2)}}

\newcommand*{\fTwo}{f^{(2)}}

\newcommand*{\xiTwo}{\xi^{(2)}}

\newcommand*{\cAThree}{\cA^{(3)}}

\newcommand*{\xiThree}{\xi^{(3)}}
\newcommand*{\fThree}{f^{(3)}}

\newcommand*{\GammaThree}{\Gamma^{(3)}}
\newcommand*{\zetaThree}{\zeta^{(3)}}
\newcommand*{\FThree}{F^{(3)}}
\newcommand*{\cFThree}{\cF^{(3)}}
\newcommand*{\cTThree}{\cT^{(3)}}
\newcommand*{\PThree}{P^{(3)}}
\newcommand*{\QThree}{Q^{(3)}}

\newcommand*{\YThree}{Y^{(3)}}

\newcommand*{\PhiThree}{\Phi^{(3)}}
\newcommand*{\psiThree}{\psi^{(3)}}
\newcommand*{\sigmaThree}{\sigma^{(3)}}

\newcommand*{\psiFour}{\psi^{(4)}}

\newcommand*{\alphaOell}{\alpha^{(0)}_{,\ell}}

\newcommand*\at[2]{\left.#1\right|_{#2}}

\newcommand*\reallywidetilde[1]{\ThisStyle{%
  \setbox0=\hbox{$\SavedStyle#1$}%
  \stackengine{-.1\LMpt}{$\SavedStyle#1$}{%
    \stretchto{\scaleto{\SavedStyle\mkern.2mu\sim}{.5467\wd0}}{.7\ht0}%
  }{O}{c}{F}{T}{S}%
}}

\def\test#1{$%
  \reallywidetilde{#1}\,
  \scriptstyle\reallywidetilde{#1}\,
  \scriptscriptstyle\reallywidetilde{#1}
$\par}
\makeatletter
\newsavebox{\@brx}
\newcommand{\llangle}[1][]{\savebox{\@brx}{\(\m@th{#1\langle}\)}%
  \mathopen{\copy\@brx\mkern2mu\kern-0.9\wd\@brx\usebox{\@brx}}}
\newcommand{\rrangle}[1][]{\savebox{\@brx}{\(\m@th{#1\rangle}\)}%
  \mathclose{\copy\@brx\mkern2mu\kern-0.9\wd\@brx\usebox{\@brx}}}
\makeatother

\shorttitle{Finite-beta near-axis expansions} 

\title{Stellarator equilibrium axis-expansion to all orders in distance from the axis
for arbitrary plasma beta}

\shortauthor{W. Sengupta, E. Rodriguez, R.Jorge, M. Landreman A. Bhattacharjee}
\author{Wrick Sengupta\aff{1}\corresp{\email{wsengupta@princeton.edu}}, Eduardo Rodriguez \aff{3}, Rogerio Jorge\aff{4,5},  Matt Landreman\aff{6}, Amitava Bhattacharjee\aff{1}}

\affiliation{
\aff{1} Department of Astrophysical Sciences, Princeton University, Princeton, NJ 08543, USA
\aff{3} Max Planck Institute for Plasma Physics, 17491 Greifswald, Germany
\aff{4} Instituto de Plasmas e Fusão Nuclear, Instituto Superior Técnico, Universidade de Lisboa, 1049-001 Lisboa, Portugal
\aff{5} Department of Physics, University of Wisconsin-Madison, Madison, Wisconsin 53706, USA
\aff{6}Institute for Research in Electronics and Applied Physics,
University of Maryland,
College Park, MD 20742, USA
}

\begin{document}

\maketitle

\begin{abstract}
A systematic theory of the asymptotic expansion of the magnetohydrodynamic (MHD) equilibrium in the distance from the magnetic axis is developed to include arbitrary smooth currents near the magnetic axis. Compared to the vacuum and the force-free system, an additional magnetic differential equation must be solved to obtain the pressure-driven currents. It is shown that there exist variables in which the rest of the MHD system closely mimics the vacuum system. Thus, a unified treatment of MHD fields is possible. The mathematical structure of the near-axis expansions to arbitrary order is examined carefully to show that the double-periodicity of physical quantities in a toroidal domain can be satisfied order by order. The essential role played by the Birkhoff–Gustavson normal form in solving the magnetic differential equations is highlighted. Several explicit examples of vacuum, force-free, and MHD equilibrium in different geometries are presented.
 
\end{abstract}

\section{Introduction \label{sec:intro}}
Understanding the magnetohydrodynamic (MHD) equilibrium in magnetically confined devices is paramount. The theory of axisymmetric MHD equilibrium, applicable to tokamaks, is well-developed at this point. The axisymmetric problem can be reduced to solving a nonlinear elliptic partial differential equation (PDE) called the Grad-Shafranov equation \citep{grad1971plasma,freidberg_idealMHD}. In contrast, 
the 3D theory cannot be reduced to a single elliptic PDE \citep{weitzner2014ideal,helander2014theory, Garabedian2012MHDstellarator}. Analytical understanding is obstructed by the intrinsic nonlinearity of the MHD system, coupled with the three-dimensional nature of the 3D equilibrium, which is generic for stellarators. These obstacles hinder an intuitive understanding of important physical effects, such as the deformations of the vacuum plasma surfaces due to pressure-driven currents or the dependence of magnetic shear on the geometry. Large-scale numerical computation has taught us much about stellarator equilibrium, but simulations can be expensive. Furthermore, the computational error can be significant, particularly near the magnetic axis \citep{panici2022_DESC_VMEC_near_axis}. Therefore, obtaining an analytical understanding of 3D stellarator MHD equilibrium is beneficial for computational studies and their applications to experiments.

In this work, we shall focus on an asymptotic expansion called the near-axis expansion (NAE) developed by several authors \citep{Mercier1964, Solovev1970,garrenBoozer1991existence,bernardin_tataronis_1985hamiltonian_NAE,weitzner2016expansions}. The basic idea is to expand the MHD equilibrium equations in a power series in the distance from the magnetic axis. The small parameter is $\kappa a$, where $\kappa$ is the curvature of the magnetic axis, and $a$ is a typical length-scale of the order of the minor radius in the direction normal to the magnetic axis. Recently, there has been a surge of interest in applying the NAE to study quasisymmetry (QS) \citep{landreman_Sengupta2018direct,landreman_Sengupta2019_2nd_order,Jorge_sengupta_landreman_2020_NF,rodriguez2023constructing_phase_space}, quasi-isodynamic (QI) stellarators \citep{plunk_Landreman_2019near_axis3,camacho_mata2022direct,rodriguez_plunck_2023higher_order_QI,camacho_mata2023helicity}, energetic particles \citep{figueiredo2023energetic}, MHD stability \citep{landreman2020magnetic_well,Kim_p_Jorge_Dorland_2021_Well_Mercier,rodriguez2023MHD_stability_shaping} and turbulence \citep{jorge_landreman2020_NAE_turb,jorge_landreman2021_ITG,rodriguez_mackenbach_2023trapped}.

The development of the NAE theory has proceeded along two approaches: direct and indirect. In the direct approach \citep{Mercier1964,Solovev1970,Jorge_sengupta_landreman_2020_NF,Jorge_Sengupta_Landreman2020NAE,bernardin_tataronis_1985hamiltonian_NAE,Duignan_Meiss2021normal_form}, one uses a coordinate system pioneered by Mercier, and expands in the distance from the magnetic axis. In the inverse approach, one expands in the toroidal flux-surface label \citep{garrenBoozer1991existence,weitzner2016expansions,landreman_Sengupta2018direct,landreman_Sengupta_Plunck2019direct,landreman_Sengupta2019_2nd_order,rodriguez2021_GBC,rodriguez2021solving_overdetermination_I,rodriguez2022weakl_QS}. 

Both the direct and the indirect approaches have distinct advantages and disadvantages. The inverse coordinate approach is best suited to address concepts such as quasisymmetry and omnigeneity, which impose constraints on the magnitude of the magnetic field. However, the direct approach offers better physical insight. This is mainly because the flux-surface shaping factors, such as ellipticity and triangularity, can be chosen independently in the direct approach but not in the indirect approach. Moreover, in the indirect approach, the lowest order involves the solution of a nonlinear Riccati equation. In the direct coordinate approach, we can avoid solving a nonlinear ODE by using Mercier's formula \citep{mercier1974lectures, Solovev1970}, which also follows from Floquet theory \citep{Duignan_Meiss2021normal_form}. The expressions for the rotational transform \citep{Mercier1964,helander2014theory} and its derivatives are also physically more intuitive in the direct approach. 


An outstanding issue in the NAE that needs to be addressed is the assumption of smooth solutions of the MHD equilibrium order by order. The mathematical analysis of the smoothness assumption is indeed quite technical \citep{burby_duignan_meiss_2021_NAE_coordinates, Duignan_Meiss2021normal_form}. The first problem we encounter is the classic problem of the occurrence of singular currents in stellarators on rational surfaces because of a lack of axisymmetry \citep{grad1967toroidal, Mercier1964, weitzner2016expansions, loizu2015jumpiota,helander2014theory}. These singularities are fundamentally tied to the magnetic differential equations (MDE) \citep{newcomb1959magnetic}, whose solution can be singular everywhere $\iota$, the rotational transform, is rational. Since rationals are dense, avoiding such singularities with non-constant $\iota$ is challenging \citep{grad1967toroidal,boozer1981plasma, weitzner2014ideal}. In the NAE expansion, thankfully, the rational singularities are governed only by $\iota_0$, the on-axis rotational transform \citep{Mercier1964, Solovev1970, Duignan_Meiss2021normal_form}.
Choosing it to be sufficiently irrational helps with the practical implementation of lower-order NAE \citep{Jorge_Sengupta_Landreman2020NAE,landreman_Sengupta2019_2nd_order}. However, since any irrational number can be approximated closely by rationals, the problem of singular divisors appearing in higher orders can only be partially resolved mathematically in this approach. An alternative is to assume that $\iota_0$ is close to a given rational number and construct special non-singular solutions \citep{weitzner2016expansions,sengupta_erin2018} or use the near-resonant normal form theory \citep{Duignan_Meiss2021normal_form}. 

We remark that there could be other potential sources of non-smoothness as well. Even in the vacuum limit, the smoothness of the solution within the NAE framework is not obvious \citep{Jorge_Sengupta_Landreman2020NAE}. When currents are present, the situation is complicated by another factor: the possibility of occurrence of logarithmic terms of the form $\rho^n (\log\rho)^m, n>m $ that, if present, make the equilibrium weakly singular near the axis ($\rho=0$) \citep{weitzner2016expansions}. Perhaps the easiest way to understand the physical origin of these logarithmic terms is to think of the plasma in the large aspect ratio picture as a thin current-carrying loop \citep{jackson1999classical_ED, freidberg_idealMHD}. The logarithmic terms are then unavoidable in the poloidal flux function \citep{freidberg_idealMHD}. However, this simplified description of plasma needs to be modified near the axis to avoid the singularities on the axis. As shown in \citep{greene_Johnson_welmer_1971tokamak}, removing such weakly singular terms is tricky, even at lower orders. These issues are expected to worsen in stellarators where the currents interact with the 3D geometry. 
In this work, we develop a systematic way to avoid such singular terms within the NAE formalism, showing that the assumption of smoothness of the NAE can be satisfied order by order. 

The current work generalizes the direct NAE approach of \citep{Jorge_Sengupta_Landreman2020NAE} by including force-free and plasma-beta-driven currents, and by discussing both the structure of the flux surface and the field line label to all orders. The outline of the paper is as follows. In Section \ref{sec:Mercier_Weitzner}, we introduce the basic formalisms developed by Mercier \citep{Mercier1964,mercier1974lectures} and Weitzner \citep{weitzner2014ideal,weitzner2016expansions} and show how they can be combined into one, which we call the Mercier-Weitzner formalism. We assume the rotational transform to be sufficiently irrational. With this assumption, we can solve the MDEs order by order without encountering resonances on rational surfaces. 
We discuss the NAE of the vacuum fields in Section \ref{sec:NAE_vacuum}, force-free fields in Section \ref{sec:NAE_FF} and finally, MHD fields with finite plasma beta in Section \ref{sec:NAE_MHD}. 

Our strategy is to construct a new variable that mimics the vacuum scalar potential but fully incorporates the effects of currents. The NAE using this variable shows a structure very close to the vacuum problem. Under the assumption of smooth fields and currents near the axis, we can show that power-series solutions to ideal MHD can be constructed to all orders. We do not attempt to prove convergence of the series. 
We show that the NAE calculations are simplified to a great extent through the use of Birkhoff–Gustavson Hamiltonian normal form variables \citep{Duignan_Meiss2021normal_form,bernardin_tataronis_1985hamiltonian_NAE}. Normal forms in the present work show up as a basis that diagonalizes the magnetic differential operator, thereby simplifying the calculations of the solutions of MDEs ubiquitous in MHD. 

In Section \ref{sec:numerical}, we present a numerical example that compares the higher-order finite-beta near-axis asymptotic expansions with a finite aspect-ratio equilibrium generated using the VMEC code \citep{hirshman1983steepest}.
We present our conclusions and open questions in Section \ref{sec:discussion}. Finally, a variety of special cases are discussed in the Appendix.

\section{The Mercier-Weitzner near-axis formalism \label{sec:Mercier_Weitzner} }
Our goal in this work is to extend the direct approach to the near-axis formalism, as developed by Mercier \citep{Mercier1964} and Solov'ev-Shafranov \citep{Solovev1970}, to include finite plasma beta. Mercier's original approach \citep{mercier1974lectures} indeed considers currents and finite beta. However, the methodology to calculate beyond the two lowest orders is not transparent. \cite{Solovev1970} extends Mercier's work significantly but primarily for vacuum fields. Effects of plasma beta, such as on the flux surface shapes, are treated through expansion subsidiary to the NAE. Only recently, a complete description of NAE in direct coordinates that included both vacuum and force-free equilibria was obtained \citep{Duignan_Meiss2021normal_form}.

Here we provide a unified framework to include vacuum, force-free as well as finite beta ideal MHD. We shall use Weitzner's approach \citep{weitzner2014ideal,weitzner2016expansions} to obtain finite-beta MHD equilibrium equations. In this approach, currents are included through the current potential obtained through the solution of an MDE. In the following, we describe the NAE in direct coordinates using the current potential approach due to Weitzner.

\subsection{The Weitzner formulation of ideal MHD with scalar pressure}
\label{sec:Weitzner_approach}
We begin with the discussion of the formalism developed in \citep{weitzner2014ideal,weitzner2016expansions} to study three-dimensional (nonsymmetric) ideal MHD equilibria with scalar pressure. We assume throughout that the magnetic field $\B$ satisfies the ideal MHD equations
\begin{align}
\dl\cdot \B =0, \quad \J=\dl \times \B , \quad \JxB = \dl p,
    \label{eq:IdealMHD}
\end{align}
where $\J$ represents the current, and $p$ is the scalar pressure. We have used units such that $\mu_0$ is set to unity. We shall assume that the magnetic field possess a set of nested toroidal flux surfaces, denoted by the flux label $\psi$, which are also iso-surfaces of the pressure. Furthermore, we assume that $\dl \psi$ is nonzero everywhere except at the magnetic axis and possibly the separatrix.

To describe the magnetic field vector, we now use the following contravariant and covariant representation 
\begin{subequations}
\begin{align}
    \B &=\dl \psi \times \dl \alpha \label{eq:Basic_Clebsch}\\
    \B &= \dl \Phi + K \dl \psi.
    \label{eq:Grad_Boozer_Weitzner}
\end{align}
    \label{eq:basic_Grad_Boozer_Weitzner_rep}
\end{subequations}
Equation \eqref{eq:Basic_Clebsch} is the standard Clebsch representation \citep{Flux_coordinates,helander2014theory} of the magnetic field with $\psi$ and $\alpha$ denoting the toroidal magnetic flux and field line label, respectively. The Clebsch form manifestly satisfies $\dl\cdot \B=0$ and $\BD\psi=0$. The second representation of the magnetic field, \eqref{eq:Grad_Boozer_Weitzner}, is the so-called Boozer-Grad representation (see Chapter 7.2 of \cite{Flux_coordinates}). We call $\Phi$ the magnetic scalar potential and $K$ the current potential, since, in the limit $K\to 0$, or in the case of $K= K(\psi)$, the curl of $\B$ \eqref{eq:Grad_Boozer_Weitzner} vanishes. The quantities $\Phi, K,\alpha$ need not be single-valued in a torus. Note also that the separation of $\Phi,K$ in the representation for $\B$ in \eqref{eq:Grad_Boozer_Weitzner} is not unique since a shift of $\Phi\to \Phi + \varphi(\psi)$ results in a shift $K\to K-\varphi'(\psi)$ without changing $\B$.

For the sake of comparison, we first look at the standard representation given by
\begin{subequations}
    \begin{align}
\B&=\dl\psi\times \dl \theta_B +\iota(\psi)\dl \phi_B \times \dl \psi \label{eq:Boozer_contra}\\ 
    \B&=I_B(\psi)\dl\theta_B + G_B(\psi)\dl\phi_B + B_\psi(\psi,\theta_B,\phi_B)\dl\psi.
    \label{eq:Boozer_cov}
\end{align}
 \label{eq:Boozer_coords}
\end{subequations}
\noindent Here, $(\psi,\theta_B,\phi_B)$ represents the standard Boozer coordinates \citep{boozer1980guiding,boozer1981establishment,boozer1981plasma} with $2\pi\psi$ equal to the toroidal flux, $\iota(\psi)$ equal to the rotational transform and $(\theta_B,\Phi_B)$ represent the poloidal and toroidal Boozer angles. The quantities $I_B(\psi),G_B(\psi),$ and $B_\psi$ are all single-valued functions. The Boozer representation is an important special case of \eqref{eq:basic_Grad_Boozer_Weitzner_rep} for the choice
\begin{align}
    \Phi= I_B(\psi)\theta_B + G_B(\psi) \phi_B, \quad K= B_\psi-\lbr I_B'(\psi)\theta_B + G_B'(\psi) \phi_B \rbr, \quad \alpha=\theta_B -\iota(\psi)\theta_B
    \label{eq:connecting_Boozer_Grad_Boozer}
\end{align}
In this work, we use the Boozer-Grad coordinates instead of the usual Boozer coordinates.  

Returning to the covariant representation of the magnetic field as given in \eqref{eq:Grad_Boozer_Weitzner}, we find that the chief advantage of this form is that the current, $\J=\dl \times \B$, has a Clebsch form 
\begin{align}
    \J=\dl K\times \dl \psi,
    \label{eq:current_Clebsch_form}
\end{align}
which manifestly satisfies
\begin{align}
    \J \cdot \dl \psi=0.
    \label{eq:J_dot_dl_psi}
\end{align}
Furthermore, the ideal MHD force balance 
\begin{align}
    \JxB=\dl p,
    \label{eq:MHD_force_balance}
\end{align}
takes a particularly simple form in terms of an MDE for $K$
\begin{align}
    \BD K = p'(\psi).
    \label{eq:MDE_K}
\end{align}
The solution to \eqref{eq:MDE_K} can be represented as 
\begin{align}
    K= \Kb(\psi,\alpha)+ G, \quad \BD \Kb =0, \quad \BD G = p'.
\end{align}
where $ \Kb(\psi,\alpha)$ is the homogeneous solution of the MDE and hence only a function of $\psi$ and possibly $\alpha$.

Let us collect the basic equations in the Weitzner formalism. These are obtained by equating the contravariant and the covariant representations of $\B$ as given in \eqref{eq:basic_Grad_Boozer_Weitzner_rep}, together with \eqref{eq:MDE_K},
\begin{subequations}
    \begin{align}
    \B =\dl \psi \times \dl \alpha &= \dl \Phi + K \dl \psi \label{eq:cov_is_contra_Weitzner}\\
     \BD K &= p'(\psi). \label{eq:force_bal_Weitzner}
\end{align}
\label{eq:basic_weitzner_eqns}
\end{subequations}
Note that \eqref{eq:cov_is_contra_Weitzner}, which we will call the basic MHD equations, is a set of three coupled nonlinear PDEs, while \eqref{eq:force_bal_Weitzner} determines $K$. Together, they form a complete set of equations for the variables  $(\Phi,\psi,\alpha,K)$. We shall now discuss the boundary conditions that must be satisfied such that \eqref{eq:basic_weitzner_eqns} has physically meaningful solutions in a torus.

In a torus, all physical quantities, such as magnetic fields and currents, must satisfy periodic boundary conditions in the two angles $(\theta,\phi)$, where $\theta$ is a poloidal angle, and $\phi$ is a toroidal angle. Unlike standard Boozer coordinates $(\theta_B,\phi_B)$, we do not require $(\theta,\phi)$ to be straight field line coordinates. This provides greater flexibility and freedom since we do not have any additional constraints on the angles. As mentioned, the functions $(\alpha,\Phi,K)$ in a toroidal domain are multi-valued. However, the multi-valued part of the $\{\Phi,\alpha,K\}$ system is not arbitrary due to the physical requirement that the magnetic field and the current must be single-valued quantities.

Following \citep{weitzner2016expansions,grad1971plasma} we observe that in order for $\B,\J$ (as given by \eqref{eq:Basic_Clebsch} and \eqref{eq:current_Clebsch_form}) to be single-valued, only the $\dl\psi$ covariant component of $\dl \alpha, \dl K$ can be multi-valued. Furthermore, from \eqref{eq:Grad_Boozer_Weitzner} we find that the secular terms of $\Phi$ and $K$ must be related to each other such that $\B$ is single-valued.

As shown in \citep{weitzner2014ideal}, single-valuedness of $\B,\J$ in a torus implies that $\Phi,\alpha$ and $K$ must be of the form
\begin{subequations}
    \begin{align}
    \Phi(\psi,\theta,\phi)&= I(\psi)\theta + G(\psi)\phi + \widetilde{\Phi}(\psi,\theta,\phi) \label{eq:sec_n_per_Phi}\\
    \alpha(\psi,\theta,\phi)&= f(\psi)(\theta - \iota(\psi)\phi) + \widetilde{\alpha}(\psi,\theta,\phi) \label{eq:sec_n_per_alpha}\\
    K(\psi,\theta,\phi)&=-I'(\psi) \theta -G'(\psi) \phi +\widetilde{K}(\psi,\theta,\phi), \label{eq:sec_n_per_K}
\end{align}
\label{eq:secular_n_periodic_Phi_alpha_K}
\end{subequations}
where, $(\widetilde{\Phi},\widetilde{\alpha},\widetilde{K})$ denotes functions periodic in both $\theta$ and $\phi$, and $\iota(\psi)$ denotes the rotational transform of the magnetic field. The form \eqref{eq:secular_n_periodic_Phi_alpha_K} builds in the nestedness of flux-surfaces. The single-valued flux functions $I(\psi),G(\psi),\iota(\psi)$ denote the same quantities as in standard Boozer coordinates \eqref{eq:connecting_Boozer_Grad_Boozer}. We briefly discuss the physical significance of these flux functions below.

Without loss of generality, we can choose $2\pi \psi$ to be the toroidal flux, as is common in stellarator literature. As shown in Appendix \ref{app:flux_choice}, with this choice, we can normalize $\alpha$ such that $f(\psi)=1$. The physical interpretation of $I(\psi),G(\psi)$ is obtained by considering the net current, $\oint \J\cdot \bm{dS}$, through a poloidal and a toroidal circuit. From \eqref{eq:current_Clebsch_form} and \eqref{eq:sec_n_per_K} it follows that $G'(\psi),I'(\psi)$ are proportional to the net poloidal and toroidal current. In the vacuum limit ($K=0,p'=0$), the net toroidal current is zero, while the net poloidal current is a constant due to external currents. Consequently, for vacuum fields, we have  $I(\psi)=0$ and $G(\psi)=G_0$ is a constant. 

Equations \eqref{eq:basic_weitzner_eqns}, subject to the conditions \eqref{eq:secular_n_periodic_Phi_alpha_K}, constitute the Weitzner formalism. The MHD equations  \eqref{eq:cov_is_contra_Weitzner}, \eqref{eq:force_bal_Weitzner} are highly nonlinear, and in a generic stellarator, analytical progress is severely hindered by lack of any apparent continuous symmetry. Furthermore, the periodicity constraint \eqref{eq:secular_n_periodic_Phi_alpha_K} imposes a nontrivial requirement. In order to gain valuable physical insight, one, therefore, resorts to asymptotic analysis in some small parameter.  

\subsection{Mercier's near-axis expansion}
\label{sec:Mercier_NAE}
Mercier developed one such asymptotic expansion scheme to study the behavior near the magnetic axis by utilizing the distance from the magnetic axis as the expansion parameter. In the following, we shall briefly describe Mercier's near-axis formalism and Mercier's coordinates \citep{Mercier1964}, which are also known as the direct coordinates \citep{Jorge_Sengupta_Landreman2020NAE}. 

\subsubsection{Mercier coordinates}
\label{sec:Mercier_coords}
The Mercier coordinates are defined with respect to the magnetic axis, which is a closed magnetic field line. Although the magnetic axis can be elliptic or hyperbolic \citep{Solovev1970,Duignan_Meiss2021normal_form}, we will only treat the elliptic case here. We will assume that the magnetic axis is a three-dimensional smooth closed curve of total length $L$, described by the Frenet-Serret frame (which we assume to exist)
\begin{align}
\dfrac{d\bm{r}_0}{d\ell}=\bm{t},\quad  \dfrac{d\bm{t}}{d\ell}=\kappa\bm{n}, \quad \dfrac{d\bm{n}}{d\ell}=\tau\bm{b}-\kappa \bm{t}, \quad
\dfrac{d\bm{b}}{d\ell}=-\tau\bm{n},
\label{eq:frenet-serret}
\end{align}
where $\bm{r}_0$ is the position vector of the magnetic axis, $\ell$ is the arclength along the axis, the functions $\kappa(\ell),\tau(\ell)$ are the curvature and torsion of the axis, and the vectors $(\bm{t},\bm{n},\bm{b})$ are the standard tangent, normal and binormal vectors in the Frenet-Serret frame. Now, let us consider a tube of radius $\rho$ with the magnetic axis as its axis. We can define a poloidal angle $\theta$ such that a point on the tube is described by
\begin{align}
    \bm{r}=\bm{r}_0 + \bm{\rho}, \quad  \bm{\rho}= x\: \bm{n}(\ell) + y \:\bm{b}(\ell), \quad x=\rho \cos \theta\:,y=\rho \sin{\theta}.
\end{align}
The coordinate system $(\rho,\theta,\ell)$ is not orthogonal when the torsion of the axis, $\tau(\ell)$, is nonzero. To obtain an orthogonal coordinate system, \cite{Mercier1964} replaced $\theta$ by
\begin{align}
     \omega =\theta +\int\tau d\ell.
     \label{eq:Mercier_omega_def}
\end{align}
The line and volume elements, $(ds^2,dV)$, of the Mercier coordinates are as follows:
\begin{align}
 ds^2\equiv |d\bm{r}|^2=(d\rho)^2+(\rho d\omega)^2+(h\: d\ell)^2, \quad dV=\rho h\: d\rho\: d\omega\: d\ell, \quad h=1-\kappa \rho \cos{\theta}.
    \label{eq:Mercier_ds2_dV_h_def}
\end{align}
In other words, the $(\rho,\omega,\ell)$ coordinates are orthogonal with $(1,\rho,h)$ as the scale factors, and the metric tensor is given by
\begin{align}
    g_{ij}=\text{diag}\lbr 1,\rho^2,h^2 \rbr, \qquad \sqrt{g}\equiv\sqrt{\text{Det}(g_{ij})}=\rho h.
\end{align}
Associated with the Mercier coordinates are the following orthonormal vectors $(\rhoh,\wh,\lh)$ such that
\begin{align}
    \rhoh\equiv \dl \rho= \cos{\theta}\, \bm{n}+\sin{\theta}\: \bm{b}, \quad  \wh\equiv \rho \dl \omega= -\sin{\theta}\,\bm{n}+\cos{\theta}\,\bm{b}, \quad \lh\equiv h \dl \ell=\bm{t},
    \label{eq:orthonormal_rhoh_wh_lh}
\end{align}

There is a certain freedom in choosing the poloidal and toroidal angles in the direct coordinates $(\rho,\omega,\ell)$. While the toroidal angle follows from the arclength $\ell$, one can generally add any arbitrary function $\delta(\ell)$, where $\delta'(\ell)$ is single-valued, to $\theta$ to obtain another poloidal angle without changing the Jacobian of transformation. Besides $(\theta,\phi=2\pi \ell/L)$ another possible choice is $(u,\phi)$, where
\begin{align}
    u=\theta+\delta(\ell)= \omega-\int \tau d\ell +\delta(\ell) ,\quad \phi = 2\pi\frac{\ell}{L}.
    \label{eq:u_definition}
\end{align}

Let us now describe the Weitzner system described in Section \ref{sec:Weitzner_approach} using the Mercier coordinates. To connect to the contravariant form of $\B$ \eqref{eq:Basic_Clebsch} in the Weitzner formalism, we shall make a choice $(\theta,\phi)$ for defining angular coordinates. To connect to the covariant form of $\B$ given by \eqref{eq:Grad_Boozer_Weitzner} in the Weitzner formalism, we shall utilize the orthonormal vectors $(\rhoh,\wh,\lh)$. Using \eqref{eq:orthonormal_rhoh_wh_lh}, we can write the magnetic field in the following component form:
\begin{align}
    &\B=B_\rho \rhoh+B_\omega \wh+ B_\ell \lh,\\
    &B_\rho= \Phi_{,\rho}+K \psi_{,\rho}, \quad B_\omega= \frac{1}{\rho}\lbr \Phi_{,\omega}+K \psi_{,\omega}\rbr, \quad B_\ell= \frac{1}{h}\lbr \Phi_{,\ell}+K \psi_{,\ell}\rbr.
    \label{eq:B_and_its_components}
\end{align}
Here and elsewhere, we shall use the notation $X_{,\ell}$ to denote the partial derivative of $X$ with respect to $\ell$ holding the other two Mercier coordinates $(\rho,\omega)$ fixed.
 
   
\subsubsection{Near-axis expansions }   
\label{sec:NAE}
The fundamental idea underlying the NAE is to solve the MHD equations using perturbation theory, where the expansion parameter scales like the distance from the magnetic axis, $\rho$. We can formally define the dimensionless expansion parameter $\ep$ as
\begin{align}
    \ep=\kappa_{\text{max}}a,
    \label{eq:exp_parameter_epsilon_def}
\end{align}
where $\kappa_{\text{max}}$ is the maximum axis curvature and $0\leq \rho < a$. With this definition, we have $\kappa \rho<1$ throughout in the domain of validity of the NAE, which is necessary for the metric $\sqrt{g}=\rho h$ to be positive everywhere. Expanding the quantities $(\psi,\alpha,\Phi,K)$ in $\epsilon$ and substituting them in the basic equilibrium equations \eqref{eq:basic_weitzner_eqns}, one can solve the MHD equations order by order in $\epsilon$. 

It is usually assumed \citep{kuo_Boozer1987numerical,garrenBoozer1991existence} that the physical quantities are sufficiently regular near the magnetic axis such that one can carry out a regular power series expansion in positive powers of $\rho$. Thus, any function of the form $\mathfrak{f}(\rho,\omega,\ell)$ is expanded as follows
\begin{align}
    \mathfrak{f}(\rho,\omega,\ell)=\sum_{n=0}^\infty (\ep \rho)^n\: \mathfrak{f}^{(n)}(\omega,\ell)
    \label{eq:NAE_power_series_gen}
\end{align}
The function $\mathfrak{f}^{(n)}(\omega,\ell)$ can be determined order by order by substituting \eqref{eq:NAE_power_series_gen} into the MHD equations \eqref{eq:basic_weitzner_eqns}. Moreover, if  $\mathfrak{f}$ has a well-defined Taylor series near the magnetic axis, we would expect $\mathfrak{f}^{(n)}$ to be a polynomial of order $n$ in $(x,y)=\rho(\cos\theta,\sin\theta)$. In \cite{Jorge_Sengupta_Landreman2020NAE,Duignan_Meiss2021normal_form}, it has been demonstrated that for vacuum fields with nested surfaces, the functions $(\Phi-G_0 \phi),\psi$ are of such form. Therefore, for such regular functions, we have
\begin{align}
    \mathfrak{f}(x,y,\ell)&=\sum_{n=0}^\infty \ep^n\: \sum_{m=0}^n \mathfrak{c}^{(n)}(\ell) x^m y^{n-m} \label{eq:NAE_power_series_analytic}\\
    &= \sum_{n=0}^\infty (\ep\rho)^n\:\sum_{m=0}^n\lbr \mathfrak{f}^{(n)}_{mc}(\ell) \cos{\lbr m \omega +\delta_m(\ell)\rbr}+\mathfrak{f}^{(n)}_{ms}(\ell) \sin{\lbr m \omega +\delta_m(\ell)\rbr}\rbr,
   \nonumber
\end{align}
where $\lbr \mathfrak{c}^{(n)}(\ell),\mathfrak{f}^{(n)}_{mc}(\ell),\mathfrak{f}^{(n)}_{ms}(\ell),\delta_m(\ell) \rbr$ are functions of $\ell$ that must be determined by substituting \eqref{eq:NAE_power_series_analytic} into \eqref{eq:basic_weitzner_eqns}. We note that even though the series \eqref{eq:NAE_power_series_analytic} looks like a regular Taylor series for $\mathfrak{f}(x,y,\ell)$ near the axis, the series may be divergent. The NAE, like many other perturbative expansions, is written formally here, with no claims regarding the convergence of the series.

We shall substitute the formal series \eqref{eq:NAE_power_series_analytic} into the basic MHD equations \eqref{eq:basic_weitzner_eqns} and collect powers of $\rho$ and the poloidal harmonics $(\cos{m \ell},\sin{m \ell})$. This then reduces the nonlinear PDE system \eqref{eq:basic_weitzner_eqns} to a nonlinear ordinary differential equation (ODE) system for $(\mathfrak{f}^{(n)}_{mc}(\ell),\mathfrak{f}^{(n)}_{ms}(\ell))$, which is a significant step forward. Moreover, \eqref{eq:NAE_power_series_analytic} already includes periodicity in $\omega$, and only the periodicity in $\ell$ remains to be imposed on the ODEs. 

Before proceeding further, we should point out some issues that can lead to a lack of regularity of the NAE. Firstly, although the periodicity in $\ell$ can always be satisfied if the on-axis rotational transform is sufficiently irrational, the coefficients $(\mathfrak{f}^{(n)}_{mc}(\ell),\mathfrak{f}^{(n)}_{ms}(\ell))$ can formally exhibit singular behavior when the on-axis rotational transform is close to a rational number. As shown in \citep{Mercier1964,Solovev1970}, the form of the singularities on a $(m,n)$ rational surface is $(\iota_0-n/m)^{-1}$, where $\iota_0$ is the on-axis rotational transform.  In this work, we shall assume that $\iota_0$ is sufficiently irrational to formally avoid such resonances in the vicinity of the axis. Secondly, 
we note that the regular power series expansion \eqref{eq:NAE_power_series_gen} is not always guaranteed to be correct. The possibility of having weak logarithmic singularities on the axis in the form of $\rho^n (\log\rho)^m, n>m>0$ has been pointed out in \cite{weitzner2016expansions}. Furthermore, functions such as $\alpha$ can not be represented in the analytic power series given by \eqref{eq:NAE_power_series_analytic}. In this work, we demonstrate that for a large class of equilibria, one can still construct a power series in $\rho$ valid to all orders in $\rho$ without encountering logarithmically singular terms. We also pay special attention to the multi-valued function $\alpha, K$ and discuss their structure. 

In the following, we discuss various limits of ideal MHD: vacuum, force-free, and finally, finite beta MHD equilibrium. We start with the vacuum limit because its mathematical structure is fundamental and the easiest to analyze. Subsequently, we will show that the force-free and MHD fields can be formulated in a form very similar to the vacuum fields.

\section{NAE of vacuum fields in Mercier-Weitzner formalism \label{sec:NAE_vacuum}}
NAE of vacuum fields in direct coordinates has been treated extensively in \citep{Jorge_Sengupta_Landreman2020NAE}. In the Mercier-Weitzner formalism, $K=0$ corresponds to the vacuum limit, and the analysis is similar to \cite{Jorge_Sengupta_Landreman2020NAE}, which also uses direct coordinates. However, a key point of departure of our present work from all other previous developments of NAE is the treatment of the Clebsch variable, which represents the field line label, $\alpha$. The primary motivation for focusing on $\alpha$ is that the rotational transform and its derivatives ( which includes the magnetic shear) are related directly to the secular part of $\alpha$ ( through \eqref{eq:sec_n_per_alpha}). Thus, a systematic construction of $\alpha$ is crucial.

Our goal in this Section is to study the structure of the Mercier-Weitzner equations in the vacuum limit. First, we show how the condition of nested flux surfaces leads to MDEs. We then explicitly employ the NAE and elucidate the properties of the solutions obtained, paying particular attention to the secular and periodic structure of the field line label, $\alpha$.

\subsection{Basic equations and need for an alternative set of equations }
\label{sec:Basic_vacuum_eqn}
In the vacuum limit, the basic equilibrium equations \eqref{eq:basic_weitzner_eqns} reduces to
\begin{align}
    \dl \psi\times \dl \alpha = \dl \Phi  \label{eq:vacuum_limit}.
\end{align}
Projecting \eqref{eq:vacuum_limit} onto the orthonormal vectors $(\rhoh,\wh,\lh)$ defined in \eqref{eq:orthonormal_rhoh_wh_lh}, we get
\begin{subequations}
\begin{align}
\lbr \psi_{,\omega}\alpha_{,\ell}- \psi_{,\ell}\alpha_{,\omega}\rbr  &=  \rho h\:\Phi_{,\rho}
    \label{eq:vacuum_Phi_rho}\\
 \left( \psi_{,\ell}\alpha_{,\rho}- \psi_{,\rho}\alpha_{,\ell}\right)  &=  \frac{h}{\rho}\Phi_{,\omega} \label{eq:vacuum_Phi_omega}
    \\
  \lbr \psi_{,\rho}\alpha_{,\omega}- \psi_{,\omega}\alpha_{,\rho}\rbr  &= \frac{\rho }{h} \Phi_{,\ell}
    \label{eq:vacuum_Phi_ell}
\end{align}
\label{eq:vacuum_system}
\end{subequations}
Equation \eqref{eq:vacuum_system} is a closed set of equations that, in principle, can be solved for the three unknowns $(\Phi,\psi,\alpha)$. However, the coupling between the various quantities makes the system \eqref{eq:vacuum_system} difficult to tackle analytically in the direct coordinate approach to NAE. Therefore, instead of \eqref{eq:vacuum_system}, we examine alternative equations that decouple $(\Phi,\psi,\alpha)$ as much as possible.

To arrive at an alternative set of equations, we now look at the consistency conditions that directly follow from \eqref{eq:vacuum_system}. To derive the first consistency condition, 
we differentiate \eqref{eq:vacuum_Phi_omega} and \eqref{eq:vacuum_Phi_ell} with respect to $\omega$ and $\ell$ respectively and add them. Using the commutativity of the cross derivatives for the functions $\psi$ and $\alpha$, we get a consistency condition
\begin{align}
 -\del_\rho\lbr \psi_{,\omega}\alpha_{,\ell}- \psi_{,\ell}\alpha_{,\omega}\rbr  = \del_\omega \lbr  \frac{h}{\rho}\Phi_{,\omega} \rbr+ \del_\ell \lbr \frac{\rho }{h} \Phi_{,\ell}\rbr
 \label{eq:pre_laplacian}
\end{align}
It follows from \eqref{eq:vacuum_Phi_rho} that \eqref{eq:pre_laplacian} is nothing but the Laplace equation for $\Phi$,
\begin{align}
    \Delta \Phi = \frac{1}{\rho h}\frac{\del}{\del \rho }\left(\rho h \frac{\del \Phi}{\del \rho}\right) +\frac{1}{\rho^2 h}\frac{\del}{\del \omega }\left( h \frac{\del \Phi}{\del \omega}\right)+ \frac{1}{ h}\frac{\del}{\del \ell }\left( \frac{1}{h} \frac{\del \Phi}{\del \ell}\right) =0,
    \label{eq:Laplace_for_Phi}
\end{align}
which also follows from $\dl \cdot \B = \dl \cdot \dl \Phi=0$.

Two other consistency conditions directly follow from \eqref{eq:vacuum_limit} and are given by 
\begin{subequations}
\begin{align}
    \BD \psi =\dl \Phi\cdot \dl \psi=0 \label{eq:vacuum_characteristic_psi}\\
    \BD \alpha =\dl \Phi\cdot \dl \alpha=0,
    \label{eq:vacuum_characteristic_alpha}
\end{align}
\label{eq:vacuum_characteristics}
\end{subequations}
One might suppose that the complete set \eqref{eq:vacuum_system} could be replaced by the new set of equations \eqref{eq:Laplace_for_Phi} and \eqref{eq:vacuum_characteristics}, which are the MDEs for $\psi$ and $\alpha$. The Laplace equation only involves $\Phi$ and the scale factors and is ideally suited for determining $\Phi$. Once $\Phi$ is known, we might determine the Clebsch variables $(\psi,\alpha)$ through \eqref{eq:vacuum_characteristic_psi} and \eqref{eq:vacuum_characteristic_alpha}. The benefit of using the characteristic equations, \eqref{eq:vacuum_characteristics}, is that $\psi$ and $\alpha$ can be decoupled. However, several mathematical intricacies need to be addressed before replacing the complete set \eqref{eq:vacuum_system}, which we now highlight. 

Firstly, \eqref{eq:vacuum_system} is a third-order system because it has three first-order derivatives in $(\rho,\omega,\ell)$, while the Laplace equation is second order and so is \eqref{eq:vacuum_characteristics} because of the two first derivatives. Therefore, the replacement of \eqref{eq:vacuum_system} by \eqref{eq:Laplace_for_Phi} and \eqref{eq:vacuum_characteristics} raises the order by one since we have taken an additional derivative of the equations, which might lead to the inclusion of spurious solutions. Furthermore, the conditions \eqref{eq:Laplace_for_Phi},\eqref{eq:vacuum_characteristic_psi} and \eqref{eq:vacuum_characteristic_alpha} are not completely independent when the equations are solved order by order. To see this, let us assume that the equations that govern the angular derivatives of $\Phi$, \eqref{eq:vacuum_Phi_omega} and \eqref{eq:vacuum_Phi_ell}, have been solved. It is then straightforward to see that the radial derivative of $\Phi$ given by \eqref{eq:vacuum_Phi_rho} and characteristic equations \eqref{eq:vacuum_characteristic_psi} and \eqref{eq:vacuum_characteristic_alpha} are not independent but are related through
\begin{align}
    \Phi_{,\rho}- \frac{1}{\rho h}\lbr \psi_{,\omega}\alpha_{,\ell}- \psi_{,\ell}\alpha_{,\omega}\rbr=\frac{1}{\psi_{,\rho}}\BD \psi =\frac{1}{\alpha_{,\rho}}\BD \alpha
    \label{eq:connecting_Phi_rho_eqn_BDpsi_BDalpha}.
\end{align}
As discussed in Appendix \ref{app:MHD_characteristics}, in order to retain the correct vacuum and MHD characteristics \citep{goedbloed1983lecture_MHD,courant_Hilbert_vol2_2008,KOfriedrichs1958nonlinearMHD,weitzner2014ideal} we cannot simply replace the complete set \eqref{eq:vacuum_system} by the consistency conditions \eqref{eq:Laplace_for_Phi}, \eqref{eq:vacuum_characteristic_psi} and \eqref{eq:vacuum_characteristic_alpha}. Fortunately, a workaround is possible that allows a partial decoupling of the variables $\Phi,\psi,\alpha$. In the following Sections, we will provide a recipe to carry out the finite-beta NAE efficiently and self-consistently to arbitrary orders by choosing the right set of fundamental equations. We will justify all the necessary mathematical steps below. 

\subsection{Alternative set of basic equations }
\label{sec:Alt_Basic_vacuum_eqn}
We first observe that the consistency condition \eqref{eq:pre_laplacian} implies that if the $\omega$ and $\ell$ components of the basic vacuum equations \eqref{eq:vacuum_limit}, i.e., \eqref{eq:vacuum_Phi_omega} and \eqref{eq:vacuum_Phi_ell}, are satisfied, the solution of the Laplace equation satisfies the $\rho$ component, \eqref{eq:vacuum_Phi_rho}, up to an arbitrary function of $(\omega,\ell)$. Therefore, \eqref{eq:vacuum_Phi_rho} needs to be imposed to eliminate this arbitrariness. As can be seen from \eqref{eq:connecting_Phi_rho_eqn_BDpsi_BDalpha} (and further discussed later on), the equation for the radial derivative of $\Phi$ \eqref{eq:vacuum_Phi_rho} is identical to the MDE for $\psi$ \eqref{eq:vacuum_characteristic_psi} order by order in $\rho$. Furthermore, it will be shown later that once  \eqref{eq:Laplace_for_Phi} and \eqref{eq:vacuum_characteristic_psi} (equivalently \eqref{eq:vacuum_Phi_rho}) are satisfied, the equations \eqref{eq:vacuum_Phi_omega} and \eqref{eq:vacuum_Phi_ell} are compatible, in the sense that either can be used to solve for $\alpha$. We will find it more straightforward to use \eqref{eq:vacuum_Phi_omega} to solve for $\alpha$ up to a function of $\ell$ by integrating with respect to $\omega$. The reason behind choosing  \eqref{eq:vacuum_Phi_omega} over \eqref{eq:vacuum_Phi_ell} is that the $\Phi$ and $\psi$ can be shown to have a finite number of poloidal harmonics which depend only on the order of expansion, whereas no such restriction exists for the toroidal harmonics. 

Finally, to determine $\alpha$ fully, we need only the poloidal average of either \eqref{eq:vacuum_Phi_ell} or \eqref{eq:vacuum_characteristic_alpha}. We use the latter because of the relevance of the MDE to calculating rotational transform and its higher derivatives. Note that the only MDE we solve is the MDE for $\psi$, \eqref{eq:vacuum_characteristic_psi}. We are not solving the MDE for $\alpha$; we are simply fixing the $\ell$ dependent part of $\alpha$ that was left undetermined because of the partial integration with respect to $\omega$. The procedure will be explained in detail in Sections \ref{eq:vacuum_lowest_orders} through \ref{sec:periodicity_n_structure_soln}.

In summary, we have seen that equating the covariant and the contravariant representations of $\B$, \eqref{eq:basic_Grad_Boozer_Weitzner_rep}, in the vacuum limit leads to a complete set of equations \eqref{eq:vacuum_system}. However, the variables $(\Phi,\psi,\alpha)$ are coupled nonlinearly, making 
\eqref{eq:vacuum_system}, not the best set of equations to start the NAE program in direct coordinates. The alternative is to solve the Laplace equation for $\Phi$ \eqref{eq:Laplace_for_Phi} and the MDE for $\psi$ \eqref{eq:vacuum_characteristic_psi}. We 
use the $\Phi_{,\omega}$ equation, \eqref{eq:vacuum_Phi_omega}, to obtain $\alpha$ up to a function of $\ell$, which is then determined from the poloidally averaged MDE for $\alpha$ \eqref{eq:vacuum_characteristic_alpha}. The Laplace equation does not introduce extraneous solutions because we are solving the MDE for $\psi$, which is equivalent to solving \eqref{eq:vacuum_Phi_rho}, as can be seen from \eqref{eq:connecting_Phi_rho_eqn_BDpsi_BDalpha}. Furthermore, the Laplace equation is also the necessary solvability condition for the function $\alpha$.


\subsection{Derivation of the vacuum NAE equations: The lowest order}
\label{eq:vacuum_lowest_orders}

We now expand the quantities $(\Phi,\psi,\alpha)$ in power series of $\rho$ in the general form
\begin{subequations}
 \begin{align}
        \Phi(\rho,\omega,\ell)&=\Phi^{(0)}(\omega,\ell)+\rho\; \Phi^{(1)}(\omega,\ell)+\rho^2 \Phi^{(2)}(\omega,\ell)+\rho^3 \Phi^{(3)}(\omega,\ell)+... \label{eq:Phi_NAE}\\
     \alpha(\rho,\omega,\ell)&=\alpha^{(0)}(\omega,\ell)+\rho\; \alpha^{(1)}(\omega,\ell)+\rho^2 \alpha^{(2)}(\omega,\ell)+\rho^3 \alpha^{(3)}(\omega,\ell)+...\label{eq:alpha_NAE}\\
     \psi(\rho,\omega,\ell)&=\rho^2 \psi^{(2)}(\omega,\ell)+\rho^3 \psi^{(3)}(\omega,\ell)+\rho^4 \psi^{(4)}(\omega,\ell)...\label{eq:psi_NAE}
    \end{align}
    \label{eq:NAE_power_series}
\end{subequations}
The analysis of the lowest-order quantities is important for the subsequent development of the analysis. Therefore, we now discuss in some detail the lowest-order structure. Firstly, note that the expansion of $\psi$ starts at $\rho^2$ since we need $\psi$ to be zero on the axis. We also need $\psi_{,\rho}$ to vanish on the axis to ensure that the poloidal field, $B_\omega=(1/\rho)\Phi_\omega\approx \psi_{,\rho}\alpha_{,\ell}$, is also zero on the magnetic axis. Furthermore, the vanishing of the poloidal field on the axis also leads to the $O(1/\rho^2)$ and $ O(1/\rho)$ of the vacuum system \eqref{eq:vacuum_Phi_ell} to be satisfied provided
\begin{align}
    \PhiO(\omega,\ell)=\Phi_0(\ell),\quad \quad \PhiOne(\omega,\ell)=0.
    \label{eq:Phi0_and_Phi1}
\end{align}
We assume that the strength of the magnetic field on the axis is positive and single-valued, i.e.,
\begin{align}
    B_0(\ell)=\Phi_0'(\ell) >0, \quad B_0(\ell +L)=B_0(\ell).
\end{align}
To make $B_0(\ell)$ single-valued $\PhiO$ must be of the form
\begin{align}
    \PhiO(\ell)=\int B_0(\ell) d\ell = \overline{B}_0 \ell + \widetilde{\Phi}_0(\ell),\qquad \overline{B}_0=\frac{1}{L}\int_0^L B_0(\ell)d\ell,
    \label{eq:form_Phi0}
\end{align}
where, $\widetilde{\Phi}_0(\ell)$ is periodic in $\ell$.

Expanding the vacuum system \eqref{eq:vacuum_system} to $O(1)$, we obtain the following coupled equations for $(\PhiTwo,\psiTwo,\alphaO)$ 
\begin{subequations}
\begin{align}
    2\PhiTwo &=\alphaOcz \psiTwo_{,\omega}-\alphaOcw \psiTwo_{,\ell} \label{eq:Phi2_PB_alpha0_psi2_eqn}\\
    2\psiTwo \alphaOcw &= \Phi_0'(\ell) \label{eq:alpha0cw}\\
    2\psiTwo \alphaOcz &= -\PhiTwo_{,\omega} \label{eq:alpha0cz},
\end{align}
\label{eq:Phi2_alpha0_psi2_system}
\end{subequations}
Multiplying \eqref{eq:Phi2_PB_alpha0_psi2_eqn} by $2\psiTwo$ and using (\ref{eq:alpha0cw},\ref{eq:alpha0cz}) we get the following linear homogeneous equation for $\psiTwo$:
\begin{align}
 \lbr \Phi_0'(\ell)\del_\ell + \PhiTwo_{,\omega} \del_\omega+4 \PhiTwo \rbr\psiTwo=0
    \label{eq:lin_eqn_psi2}
\end{align}
To solve \eqref{eq:lin_eqn_psi2} we need $\PhiTwo$. The compatibility condition $\del_\ell\alphaOcw=\del_\omega\alphaOcz$ with the use of (\ref{eq:alpha0cw},\ref{eq:alpha0cz}) leads to the following Poisson equation for $\PhiTwo$: 
\begin{align}
   ( \del^2_{\omega}+2^2)\PhiTwo+\Phi''_0=0.
   \label{eq:Phi2_laplacian}
\end{align}
Since \eqref{eq:Phi2_laplacian} is a simple harmonic oscillator, its solution can be easily seen to be of the form
\begin{align}
    \PhiTwo= -\frac{1}{4}\Phi''_0(\ell) + \Phi_0'(\ell) \lbr \fTwocTwo(\ell)\cos{(2u(\omega,\ell))} +\fTwosTwo(\ell)\sin{(2u(\omega,\ell))} \rbr,
    \label{eq:phi2_expression}
\end{align}
where, $\fTwocTwo(\ell),\fTwosTwo(\ell)$ are the ``integration constants" with respect to the $\del_\omega$ operator. We have also chosen $u$, defined by \eqref{eq:u_definition}, as the helical angle. Note that from the form of $\PhiO$ as given in \eqref{eq:form_Phi0}, $\PhiTwo$ can be made single-valued in $\ell$ by ensuring that the integration constants are periodic.
The function $\PhiTwo_{,\omega}$ obtained from \eqref{eq:phi2_expression} is harmonic with zero net poloidal average. 

The solution of the equation for $\psiTwo$, \eqref{eq:lin_eqn_psi2}, is of the form
\begin{align}
    \psiTwo= \Phi_0'(\ell)(a(\ell)+b(\ell)\cos{(2 u)}),
    \label{eq:psi2_expression}
\end{align}
which can be checked by substituting \eqref{eq:psi2_expression} in \eqref{eq:lin_eqn_psi2} and collecting the poloidal harmonics $1,\cos{2u},\sin{2u}$. Equating the coefficients of the various poloidal harmonics to zero, we get
\begin{align}
    a(\ell)=c_0 \cosh{\eta(\ell)}, \quad b(\ell)=c_0 \sinh{\eta(\ell)},
    \label{eq:a_b_definitions}
\end{align}
together with
\begin{align}
    \fTwocTwo(\ell)= -\frac{b'(\ell)}{4 a(\ell)}=-\frac{1}{4}\eta'(\ell),\quad \fTwosTwo(\ell)= \frac{b(\ell)}{2a(\ell)}\lbr \delta'(\ell) -\tau(\ell)\rbr.
    \label{eq:f2c2_f2s2_definitions}
\end{align}
Here $\eta(\ell)$ is an arbitrary function of $\ell$. We require that 
$\lbr \fTwocTwo(\ell),\fTwosTwo(\ell) \rbr$ are single-valued functions by requiring $\eta(\ell),\delta'(\ell)$ to be single-valued and periodic in $\ell$. With this choice, we make both $\PhiTwo, \psiTwo$ single-valued functions. Later, we determine the constant $c_0$ by imposing the choice of $\psi$ as the toroidal flux. To see that this is indeed the only solution, we can divide the equations for the $\omega,\ell$ derivatives of $\alphaO$, \eqref{eq:alpha0cw} and \eqref{eq:alpha0cz}, by $2\psiTwo$ and eliminate $\alphaO$ through cross-differentiation. The resultant consistency condition
\begin{align}
    \del_\ell \lbr \frac{\Phi_0'}{\psiTwo}\rbr+ \del_\omega \lbr \frac{\PhiTwo_{,\omega}}{\psiTwo}\rbr=0,
    \label{eq:for_1bypsi2}
\end{align}
is a linear PDE for $(1/\psiTwo)$ with periodic boundary conditions in $\omega,\ell$. Since the expression for $\psiTwo$ satisfies \eqref{eq:for_1bypsi2} and the periodic boundary conditions, it must be the unique solution.

For the physical interpretation of the various functions $\eta(\ell),\delta(\ell)$, we refer to the excellent review article \citep{helander2014theory}. In short, the flux-surface shape obtained from the equation for $\psiTwo$ \eqref{eq:psi2_expression} is a rotating ellipse. The function $\eta(\ell)$ controls the ellipticity, while $\delta(\ell)$ measures its angle of rotation respect to the Frenet frame. We note that the choices of the free-functions in $\fTwocTwo,\fTwosTwo$ in \eqref{eq:f2c2_f2s2_definitions} were made in terms of the shaping parameters of the lowest order flux-surface. Later, we make choices such that the interpretation of higher-order free-functions $\Phi^{(n+2)}_{c},\Phi^{(n+2)}_{s}$ can be made in terms of higher-order flux surface shapes $\psi^{(n+2)}_{c},\psi^{(n+2)}_{s}$.

It is possible to find special coordinates such that $\psiTwo$ is a circle in these coordinates. Since $\Phi_0'=B_0>0$ and $a(\ell)>b(\ell)\geq 0$, it follows that $\psiTwo$ is non-vanishing. As a result, we can define ``rotating coordinates" \citep{Solovev1970} or ``normal form" coordinates
\begin{align}
    X_{\cN}= e^{\eta/2}\rho \cos u , \quad  Y_{\cN}= e^{-\eta/2}\rho \sin u,
    \label{eq:Rotating_coords}
\end{align}
such that $\psiTwo$, given by
\begin{align}
    \psiTwo \rho^2= \frac{1}{2}\lbr X^2_{\cN}+Y_{\cN}^2\rbr \Phi_0',
\label{eq:psi2_normal_form}
\end{align}
is a circle in the $(x_\cN\equiv X_\cN/\rho,y_\cN\equiv Y_\cN/\rho)$ coordinates. The benefits of using the ``normal form" coordinates have been noted in \citep{Duignan_Meiss2021normal_form}. As shown in Appendix \ref{app:general_first_order_vacuum}, these coordinates are particularly useful in solving the coupled ODEs that result from the MDE for higher-order flux-surface shapes. 

With $\PhiTwo$ and $\psiTwo$ in hand let us focus on obtaining $\alphaO$. Substituting the expressions for $\psiTwo$, \eqref{eq:psi2_expression}, together with the expressions for $a,b$, \eqref{eq:a_b_definitions}, in the equation for $\alphaO$, \eqref{eq:alpha0cw}, and integrating with respect to $\omega$ we get
\begin{align}
    \alphaO=\frac{1}{2c_0}\arctan{\lbr e^{-\eta(\ell)}\tan{u}\rbr}+\mathfrak{a}^{(0)}(\ell).
    \label{eq:pre_alpha0_expression_c0}
\end{align}
To determine the unknown function $\mathfrak{a}^{(0)}(\ell)$ we only need the poloidally averaged \eqref{eq:alpha0cz}. Since $\PhiTwo_{,\omega}$ has zero average, the poloidally averaged $\alphaO_{,\ell}$ equation \eqref{eq:alpha0cz} reads
\begin{align}
    \oint \frac{du}{2\pi}\lbr 2\psiTwo \alphaOcz \rbr=0.
    \label{eq:pol_avg_alpha0cz}
\end{align}
Using the expressions for $\alphaO,\psiTwo$ as given in \eqref{eq:pre_alpha0_expression_c0} and \eqref{eq:psi2_expression}, we can show that \eqref{eq:pol_avg_alpha0cz} yields 
\begin{align}
    -2a(\ell)\: {\mathfrak{a}^{(0)}}'(\ell)=\del_\ell u=\delta'(\ell)-\tau(\ell)
    \label{eq:a0p_expr}
\end{align}
Therefore,
\begin{align}
    \alphaO=\frac{1}{2c_0}\arctan{\lbr e^{-\eta(\ell)}\tan{u}\rbr}-\int d\ell\: \frac{\delta'(\ell)-\tau(\ell)}{2 c_0 \cosh{\eta(\ell)}}.
    \label{eq:alpha0_expression_c0}
\end{align}

Let us briefly summarize the findings of NAE up to $O(1)$. From the requirement that the poloidal magnetic field and the flux surface label vanish on the axis, we can determine $\Phi^{(0)},\PhiOne$ to be as given in \eqref{eq:Phi0_and_Phi1}.
The vacuum system to $O(1)$ gives three coupled equations for $\PhiTwo,\psiTwo,\alphaO$. We have shown that these equations can be decoupled to yield a Poisson equation for $\phiTwo$,\eqref{eq:Phi2_laplacian} and an MDE for $\psiTwo$, \eqref{eq:lin_eqn_psi2}. The solutions $\PhiTwo,\psiTwo$, given in \eqref{eq:phi2_expression} and \eqref{eq:psi2_expression}, can then be used to determine $\alphaO$, which is given in \eqref{eq:alpha0_expression_c0} up to an overall constant. With the help of these lowest-order expressions, we now carry out the expansion to $O(n)$ in the next Section, where we demonstrate that at $O(\rho^n)$, an analogous decoupling of the higher-order quantities $(\Phi^{(n+2)},\psi^{(n+2)},\alpha^{(n)})$ is possible.

\subsection{Derivation of the vacuum NAE equations: $n^{\text{th}}$ order}
\label{sec:series_exp_Order_n}
The system \eqref{eq:vacuum_system} to $O(n)$ leads to the following closed set of equations
\begin{subequations}
\begin{align}
    \lbr {\alphaOcw} \del_{\ell}-{\alphaOcz}\del_{\omega}\rbr\psi^{(n+2)}+ \lbr {\alpha^{(n)}_{,\omega}} \del_{\ell}-{\alpha^{(n)}_{,\ell}}\del_{\omega}\rbr\psi^{(2)}=-(n+2) \Phi^{(n+2)}+...
    \label{eq:phi_psi_eqn_order_n}\\
      \lbr 2\psiTwo \rbr^{n/2+1} \del_{\omega}\lbr\dfrac{\alpha^{(n)}}{\lbr 2\psiTwo \rbr^{n/2}}\rbr = {\alphaOcw}\lbr- (n+2)\:\psi^{(n+2)} \rbr +...\label{eq:vac_omega_order_n}\\
     -\lbr 2\psiTwo \rbr^{n/2+1} \del_{\ell}\lbr\dfrac{\alpha^{(n)}}{\lbr 2\psiTwo \rbr^{n/2}}\rbr = {\alphaOcz}\lbr (n+2)\:\psi^{(n+2)} \rbr+\Phi^{(n+2)}_{,\omega}+...  \label{eq:vac_ell_order_n}
     \end{align}
     \label{eq:vacuum_order_n_system}
\end{subequations}   
where $...$ refers to terms involving the lower-order known quantities. 

Although \eqref{eq:vacuum_order_n_system} is linear in $(\Phi^{(n+2)},\psi^{(n+2)},\alpha^{(n)})$, it still couples these functions in a complicated manner. As discussed in Section \ref{sec:Basic_vacuum_eqn}, an alternate approach to solving the system \eqref{eq:vacuum_order_n_system} is the system comprising of the Poisson equation for $\Phi^{(n+2)}$, the MDE for $\psi^{(n+2)}$, equation \eqref{eq:vac_omega_order_n}, and finally, the averaged MDE for $\alpha^{(n)}$ to determine $\alpha^{(n)}$ completely. In this Section, we show how these alternative equations are equivalent to the complete set \eqref{eq:vacuum_order_n_system}.


Expanding the equation relating the $\Phi_{,\rho}$ equation and the MDEs for $\psi,\alpha $, \eqref{eq:connecting_Phi_rho_eqn_BDpsi_BDalpha}, to $O(n)$ we observe that the equation for $\psi^{(n+2)}$ as given in \eqref{eq:phi_psi_eqn_order_n} is equivalent to the $O(n)$ MDE for $\psi^{(n+2)}$. However, the solution of the MDE for $\psi^{(n+2)}$ does not uniquely determine  $\psi^{(n+2)}$ since a homogeneous solution that satisfies $\BD\psi$ to $O(n+2)$ can always be added to it. We show later that imposing the choice of $2\pi\psi$ to be the toroidal flux, order by order, determines the homogeneous piece of $\psi$.

To solve the MDE for $\psi^{(n+2)}$ or equivalently \eqref{eq:phi_psi_eqn_order_n} we need $\Phi^{(n+2)}$ as evident from the right-hand side of \eqref{eq:phi_psi_eqn_order_n}. We have demonstrated in Section \ref{sec:Basic_vacuum_eqn} that the Laplace equation for $\Phi$ arises as an exact compatibility condition of \eqref{eq:vacuum_Phi_omega} and \eqref{eq:vacuum_Phi_ell}, provided \eqref{eq:vacuum_Phi_rho} holds. The same procedure works in higher orders as well. To $O(n)$, elimination of $\alpha^{(n)}$ from \eqref{eq:vac_omega_order_n} and \eqref{eq:vac_ell_order_n} through cross differentiation, i.e., $\del_\ell$\eqref{eq:vac_ell_order_n}$+ \del_\omega$ \eqref{eq:vac_omega_order_n}, followed by elimination of the derivatives of $\psi^{(n+2)}$ using \eqref{eq:phi_psi_eqn_order_n}, leads to an equation for $\Phi^{(n+2)}$ of the form
\begin{align}
    \lbr \del^2_\omega +(n+2)^2 \rbr \Phi^{(n+2)}=F^{(n+2)}_\Phi,
     \label{eq:Laplace_Phinp2_dots}
\end{align}
where $F^{(n+2)}_\Phi$ denotes known inhomogeneous forcing terms from lower order quantities. 
Note that \eqref{eq:Laplace_Phinp2_dots} is simply the Laplace equation for $\Phi$ to $O(n+2)$ that leads to a Poisson equation for $\Phi^{(n+2)}$.

Now that the equation for its angular derivatives \eqref{eq:vac_omega_order_n} and \eqref{eq:vac_ell_order_n} are compatible thanks to \eqref{eq:Laplace_Phinp2_dots}, we use \eqref{eq:vac_omega_order_n} to solve for $\lbr \alpha^{(n)}\rbr/\lbr 2\psiTwo\rbr^{n/2}$ up to an arbitrary function of $\ell$ say $\mathfrak{\overline{a}}^{(n)}(\ell)$. To determine $\mathfrak{\overline{a}}^{(n)}(\ell)$ we need the poloidally averaged MDE for $\alpha^{(n)}$. The MDE for $\alpha^{(n)}$ can be obtained through an algebraic combination of \eqref{eq:vac_omega_order_n} and \eqref{eq:vac_ell_order_n} that eliminates $\psi^{(n+2)}$. It reads
\begin{align}
    \lbr 2\psiTwo\rbr^{n/2+1}\lbr {\alphaOcw} \del_{\ell}-{\alphaOcz}\del_{\omega}\rbr \lbr\dfrac{\alpha^{(n)}}{\lbr 2\psiTwo\rbr^{n/2}}\rbr = -{\alphaOcw}\lbr \Phi^{(n+2)}_{,\omega}+\dots \rbr+\dots
    \label{eq:vac_MDE_alpha_order_n}
\end{align}
Thus, we have derived an alternate set of equations which decouples $\Phi^{(n+2)},\psi^{(n+2)},\alpha^{(n)}$. The next step is to impose the periodicity constraint \eqref{eq:secular_n_periodic_Phi_alpha_K}. To understand the nature of the constraints \eqref{eq:secular_n_periodic_Phi_alpha_K} in a torus, it turns out that inverse coordinates, where $\psi$ is used as the radial coordinate, are ideally suited. Connecting the direct and the inverse coordinates enables us to establish a direct relation between higher derivatives of the rotational transform, $\iota$, and the secular pieces of $\alpha^{(n)}$. Therefore, in Section \ref{sec:map_to_inverse_coords}, we discuss the map between the direct and the inverse to understand the periodicity constraints better.

In what follows, our strategy would be to solve the following equations for $(\Phi^{(n+2)},\psi^{(n+2)},\alpha^{(n)})$ 
\begin{enumerate}[I.    ]
    \item Poisson equation \eqref{eq:Laplace_Phinp2_dots} for $\Phi^{(n+2)}$
    \item The MDE for $\psi^{(n+2)}$ \eqref{eq:phi_psi_eqn_order_n}
    \item The $\alpha_{,\omega}$ equation \eqref{eq:vac_omega_order_n} to solve for $\alpha^{(n)}$ up to a function $\mathfrak{\overline{a}}^{(n)}(\ell)$
    \item Poloidal $\omega$ average of the MDE for $\alpha$ \eqref{eq:vac_MDE_alpha_order_n} to determine $\mathfrak{\overline{a}}^{(n)}(\ell)$.
\end{enumerate}
Adding currents and plasma beta does not appreciably change the overall structure, as we show later. Therefore, a thorough understanding of the mathematical structure of the vacuum system (I-IV) and its solution is of merit and will be discussed in detail in the next few Sections. 

We discuss boundary conditions for Steps (I-IV), derived from the periodicity requirements of physical quantities in the poloidal and toroidal angles, in Section \eqref{sec:map_to_inverse_coords}. Then we discuss the solutions to the Poisson equation \eqref{eq:Laplace_Phinp2_dots} in Section \ref{sec:Phi_structure}, the MDE for $\psi^{(n+2)}$ in Section \ref{sec:sec_vacuum_MDE_psi} and in Section \ref{sec:Phi_ell_alpha_per} we provide the details of obtaining $\alpha$ following Steps III and IV. 

\subsection{Derivation of the vacuum NAE equations: boundary conditions in toroidal geometry and periodicity constraints}
\label{sec:map_to_inverse_coords}

To obtain solutions of the equations (I-IV) in a torus, we need to impose double-periodicity in the two angles and regularity in the radial coordinate near the axis. In Section \ref{sec:Mercier_Weitzner}, we presented the secular and periodic pieces of $\Phi,\alpha$ in equation \eqref{eq:secular_n_periodic_Phi_alpha_K} in the inverse coordinates. The structure of the functions of $\Phi,\psi,\alpha$ follows from the physical requirement that $\J,\B$ be periodic in a torus. Since the Mercier direct coordinates do not use the flux-surface label as a coordinate, separating secular and periodic parts of the multi-valued functions $\Phi,\alpha$ is challenging. In particular, the form of $\alpha$ is crucial for understanding the MDEs. Hence, before discussing how to solve the equations, we need to understand the periodicity constraints in the direct coordinates.  


With the choice of $2\pi \psi$ to be the toroidal flux, we get the following expression for vacuum $(\alpha,\Phi)$ from \eqref{eq:secular_n_periodic_Phi_alpha_K}:
\begin{align}
\alpha=\theta - \iota(\psi) \phi +\widetilde{\alpha}(\psi,\theta,\phi), \qquad \Phi=G_0 \phi +\widetilde{\Phi}(\psi,\theta,\phi).
    \label{eq:alpha_phi_form_inverse_coord}
\end{align}
To work out the near-axis expansion in inverse Mercier coordinates $(\psi,\omega,\phi=\ell/ L)$ one needs to invert the function $\psi(\rho,\omega,\ell)$ to obtain $\rho(\psi,\omega,\ell)$. Equivalently one needs to invert the NAE for $\psi$ \eqref{eq:psi_NAE} to obtain an expression for $\rho$ as a power series in $\sqrt{2\psi}$. We can asymptotically invert the power series of $\psi$ in $\rho$ to obtain
\begin{align}
    \rho=\sqrt{\frac{\psi}{\psiTwo}}-\frac{\psiThree}{2\psiTwo}\lbr \frac{\psi}{\psiTwo}\rbr -\frac{1}{8}\lbr 4\frac{\psiFour}{\psiTwo}-5\lbr \frac{\psiThree}{\psiTwo}\rbr\rbr \lbr\frac{\psi}{\psiTwo}\rbr^{3/2}+O(\psi^2).
    \label{eq:rho_of_psi}
\end{align}
Let us first focus on the field-line label $\alpha$. The near-axis expansion of $\alpha$ in the inverse (denoted by subscripts) and the direct (denoted by superscripts) coordinates,
\begin{subequations}
    \begin{align}
     \alpha(\rho,\omega,\ell)&= \alphaO(\omega,\ell) +\rho \alphaOne(\omega,\ell)+\rho^2 \alphaTwo(\omega,\ell)+O(\rho^3),\\
    \alpha(\psi,\omega,\ell)&=\alpha_0(\omega,\ell) +\sqrt{\psi} \alpha_1(\omega,\ell)+\psi\: \alpha_2(\omega,\ell)+O(\psi^{3/2}),
\end{align}
 \label{eq:alpha_form_direct_indirect}
\end{subequations}
are, therefore, connected through the relations
\begin{align}
    \alpha_0&= \alphaO,\quad \alpha_1= \frac{\alphaOne}{\sqrt{\psiTwo}}, \quad \alpha_2=\frac{\alphaTwo}{\psiTwo}-\frac{1}{2}\frac{\alphaOne}{\sqrt{\psiTwo}}\frac{\psiThree}{\lbr \psiTwo\rbr^{3/2}},\\
    \alpha_n&= \frac{\alpha^{(n)}}{(\psiTwo)^{n/2}}+\sum_{m=1}^{n-1} \frac{\alpha^{(m)}}{\lbr \psiTwo \rbr^{m/2}} \mathcal{P}_{m(n-1)},\nonumber
    \label{eq:connecting_alpha_direct_indirect}
\end{align}
where, $\cP_{m(n-1)}$ is a polynomial of order $(n-1)$ whose arguments are $\lbr \psi^{(j)}/\lbr \psiTwo\rbr^{j/2} \rbr$ with $2\leq j\leq m+1$.
Expanding $\iota(\psi)$ in a Taylor series in $\psi$ near the axis,
\begin{align}
    \iota(\psi)= \iota_0 +\iota_2 \psi + \dots +\iota_{2k}\psi^{(2k)} +O(\psi^{(2k+1)}),
\end{align}
and using the form of $\alpha$ in the inverse coordinates as given in \eqref{eq:secular_n_periodic_Phi_alpha_K}, we find that
\begin{align}
    \alpha_0=\theta-2\pi\iota_0 \frac{\ell}{ L}+\widetilde{\alpha}_0, \quad \alpha_1 = \widetilde{\alpha}_1, \quad \alpha_2 = -2\pi\iota_2 \frac{\ell}{ L} + \widetilde{\alpha}_2,\quad \alpha_{2k}= -2\pi\iota_{2k} \frac{\ell}{L}+\widetilde{\alpha}_{2k},
    \label{eq:alpha012_structure}
\end{align}
where $\widetilde{\alpha}_k, k=0,1,2$ denotes functions periodic in the angles. Averaging the $\omega,\ell$ derivatives of $\alpha_{2k}$ from  \eqref{eq:alpha012_structure} over $\omega$ and $\ell$, and exploiting the periodicity of $\widetilde{\alpha}_k$, we find that
\begin{align}
\oint \frac{d\omega}{2\pi} \del_\omega \alpha_{2k}=\delta_{2k},\qquad
2\pi\iota_{2k}= -\oint \frac{d\omega}{2\pi}\int_0^L d\ell\: \del_\ell \lbr \alpha_{2k}\rbr,
\label{eq:pre_iota2k_eqn}
\end{align}
where, $\delta_{2k}=0$ for $k>0$ and $1$ for $k=0$.

We are now in a position to determine the constant $c_0$ in \eqref{eq:a_b_definitions}. To do so, we impose the choice of $2\pi\psi$ as the toroidal flux (equivalently $f(\psi)=1$ in \eqref{eq:sec_n_per_alpha}), order by order, by insisting that the secular $\theta$ piece only appears in $\alphaO$ as shown in \eqref{eq:alpha012_structure}. Here, we note that \eqref{eq:alpha012_structure} implies that the average of $ \alphaOcw$ must be unity.

Using the form of $\alphaO$ obtained in \eqref{eq:alpha0_expression_c0} and averaging over $\omega$, we obtain the average $\alphaOcw$ in the following form:
\begin{align}
    \oint \frac{d\omega}{2\pi} \alphaOcw= \frac{1}{2a(\ell)} \oint \frac{d\omega}{2\pi}\frac{1}{1+\varepsilon(\ell) \cos{2u}}.
    \label{eq:int_alphacw_is_1}
\end{align}
Here, we have used \eqref{eq:alpha0cw} for $\alphaO_{,\omega}$, \eqref{eq:psi2_expression} for the functional form of $\psiTwo$, and \eqref{eq:a_b_definitions} to define
\begin{align}
    \vep(\ell) \equiv \frac{b(\ell)}{a(\ell)}= \tanh{\eta(\ell)}.
\end{align}
Replacing the $\omega$ integral by a $u$ integral, using the identity
\begin{align}
    \oint \frac{du}{2\pi}\frac{1}{1+\varepsilon \cos{2u}}=\frac{1}{\sqrt{1-\varepsilon^2}}=\cosh{\eta(\ell)},
    \label{eq:Useful_integral_1}
\end{align}
and finally imposing the condition that the average of $\alphaOcw$ must be unity, we find that
\begin{align}
    c_0=\frac{1}{2}, \quad a(\ell)= \cosh\eta(\ell).
    \label{eq:c0_expression}
\end{align}
Therefore, the expression for $\alphaO$, \eqref{eq:alpha0_expression_c0}, now reads
\begin{align}
    \alphaO=\arctan{\lbr e^{-\eta(\ell)}\tan{u}\rbr}-\int d\ell\: \frac{\delta'(\ell)-\tau(\ell)}{2a(\ell)}.
    \label{eq:alpha0_expression}
\end{align}
The on-axis rotational transform can be obtained from \eqref{eq:alpha0_expression} by utilizing
\begin{align}
    \iota_0= \alpha(\ell,\theta)-\alpha(\ell+L,\theta)+N = \frac{v(L)-(\delta(L)-\delta(0))}{2\pi}+N,
\end{align}
where we have defined
\begin{align}
    v(\ell)=\int_0^\ell ds \frac{u'(s)}{\cosh{\eta(s)}}= \int_0^\ell ds \frac{\delta'(s)-\tau(s)}{\cosh{\eta(s)}} .
\end{align}
The extra integer $N$ comes from the rotation of the Frenet frame \citep{helander2014theory}.

We now explicitly derive the periodicity conditions in the direct coordinates. Firstly, every $\psi^{(m+2)}, m=0,1,2,..$ needs to be single-valued so that the flux-surface label $\psi$ is single-valued. Secondly, by substituting the relation between the components of $\alpha$ in the inverse representation $\alpha_n$ and the direct representation $\alpha^{(n)}$ as given in \eqref{eq:alpha012_structure} into the identities \eqref{eq:pre_iota2k_eqn}, we get
\begin{align}
   &\oint \frac{d\omega}{2\pi} \del_\omega\lbr \frac{\alpha^{(n)}}{(\psiTwo)^{n/2}}\rbr=0, \qquad (n>0) \nonumber\\
   &\oint \frac{d\omega}{2\pi}\oint\frac{d\ell}{2\pi}\del_\ell \lbr \frac{\alpha^{(n)}}{(\psiTwo)^{n/2}} +\sum_{m=1}^{n-1} \frac{\alpha^{(m)}}{\lbr \psiTwo \rbr^{m/2}} \mathcal{P}_{m(n-1)}\rbr = -\iota_{2k} \quad (n=2k). 
   \label{eq:integral_constraints_alpha}
\end{align}
Returning to the discussion of the multi-valued structure of $\Phi$, we note that the analysis is almost identical to that of $\alpha$. The analogs of the relation between inverse and direct components \eqref{eq:alpha_form_direct_indirect}, and the order-by-order form in inverse-coordinates \eqref{eq:connecting_alpha_direct_indirect} for $\Phi$ are obtained by simply replacing $\alpha$ by $\Phi$. An important difference between $\Phi$ and $\alpha$ occurs because in vacuum $G_0$ in \eqref{eq:alpha_phi_form_inverse_coord} is a constant. Therefore, we have  
\begin{align}
    \Phi_0=G_0 \frac{\ell}{ L}+\widetilde{\Phi}_0, \quad \Phi_{n}=\widetilde{\Phi}_{n} \quad \text{for all $n>0$},
    \label{eq:Phi012_structure}
\end{align}
which implies that except for $\PhiO$, all other $\Phi^{(n)}$ are periodic since $\psiTwo$ is periodic. Single-valuedness of $\Phi^{(n)}, n>0$ follows from the Laplace equation as will be shown in Section \ref{sec:Phi_structure}. 

To summarize, we have deduced the periodicity constraints on $(\Phi^{(n+2)},\psi^{(n+2)},\alpha^{(n)})$ in this Section by connecting to the inverse coordinates. The flux label $\psi^{(n+2)}$ must be periodic in both angles on a torus. In the inverse coordinates, the general periodicity requirement on $\Phi,\alpha$ is given by \eqref{eq:alpha_phi_form_inverse_coord} and order by order the requirement is \eqref{eq:alpha012_structure} together with \eqref{eq:Phi012_structure}. In the direct coordinates, the periodicity constraints are given by \eqref{eq:integral_constraints_alpha} and single-valuedness of $\Phi^{(n+2)}$ for all $n>0$.


\subsection{ Solutions of the vacuum NAE equations to $O(n)$}
\label{sec:periodicity_n_structure_soln}

Our goal now is to analyze the harmonic contents of  $(\Phi^{(n+2)},\psi^{(n+2)},\alpha^{(n)})$. The structure of the vacuum potential $\Phi$ was analyzed in detail in \citep{Jorge_Sengupta_Landreman2020NAE}. We briefly summarize some of the relevant results here. To analyze the harmonic content of $\psi^{(n+2)}$ and $\alpha^{(n)}$, we study the structure and properties of MDEs in the next Sections.

\subsubsection{Solution of Step I : The Poisson equation for $\Phi^{(n+2)}$}
\label{sec:Phi_structure}

The Laplace equation to $O(n)$ is 
\begin{align}
 &\lbr \del^2_{\omega}+(n+2)^2 \rbr\Phi^{(n+2)}\label{eq:vac_Laplacian_order_n}\\ &=-\sum_{m=0}^{n} \kappa^{m} \cos^{m} \theta\left[\kappa\left(\del_\omega \Phi^{(n-m+1)} \sin \theta- (n-m+1)\Phi^{(n-m+1)}\cos \theta\right)\right.\nonumber \\
    & \qquad\qquad\qquad\qquad\quad+\left.(m+1)\del^2_{\ell}\Phi^{(n-m)}+ \frac{(m+1)(m+2)}{2}\del_\ell(\kappa \cos \theta) \del_\ell\Phi^{(n-m-1)}\right]. \nonumber
\end{align}
As shown in \citep{Jorge_Sengupta_Landreman2020NAE}, the right side of \eqref{eq:vac_Laplacian_order_n} has at most harmonic terms of order $(n)$. If the $(n+2)$-th poloidal harmonics were present in the forcing term of \eqref{eq:vac_Laplacian_order_n}, the inversion of   \eqref{eq:vac_Laplacian_order_n} would lead to linear (hence secular) $\omega$ terms in $\Phi^{(n+2)}$. Absence of $(n+2)$ forcing poloidal harmonics implies that $\Phi^{(n+2)}$ has poloidal up to $n+2$, with the coefficients of $(n+2)^{\text{th}}$ harmonics being free.


Therefore, we can write down the solution of $\Phi^{(n+2)}$ as
\begin{align}
    \Phi^{(n+2)}= \sum_j  \lbr \Phi^{(n+2)}_{cj}(\ell)\cos(j u)+ \Phi^{(n+2)}_{sj}(\ell)\sin(j u)\rbr,
    \label{eq:Phi_nplus2_form}
\end{align}
where, $u$ is defined by \eqref{eq:u_definition}. For even $n$, $j=2r$ and $r$ varies from $0$ to $(n/2+1)$. For odd $n$, $j=2r+1$ and $r$ varies from $0$ to $(n+1)/2$. The functions $\Phi^{(n+2)}_{c(n+2)}(\ell),\Phi^{(n+2)}_{s(n+2)}(\ell)$ are the ``free functions," and the rest are completely determined by the lower order quantities. 

We now ensure that derivatives of $\Phi^{(n+2)}$ are single-valued since they are physically related directly to the components of the magnetic field. The right-hand-side is manifestly periodic in the poloidal angle. Therefore, periodicity in the poloidal angle is maintained order by order. Let us now check for periodicity in $\ell$. The presence of $\Phi^{(n-m+1)}$ without derivatives in the forcing term on the right-hand side of \eqref{eq:Phi_nplus2_form} might suggest that $\Phi^{(n+2)}$ is not periodic since $\PhiO$ is multi-valued in $\ell$ after all. However, $\PhiO$ never appears without any derivative in the forcing term. To have $\PhiO$, we must have $m=n+1$, which is outside the summation range. Furthermore, $\Phi^{(1)}=0$ implies that for any $n$, the lowest nontrivial term of the form $\Phi^{(n-m+1)}$ is $\Phi^{(2)}$ for $m=n-1$, which is periodic. Therefore, the forcing term in \eqref{eq:vac_Laplacian_order_n} is a linear superposition of derivatives of $(\PhiO,\PhiTwo,...,\Phi^{(m)})$ and $\Phi^{m}$ terms with $m>0$. By induction $\Phi^{(n+2)}$ must be single-valued for all $n>0$ as required by \eqref{eq:Phi012_structure}.

This completes our solution of Step I. The solution of Step II is more involved and will be developed over the next few Sections.
  
\subsubsection{Solution of Step II : Analysis of the general MDE}
\label{sec:sec_vacuum_MDE}
To solve the MDE for $\psi$ in Step II, we need to discuss some properties of the MDE. In this Section, we shall derive these properties for a general vacuum MDE.

The MDE \citep{newcomb1959magnetic} for a quantity $\chi$ takes the general form
\begin{align}
    \BD \chi +F_\chi=0,\qquad \BD\equiv B_\rho \del_\rho + B_\omega \frac{1}{\rho}\del_\omega + B_\ell \frac{1}{h}\del_\ell
    \label{eq:generic_MDE}
\end{align}
where $F_\chi$ is the forcing (or the inhomogeneous) term. Expressing the components of $\B$ in \eqref{eq:generic_MDE} in terms of derivatives of the scalar potential $\Phi$, the magnetic differential operator (MDO) can be written as
\begin{align}
    \BD= \Phi_{,\rho} \del_\rho + \frac{1}{\rho^2}\Phi_{,\omega}\del_\omega +\frac{1}{h^2}\Phi_{,\ell}\del_\ell .
    \label{eq:MDO_general}
\end{align}
Expanding in powers of $\rho$ and using the expressions for $\PhiO, \PhiOne$ from \eqref{eq:Phi0_and_Phi1}, we find that \eqref{eq:generic_MDE} takes the form
\begin{align}
    (\BD)^{(n)}_0 \chi^{(n)} +\mathfrak{f}^{(n)}_\chi=0, \quad 
    \label{eq:MDE_standard_form}
\end{align}
where,
\begin{align}
    (\BD)^{(n)}_0\equiv \lbr \Phi_0'(\ell)\del_\ell + \PhiTwo_{,\omega} \del_\omega+2 n \PhiTwo \rbr.
    \label{eq:MDO_lowest}
\end{align}
The term $\mathfrak{f}^{(n)}_\chi$ consists of $\chi^{(n-1)}$ and other lower order quantities obtained from both $F_\chi$ and $\BD \chi$ expanded to $O(n)$. Note also that the definition of the lowest order MDO depends on the order $n$ due to the $\del_\rho$ operator in \eqref{eq:MDO_general}. Clearly, for $n=2$, $\psiTwo$ is a homogeneous solution of \eqref{eq:MDE_standard_form}, as claimed earlier.

Unsurprisingly, we will find the MDO \eqref{eq:MDO_lowest} repeatedly in this work. We shall therefore record a few essential properties of the MDO before we continue with the analysis of the MDE \eqref{eq:MDE_standard_form}.

The first important property to note is that an alternate form of the MDO is given by
\begin{align}
    (\BD)^{(n)}_0\lbr \cdot \rbr= 2\lbr \psiTwo\rbr^{n/2+1}\left\{\alphaO,\frac{1}{ \lbr \psiTwo\rbr^{n/2}}\lbr \cdot \rbr \right\}_{(\omega,\ell)},
    \label{eq:MDO_PB_form}
\end{align}
where, $\{f,g\}_{(\omega,\ell)}$ denotes a standard Poisson of bracket of functions $f$ and $g$  with respect to $(\omega,\ell)$. The form \eqref{eq:MDO_PB_form} follows from the definition of MDO \eqref{eq:MDO_lowest}, the equation for $\psiTwo$ \eqref{eq:lin_eqn_psi2}, and the relations \eqref{eq:Phi2_alpha0_psi2_system} that are satisfied by $(\PhiTwo,\alphaO,\psiTwo)$. 

The Poisson-bracket form of the MDO, \eqref{eq:MDO_PB_form}, allows us to readily find the homogeneous solutions of the MDE \eqref{eq:MDE_standard_form}. Thus, if $\chi^{(n)}_h$ is a homogeneous solution of \eqref{eq:MDE_standard_form}, it must be of the form
\begin{align}
    \chi^{(n)}_h=\lbr \psiTwo\rbr^{n/2}\mathcal{C}_h\lbr \alphaO\rbr
\end{align}
When the rotational transform is irrational, $\mathcal{C}_h$ must be a constant if $\chi^{(n)}_h$ represents a physical quantity. 

The next important property of the  MDO \eqref{eq:MDO_PB_form} is that it possesses an annihilator (an operator to whose kernel $(\BD)^{(n)}_0$ belongs). To obtain the annihilator, we use the following property of the Poisson bracket
\begin{align}
   \left\{\alphaO,\chi\right\}_{(\omega,\ell)} \equiv \alphaOcw \chi_{,\ell}-\alphaOcz \chi_{,\omega}= \del_\ell \lbr \alphaOcw \chi \rbr-\del_\omega \lbr \alphaOcz \chi \rbr.
   \label{eq:PB_alpha0_chi}
\end{align}
Here we have used the fact that $\alphaOcw,\alphaOcz$ are single-valued in order to commute the partial derivatives $\del_\ell,\del_\omega$ acting on $\alphaO$ in \eqref{eq:PB_alpha0_chi}. Averaging \eqref{eq:PB_alpha0_chi} over $(\omega,\ell)$ we get
\begin{align}
    \oint d\ell\oint \frac{d\omega}{2\pi} \left\{\alphaO,\chi \right\}_{(\omega,\ell)} =0,
    \label{eq:PB_periodic_is_0}
\end{align}
provided $\chi$ is single-valued. Therefore, the operator 
\begin{align}
    \langle \lbr \cdot \rbr \rangle_{(n)} = \oint \frac{d\ell}{2\pi} \oint \frac{d\omega}{2\pi}\frac{1}{ \lbr \psiTwo\rbr^{n/2+1}} \lbr \cdot \rbr
    \label{eq:ann_MDO}
\end{align}
is an annihilator for the MDO \eqref{eq:MDO_PB_form}. Note that the factor of $2\pi$ below $d\ell$ is because \eqref{eq:MDO_PB_form} has a derivative with respect to $\ell$. When written in terms of $\phi=2\pi\ell/L$, the factor of $2\pi$ turns out to be the usual measure.

Finally, we must discuss the poloidal harmonics generated when the MDO acts on the $n^{\text{th}}$ harmonics. This property is essential in order to solve the MDE order by order. In the inverse coordinate approach, one can use analyticity to show that there is an upper bound on the poloidal harmonics at each order. In the direct coordinate approach, the upper bound on the poloidal harmonics is not evident since various powers of $\lbr \psiTwo\rbr$ appear in the MDO \eqref{eq:MDO_PB_form} and its annihilator \eqref{eq:ann_MDO}. 

To deduce the poloidal harmonic content of $\chi^{(n)}$ that satisfies the MDE \eqref{eq:MDE_standard_form}, we will need another important property of the MDO \eqref{eq:MDO_lowest}. Let us consider a generic periodic function $\chi^{(n)}_j$ of the form 
\begin{align}
   \chi^{(n)}_j= \chi^{(n)}_{cj}(\ell) \cos{(j u(\omega,\ell))}+ \chi^{(n)}_{sj}(\ell) \sin{(j u(\omega,\ell))},
   \label{eq:chi_j_n_definition}
\end{align}
with $u(\omega,\ell)$ defined in \eqref{eq:u_definition}. The integer $j$ such that $2\leq j \leq n$, is even (odd) if n is even (odd). Using \eqref{eq:phi2_expression} and \eqref{eq:psi2_expression} we can show that
\begin{subequations}
\begin{align}
    &(\BD)^{(n)}_0 \chi^{(n)}_j = C^{(n)}_{cj}(\ell) \cos{j u}+C^{(n)}_{sj}(\ell) \sin{j u} +(n-j) C^{(n)}_{(+)}(\omega,\ell)+(n+j) C^{(n)}_{(-)}(\omega,\ell)\\
   &C^{(n)}_{cj}(\ell)= \Phi_0' \lbr \frac{d\chi^{(n)}_{cj}}{d\ell} +n(\delta'-\tau)\chi^{(n)}_{sj}-\frac{n}{2}\frac{\Phi''_0}{\Phi_0'}\chi^{(n)}_{cj}\rbr\\
    &C^{(n)}_{sj}(\ell)= \Phi_0' \lbr \frac{d\chi^{(n)}_{sj}}{d\ell} -n(\delta'-\tau)\chi^{(n)}_{cj}-\frac{n}{2}\frac{\Phi''_0}{\Phi_0'}\chi^{(n)}_{sj}\rbr\\
    &C^{(n)}_{(+)}=\Phi_0' \left[\lbr \fTwocTwo\chi^{(n)}_{cj}- \fTwosTwo\chi^{(n)}_{sj} \rbr\cos{(j+2)u}+\lbr \fTwocTwo\chi^{(n)}_{sj}+ \fTwosTwo\chi^{(n)}_{cj} \rbr\sin{(j+2)u} \right]\\
    &C^{(n)}_{(-)}=\Phi_0' \left[\lbr \fTwocTwo\chi^{(n)}_{cj}+ \fTwosTwo\chi^{(n)}_{sj} \rbr\cos{(j-2)u} -\lbr \fTwosTwo\chi^{(n)}_{cj}- \fTwocTwo\chi^{(n)}_{sj} \rbr\sin{(j-2)u}\right].
\end{align}
\label{eq:BD0_n_chi_j_n}
\end{subequations}

We note, in particular, that for $j=n$, there are no terms with $(n+2)$ poloidal harmonics. Thus, for a function of the form \eqref{eq:chi_j_n_definition}, i.e. with $j^{\text{th}}$ poloidal harmonics in $u$, where $j$ has the same parity as $n$ and $0\leq j\leq n$, $(\BD)^{(n)}_0 \chi^{(n)}_j$ is a function with poloidal harmonics strictly between $j-2$ and $\min{(j+2,n)}$. 

Now let us define a function $\chi^{(n)}=\sum_j \chi^{(n)}_j$ such that
\begin{align}
    \chi^{(n)}=
    \begin{cases}
    \sum_{r=0}^{ n/2}\chi^{(n)}_{2r} , \quad \quad\quad \text{n is even}\\ \\
    \sum_{r=0}^{(n-1)/2 }\chi^{(n)}_{2r+1} , \quad \text{n is odd},
    \end{cases}
    \label{eq:chi_n_definition}
\end{align}
where, $\chi^{(n)}_j$'s for all $j,n$ are of the form \eqref{eq:chi_j_n_definition}, $(\BD)^{(n)}_0\chi^{(n)}$ will have even (odd) poloidal harmonics only up to $n$ for even (odd) values of $n$.
If we are given a function $\mathfrak{f}^{(n)}_\chi$, such that
\begin{align}
   \mathfrak{f}^{(n)}_\chi= 
    \begin{cases}
    \sum_{r=0}^{ n/2} \lbr F^{(n)}_{c(2r)} \cos{2r u} + F^{(n)}_{s(2r)} \sin{2r u} \rbr , \quad \quad\quad \text{n is even}\\ \\
    \sum_{r=0}^{(n-1)/2 }\lbr F^{(n)}_{c(2r+1)} \cos{(2r+1) u} + F^{(n)}_{s(2r+1)} \sin{(2r+1) u} \rbr , \quad \text{n is odd},
    \end{cases}
    \label{eq:F_chi_n_definition}
\end{align}
the solution of 
\begin{align}
    (\BD)^{(n)}_0 \chi^{(n)}+\mathfrak{f}^{(n)}_\chi=0
    \label{eq:MDE_chi_n}
\end{align}
 is obtained by solving the coupled ODEs obtained by equating to zero the sum of the $(\BD)^{(n)}_0 \chi^{(n)}$ terms given in \eqref{eq:BD0_n_chi_j_n} and the $\mathfrak{f}^{(n)}_\chi$ terms given in \eqref{eq:F_chi_n_definition}.  
 
We want to emphasize that without the property that no $(n+2)$ poloidal harmonics are generated when the MDO acts on the $n^{\text{th}}$ harmonics, there would be no guarantee of an upper bound on the number of poloidal harmonics in $u$.
Thanks to this property, a forcing term $\mathfrak{f}^{(n)}_\chi$ that has $n$ poloidal harmonics will lead to a solution $\chi^{(n)}$, which also has at most $n$ poloidal harmonics.

\subsubsection{Solution of Step II : Solution of the MDE for $\psi^{(n+2)}$}
\label{sec:sec_vacuum_MDE_psi}
As we discussed in Section \ref{sec:series_exp_Order_n}, equation \eqref{eq:phi_psi_eqn_order_n} can be rewritten as a MDE for $\psi^{(n+2)}$ by eliminating $\alpha^{(n)}$ algebraically using  \eqref{eq:vac_omega_order_n} and \eqref{eq:vac_ell_order_n}. To show that explicitly,
we need the following identity obtained from the $n^{th}$ order NAE vacuum equations \eqref{eq:vac_omega_order_n}, \eqref{eq:vac_ell_order_n}
\begin{align}
\lbr 2\psiTwo\rbr\{\alpha^{(n)},\psi^{(2)}\}_{(\omega,\ell)}&=(n+2)\psi^{(n+2)}\{\alphaO,\psiTwo\}_{(\omega,\ell)}+\psiTwo_{,\omega}\Phi^{(n+2)}_{,\omega}\dots
    \label{eq:PB_alpha_n_psi2}\\
    &=2(n+2)\psi^{(n+2)}\PhiTwo +\psiTwo_{,\omega}\Phi^{(n+2)}_{,\omega}\dots ,\nonumber
\end{align}
where, in the last step, we have used the $O(1)$ relation  \eqref{eq:Phi2_alpha0_psi2_system} that gives the Poisson bracket of $\alphaO,\psiTwo$ in terms of $\PhiTwo$.

Multiplying \eqref{eq:phi_psi_eqn_order_n} by $2\psiTwo$, using \eqref{eq:Phi2_alpha0_psi2_system} and \eqref{eq:PB_alpha_n_psi2}, we now rewrite \eqref{eq:phi_psi_eqn_order_n}  as
\begin{align}
\lbr \Phi_0'(\ell)\del_\ell + \PhiTwo_{,\omega} \del_\omega+ 2(n+2) \PhiTwo \rbr\psi^{(n+2)}+ \lbr 2\psiTwo (n+2)\Phi^{(n+2)}+ \psiTwo_{,\omega}\Phi^{(n+2)}_{,\omega}\rbr+...=0.
    \label{eq:MDE_psi_order_n+2}
\end{align}
which is clearly of the form 
\begin{align}
    (\BD)^{(n+2)}_0 \psi^{(n+2)}+ F^{(n+2)}_\psi=0, \quad F^{(n+2)}_\psi= \lbr 2\psiTwo (n+2)\Phi^{(n+2)}+ \psiTwo_{,\omega}\Phi^{(n+2)}_{,\omega}\rbr+...
    \label{eq:MDE_psi_n+2}
\end{align}
where, $(\BD)^{(n+2)}_0$ is defined by \eqref{eq:MDO_lowest}.

Firstly, following the discussion in Section \ref{sec:sec_vacuum_MDE}, we note that \eqref{eq:MDE_psi_n+2} determines $\psi^{(n+2)}$ only up to the homogeneous solution, 
\begin{align}
    {\psi_H}^{(n+2)}= \lbr {2\psiTwo}\rbr^\frac{n+2}{2} \mathcal{Y}_H^{(n+2)}(\alphaO).
\end{align}
For irrational rotational transform,  $\mathcal{Y}_H^{(n+2)}$ must be a constant. The value of the constant will be determined in a subsequent Section by imposing the constraint that $2\pi\psi$ is the net toroidal flux.

To determine the particular solution, $\psi_P^{(n+2)}$, of the MDE \eqref{eq:MDE_psi_n+2}, we first need to investigate the harmonic structure of the forcing term $F^{(n+2)}_\psi$.
Let us convince ourselves that $F^{(n+2)}_\psi$ does not have any harmonics higher than $(n+2)$. We'll start with the highest harmonics of $\Phi^{(n+2)}$ beating with $\psiTwo$ according to \eqref{eq:MDE_psi_n+2}. Straightforward algebra using \eqref{eq:psi2_expression} shows that
\begin{multline}
   2\psiTwo (n+2)\Phi^{(n+2)}+ \psiTwo_{,\omega}\Phi^{(n+2)}_{,\omega}= \\ 2(n+2)\Phi_0'\left[ a(\ell) \lbr \Phi^{(n+2)}_{c(n+2)}\cos{(n+2)u}+ \Phi^{(n+2)}_{c(n+2)}\sin{(n+2)u} \rbr \right.  \\
    \left.+b(\ell)\lbr \Phi^{(n+2)}_{c(n+2)}\cos{n u}+ \Phi^{(n+2)}_{s(n+2)}\sin{n u} \rbr \right]+...
    \label{eq:F_psi_n+2}
\end{multline}
The rest of the terms are similar, as shown inductively in Appendix \ref{app:MDE_structure}. Given this form for the forcing term,  the solution of $\psi^{(n+2)}$ can then be deduced from the solution of $\chi^{(n)}$ as given in \eqref{eq:chi_n_definition} with $n$ replaced by $(n+2)$. 

We shall now briefly analyze the coupled ODEs that result from equating the poloidal harmonics of the MDE \eqref{eq:MDE_psi_n+2}. A more complete description will be presented in the next Section. Our goal in the remainder of this Section is to motivate a choice for the free-functions, $ \Phi^{(n+2)}_{c(n+2)}, \Phi^{(n+2)}_{s(n+2)}$, that appear in the forcing term \eqref{eq:F_psi_n+2}, in terms of the $O(n+2)$ flux-surface shaping functions $\psi^{(n+2)}_{c(n+2)}, \psi^{(n+2)}_{s(n+2)}$.

We first note that using the expressions for $\PhiTwo$ as given in \eqref{eq:phi2_expression} together with the definitions \eqref{eq:f2c2_f2s2_definitions}, the MDO $(\BD)^{(n+2)}_0$ defined in \eqref{eq:MDO_lowest} can be explicitly written out as follows:
\begin{align}
   (\BD)^{(n+2)}_0=\Phi_0' \left( \del_\ell -\frac{n+2}{2}\frac{\Phi_0''}{\Phi_0'}+\Omega_0 \cL_1 + \frac{\eta'}{2}\cL_2 \right).
   \label{eq:BDnp2_detail}
\end{align}
Here,
\begin{align}
    \Omega_0 = \frac{u'}{\cosh\eta} , \quad u'=\delta-\tau,
\end{align}
and the operators $\cL_1,\cL_2$ are given by
\begin{align}
   \cL_1=\sinh\eta \lbr \cos{2u}\:\del_\omega +  (n+2)\sin{2u}\rbr, \quad
   \cL_2= \lbr \sin{2u}\:\del_\omega - (n+2)\cos{2u}\rbr.
   \label{eq:BDnp2_components}
\end{align}
Using \eqref{eq:BDnp2_detail} to solve the MDE \eqref{eq:MDE_psi_n+2}, we note that the $\Phi_0''$ term in  \eqref{eq:BDnp2_detail} can be removed by defining $Y^{(n+2)}$ such that
\begin{align}
    \psi^{(n+2)}&=\lbr \Phi_0'\rbr^\frac{n+2}{2} Y^{(n+2)} \label{eq:psi_and_Y}\\
    Y^{(n+2)}&= \sum_{m=0,1}^{n}Y^{(n+2)}_{c(m+2)}\cos{(m+2)u}+ Y^{(n+2)}_{s(m+2)}\sin{(m+2)u}.\nonumber
\end{align}
The lower limit on the sum is zero (one) for even (odd) orders. It is straightforward to show that the operators $\cL_1,\cL_2$ , in general, couple the $m^{th}$harmonics to $m\pm 2$ harmonics. This leads to a complicated set of coupled linear ODEs for the variables $Y^{(n+2)}_{c m},Y^{(n+2)}_{s m}$. However, $\cL_1,\cL_2$ acting on the $(n+2)$ harmonics of $\psi^{(n+2)}$ yields only harmonics of order $n$. From \eqref{eq:Phi_nplus2_form} we can deduce that the free-functions $ \Phi^{(n+2)}_{c(n+2)}, \Phi^{(n+2)}_{s(n+2)}$ also appear with the $(n+2)$ harmonics. We shall now focus only on the $ \Phi^{(n+2)}_{c(n+2)}, \Phi^{(n+2)}_{s(n+2)}$ equations. Substituting the forms of $(\BD)^{(n+2)}_0$ and $\psi^{(n+2)}$ , \eqref{eq:BDnp2_detail} and \eqref{eq:psi_and_Y} we find that
\begin{align}
    (\BD)^{(n+2)}_0 Y^{(n+2)} = \lbr \Phi_0' \rbr^{\frac{n+2}{2}}\:  \Phi_0'\frac{\del}{\del \ell}\lbr Y^{(n+2)}_{c(n+2)} \cos{(n+2)u}+Y^{(n+2)}_{s(n+2)}\sin{(n+2)u} \rbr+...
    \label{eq:BDYn+2_exp}
\end{align}
where, $\dots$ represent the rest of the terms including the $(n+2)$ harmonics that are generated by $\cL_1,\cL_2$ acting on  $ \Phi^{(n+2)}_{c n}, \Phi^{(n+2)}_{s n}$. The forcing terms with $(n+2)$ harmonics are of the form \eqref{eq:F_psi_n+2}. We shall now make a choice for the free-functions $ \Phi^{(n+2)}_{c(n+2)}, \Phi^{(n+2)}_{s(n+2)}$ that would allow us to solve the MDE \eqref{eq:MDE_psi_n+2} partially to obtain  $ Y^{(n+2)}_{c(n+2)}, Y^{(n+2)}_{s(n+2)}$. (The solution is partial because the coupling to 
$ \Phi^{(n+2)}_{c n}, \Phi^{(n+2)}_{s n}$ is not included.)

Motivated by the particular choice for the free functions in \citep{Solovev1970}, we now choose the free function as
\begin{align}
\Phi^{(n+2)}_{\text{free}}&\equiv \lbr \Phi^{(n+2)}_{c(n+2)}\cos{(n+2)u}+ \Phi^{(n+2)}_{c(n+2)}\sin{(n+2)u} \rbr\nonumber\\
&=-\frac{\lbr \Phi_0'(\ell)\rbr^{(n+2)/2}}{2a(\ell)(n+2)^2}\frac{\del}{\del \ell}\lbr Q^{(n+2)}(\ell)\cos{(n+2)u}+P^{(n+2)}(\ell)\sin{(n+2)u} \rbr.
    \label{eq:choice_free_function}
\end{align}
Note that our choice differs from \citep{Solovev1970} by an extra factor of ${\Phi_0'}^{n/2}/2a(\ell)$. 

The MDE for $\psi^{(n+2)}$ \eqref{eq:MDE_psi_n+2} upon using the expressions \eqref{eq:BDYn+2_exp} and \eqref{eq:choice_free_function} then leads to 
\begin{align}
Y^{(n+2)}_{s(n+2)}=\frac{1}{n+2}P^{(n+2)}+\dots,\quad Y^{(n+2)}_{c(n+2)}=\frac{1}{n+2}Q^{(n+2)}+\dots,
    \label{eq:constant_state_psi_n+2}
\end{align}
where, $\dots$ represent the terms obtained from $Y^{(n+2)}_{c n},Y^{(n+2)}_{s n}$ that we have ignored. Thus, at each order, $(n+2)$, the choice of free function \eqref{eq:choice_free_function} corresponds to a choice of the $(n+2)^{\text{th}}$ poloidal of the flux-surface shape. 

\subsubsection{Solution of Step II : Decoupling the ODEs obtained from the MDE for $\psi^{(n+2)}$}
\label{sec:sec_Step2_normal_form}
In the last Section, we showed how the MDE for $\psi^{(n+2)}$ can be expanded in a finite number of poloidal Fourier harmonics, given in \eqref{eq:psi_and_Y}. Collecting the various harmonics, the MDE for $\psi^{(n+2)}$, which is a PDE, can be reduced to a coupled set of ODEs for the amplitudes, $Y^{(n+2)}_{c m},Y^{(n+2)}_{s m}$, where $0\leq m \leq n+2$. The coupling between $m$ and $m\pm 2$ harmonics can be seen from \eqref{eq:BD0_n_chi_j_n} to involve the quantities $\fTwocTwo,\fTwosTwo$. We now address how to decouple these ODEs. To do so, we need to discuss the normal form representation of $\psi^{(n+2)}$ instead of the poloidal harmonic representation.

The poloidal harmonic structure of $\Phi^{(n+2)}$ and $\psi^{(n+2)}$ are the same order by order, both having a maximum poloidal frequency of $(n+2)$ at order $(n+2)$. This observation is directly related to the ``analyticity" (in the sense of regularity near the axis) of vacuum magnetic fields assumed by Mercier and others. Rigorous proof of the analyticity condition for $\Phi,\psi$ can be found in \citep{Duignan_Meiss2021normal_form,Jorge_Sengupta_Landreman2020NAE}. More concretely, ``analyticity" implies \citep{Duignan_Meiss2021normal_form,Solovev1970}
\begin{align}
    \psi^{(n+2)}\rho^{n+2}=\mathcal{P}_{(n+2)}\lbr X_\cN,Y_\cN \rbr,
    \label{eq:analyiticity_psi}
\end{align}
where, $\mathcal{P}_{(n+2)}$ is a homogeneous polynomial of order $(n+2)$ in $(X_\cN,Y_\cN)$ with coefficients that are functions of $\ell$. Here, the normal form variables $(X_\cN,Y_\cN)$ satisfy \eqref{eq:Rotating_coords}. At $O(n)$, the polynomial $\mathcal{P}_{(n+2)}$ can have at most $(n+2)$ poloidal harmonics  since the normal form variables only have first harmonics. 

In the following, we use the following variables obtained from the normal form variables $(X_\cN,Y_\cN)$
\begin{align}
    x_\cN\equiv\frac{X_\cN}{\rho}= e^{\eta/2}\cos u, \qquad y_\cN\equiv\frac{Y_\cN}{\rho}=e^{-\eta/2}\sin u.
    \label{eq:norm_normal_form}
\end{align}
From the expression of $\psiTwo$ in normal form variables \eqref{eq:psi2_normal_form}, we find that
\begin{align}
    x_\cN= \sqrt{\frac{2\psiTwo}{\Phi_0'}}\cos\Theta, \quad  y_\cN= \sqrt{\frac{2\psiTwo}{\Phi_0'}}\sin\Theta, \quad \tan\Theta = e^{-\eta}\tan u.
     \label{eq:xN_yN_def}
\end{align}
Following \citep{Duignan_Meiss2021normal_form},we now employ complex variables 
\begin{align}
    z_\cN=x_\cN + i y_\cN = \sqrt{\frac{2\psiTwo}{\Phi_0'}}e^{i\Theta}, \qquad \overline{z}_\cN=x_\cN-i y_\cN = \sqrt{\frac{2\psiTwo}{\Phi_0'}}e^{-i\Theta},
    \label{eq:complex_zN_def}
\end{align}
to rewrite the expression of $\psi^{(n+2)}$ in the complex normal form coordinates \eqref{eq:analyiticity_psi} as
\begin{align}
    \psi^{(n+2)}= \sum_{m=0}^{n+2} \Psi^{(n+2)}_{m (n+2-m)}(\ell)\: \overline{z}_\cN^m {z_\cN}^{(n+2-m)}.
    \label{eq:psin+2_normal_form}
\end{align}
The connection between the amplitudes $\Psi^{(n+2)}_{m r}(\ell)$ and the amplitudes $\psi^{(n+2)}_{cm},\psi^{(n+2)}_{sm}$ follows from the definitions \eqref{eq:norm_normal_form} and \eqref{eq:complex_zN_def}. Note that the reality condition of $\psi^{(n+2)}$ implies that $\Psi^{(n+2)}_{m (n+2-m)}= {\overline{\Psi}}^{(n+2)}_{(n+2-m) m}$.

We now substitute the complex normal form expression for $\psi^{(n+2)}$ \eqref{eq:psin+2_normal_form} into the MDE for $\psi^{(n+2)}$ \eqref{eq:MDE_psi_n+2}. For that we first need to evaluate $(\BD)^{(n+2)}_0 \psi^{(n+2)}$. 

In Appendix \ref{app:normal_form}, we prove the following identities
\begin{subequations}
    \begin{align}
    &(\BD)^{(n+2)}_0 \zeta_{m\cN}^{(n+2)} = \Phi_0' \lbr i \Omega_0 (n+2-2m)-\frac{n+2}{2}\frac{\Phi_0''}{\Phi_0'}\rbr \zeta_{m\cN}^{(n+2)},\quad \zeta_{m\cN}^{(n+2)}=\overline{z}_\cN^m\: {z}^{n+2-m}_\cN \label{eq:BDnorm_form_id_1} \\
    &(\BD)^{(n+2)}_0 \lbr \cZ_m(\ell)\zeta_{m\cN}^{(n+2)} \rbr= \Phi_0' \lbr \cZ_m'+ \lbr i \Omega_0 (n+2-2m)-\frac{n+2}{2}\frac{\Phi_0''}{\Phi_0'}\rbr \cZ_m\rbr\zeta_{m\cN}^{(n+2)} \label{eq:BDnorm_form_id_2},
\end{align}
 \label{eq:BDn+2_normal_form_basis}
\end{subequations}
which show that the complex normal form basis diagonalizes the MDO. As before, we get rid of the $\Phi_0''$ term by defining $Y^{(n+2)}={\Phi_0'}^{-(n+2)/2}\psi^{(n+2)}$, equivalently by $\cZ_m\to \cZ_m {\Phi_0'}^{(n+2)/2}$ in \eqref{eq:BDn+2_normal_form_basis}.

Therefore, the substitution of the following form of $ \psi^{(n+2)}$, 
\begin{align}
    \psi^{(n+2)}=\lbr\Phi_0' \rbr^\frac{n+2}{2} \sum_{m=0}^{n+2} \cZ^{(n+2)}_{m (n+2-m)}(\ell)\: \overline{z}_\cN^m {z_\cN}^{(n+2-m)}, \quad \cZ^{(n+2)}_{m (n+2-m)}={\overline{\cZ}}^{(n+2)}_{(n+2-m) m},
\end{align}
into the MDE \eqref{eq:MDE_psi_n+2} leads to a system of $(n+1)$ decoupled complex ODEs of the form
\begin{align}
   \dfrac{d}{d\ell} {\cZ_{m (n+2-m)}^{(n+2)}}  + i \Omega_0 (n+2-2m) \cZ^{(n+2)}_{m (n+2-m)}+ \cF^{(n+2)}_{\cZ_m}=0,
   \label{eq:reduced_MDE}
\end{align}
where $\cF^{(n+2)}_{\cZ_m}$ is an appropriate forcing term. We refer to \eqref{eq:reduced_MDE} as a reduced MDE. The above results also holds for $m=0,(n+2)=1$ as shown in Appendix \ref{app:normal_form}.

The complex normal form \eqref{eq:analyiticity_psi} for $ \psi^{(n+2)}$ is, therefore, of great help in systematically solving the MDE for $\psi$ by reducing it to a system of decoupled ODEs \eqref{eq:reduced_MDE}. In Appendix \ref{app:general_first_order_vacuum} we have shown explicitly how the MDE for $\psiThree$ can be solved by leveraging the normal form structure.

\subsubsection{Solutions of Steps III and IV : Determination of $\alpha$}
\label{sec:Phi_ell_alpha_per}
Our main goal here is to calculate $\alpha^{(n)}$ given the solutions of $\Phi^{(n+2)},\psi^{(n+2)}$, combining Steps III and IV, outlined in Section \ref{sec:series_exp_Order_n}. Step III consists of integrating the equation for $\alpha_{,\omega}$ with respect to $\omega$ to determine $\alpha^{(n)}$ up to a function $\mathfrak{a}(\ell)$. Step IV involves poloidally averaging the MDE for $\alpha^{(n)}$ to get an equation for $\mathfrak{a}(\ell)$.

We start with Step III. Power series expansion of \eqref{eq:vacuum_Phi_ell}, rewritten as
\begin{align}
    \psi_{,\rho}\alpha_{,\omega}-\psi_{,\omega}\alpha_{,\rho}-\frac{\rho}{h}\Phi_{,\ell}=0,
    \tag{\ref{eq:vacuum_Phi_ell}}
\end{align}
at each order, $n$, leads to
\begin{align}
    \sum_{m=0}^n
    \lbr 
    (n-m)  \psi^{(m+2)}_{,\omega}\alpha^{(n-m)}-
    (m+2)  \psi^{(m+2)}\alpha^{(n-m)}_{,\omega} +\lbr \kappa \cos \theta \rbr^{(n-m)} \Phi^{(m)}_{,\ell}
    \rbr =0.
    \label{eq:phi_ell_sum_m}
\end{align}
The function $1/h=(1-\kappa\cos\theta)^{-1}$ has been expanded in powers of $\rho$ as well. 

 We shall now split $\psi^{(n+2)}$ into the homogeneous $\psi^{(n+2)}_H$ solution and the particular solution $\psi^{(n+2)}_P$,
\begin{align}
   \psi^{(n+2)}= \psi^{(n+2)}_P+{\psi_H}^{(n+2)}, \quad {\psi_H}^{(n+2)}= \lbr 2\psiTwo\rbr^{n/2+1}\cY^{(n+2)}_H.
\end{align}
Here, we assume that the on-axis rotational transform is irrational, so $\cY^{(n+2)}_H$ is a constant. Separating the $m=0$ and $m=n$ terms in \eqref{eq:phi_ell_sum_m} from the rest and using the lowest order relation \eqref{eq:Phi2_alpha0_psi2_system} we obtain
\begin{align}
    \del_\omega\lbr\frac{\alpha^{(n)}}{\lbr 2\psiTwo\rbr^{n/2}} \rbr= \frac{\Phi^{(n)}_{,\ell} +\Sigma^{(n)}_1}{\lbr 2\psiTwo\rbr^{n/2+1}}-(n+2)\lbr \frac{\Phi_0'\psi^{(n+2)}_P}{\lbr 2\psiTwo\rbr^{n/2+2}} +\alphaOcw \cY^{(n+2)}_H\rbr,
    \label{eq:del_omega_alpha_n}
\end{align}
where, $\lbr \Sigma^{(n)}_1-(\kappa \cos \theta)^n\Phi_0'\rbr$ is the sum of terms in \eqref{eq:phi_ell_sum_m} from $m=1$ to $(n-1)$.
Imposing the poloidal integral constraint on $\alpha^{(n)}$ as given in \eqref{eq:integral_constraints_alpha} and equating the average of $\alphaOcw$ to unity, as shown in \eqref{eq:int_alphacw_is_1}, we find that
\begin{align}
    \cY^{(n+2)}_H= \oint\frac{d\omega}{\lbr 2\psiTwo\rbr^{n/2+1}}\frac{\lbr \Phi^{(n)}_{,\ell} +\Sigma^{(n)}_1\rbr}{(n+2)} -\Phi_0'\oint\frac{d\omega}{\lbr 2\psiTwo\rbr^{n/2+2}}\psi^{(n+2)}_P.
    \label{eq:YHnp2_expr}
\end{align}
We can now integrate \eqref{eq:del_omega_alpha_n} with respect to $\omega$ to obtain 
\begin{align}
     \lbr\frac{\alpha^{(n)}}{\lbr 2\psiTwo\rbr^{n/2}} \rbr=\overline{\mathfrak{a}}^{(n)}(\ell)+\widetilde{\mathfrak{a}}^{(n)}(\omega,\ell),
     \label{eq:abar_atilde_defn}
\end{align}
where, 
\begin{align}
   \widetilde{\mathfrak{a}}^{(n)}(\omega,\ell)= \int_0^\omega d\omega' \lbr  \frac{\Phi^{(n)}_{,\ell} +\Sigma^{(n)}_1}{\lbr 2\psiTwo\rbr^{n/2+1}}-(n+2)\Phi_0'\frac{\psi^{(n+2)}_P+\psi^{(n+2)}_H}{\lbr 2\psiTwo\rbr^{n/2+2}}\rbr, \quad \oint \frac{d\omega'}{2\pi}\:\:\widetilde{\mathfrak{a}}(\omega',\ell)=0.
\end{align}
Note that the average of $\widetilde{\mathfrak{a}}$ is zero because of the condition \eqref{eq:YHnp2_expr} that determines the homogeneous solution $\psi^{(n+2)}_H$. This completes Step III.

We now proceed with Step IV. To determine $\overline{\mathfrak{a}}^{(n)}(\ell)$, we use the Poisson bracket form of the MDO \eqref{eq:MDO_PB_form} 
\begin{align}
  \lbr \alphaOcw\del_\ell -\alphaOcz\del_\omega\rbr \lbr \frac{\alpha^{(n)}}{\lbr 2\psiTwo \rbr^{n/2}}\rbr +\frac{F^{(n+2)}_\alpha}{\lbr 2\psiTwo\rbr^{n/2+1}}=0,
  \label{eq:MDE_alphan_PB_form}
\end{align}
which was derived earlier in \eqref{eq:vac_MDE_alpha_order_n}. 

Writing \eqref{eq:MDE_alphan_PB_form} in the form \eqref{eq:PB_alpha0_chi} and integrating with respect to $\omega$ between $(0,2\pi)$ we obtain the poloidally averaged MDE
\begin{align}
    \del_\ell \lbr  \oint \frac{d\omega}{2\pi} \alphaOcw  \lbr \frac{\alpha^{(n)}}{\lbr 2\psiTwo \rbr^{n/2}}\rbr \rbr +  \oint \frac{d\omega}{2\pi} \lbr \frac{F^{(n+2)}_\alpha}{\lbr 2\psiTwo\rbr^{n/2+1}} \rbr=0.
\end{align}
We note here that the derivative in $\omega$ vanished because both $\alphaOcw$ and $\lbr \alpha^{(n)}/\lbr 2\psiTwo\rbr^{n/2}\rbr$ are single-valued in $\omega$.

From \eqref{eq:abar_atilde_defn} and \eqref{eq:int_alphacw_is_1} we then obtain the following equation for $\overline{\mathfrak{a}}^{(n)}(\ell)$
\begin{align}
    {\overline{\mathfrak{a}}^{(n)}}'(\ell) + \del_\ell  \oint \frac{d\omega}{2\pi}\:\frac{\Phi_0'}{2\psiTwo}\widetilde{\mathfrak{a}}^{(n)}(\omega,\ell) + \oint \frac{d\omega}{2\pi} \lbr \frac{F^{(n+2)}_\alpha}{\lbr 2\psiTwo\rbr^{n/2+1}} \rbr=0
    \label{eq:abar_n}
\end{align}
The solution of \eqref{eq:abar_n} can be obtained through direct integration with respect to $\ell$. Note that ${\overline{\mathfrak{a}}^{(n)}}'(\ell)$ is not periodic in general since its average is related to the higher order magnetic-shear as given in \eqref{eq:integral_constraints_alpha}. 

We reiterate that the form of $\alpha$ in inverse-coordinates, as given in \eqref{eq:alpha_phi_form_inverse_coord}, is well-known. In particular, the secular terms  $\theta-\iota(\psi)\phi$ remain separate from the periodic components of $\alpha$. However, in direct coordinates, the secular and the periodic terms are mixed because the $\iota(\psi)\phi$ term is now expanded in a power series of $\rho$. This explains the need to take special care in order to obtain $\alpha$ in direct coordinates.

\subsubsection{Summary of Steps I-IV}
We now briefly summarize the main results obtained in Section \ref{sec:periodicity_n_structure_soln}. We have individually analyzed the structure of $(\Phi,\psi,\alpha)$, proving that the necessary periodicity conditions \eqref{eq:alpha_phi_form_inverse_coord} for obtaining physically meaningful solutions can be satisfied order by order. In Section \eqref{sec:Phi_structure} we have discussed the $O(n+2)$ Laplace equation and the structure of its solution $\Phi^{(n+2)}$. We have also given a prescription, \eqref{eq:choice_free_function},  with which the free functions (the homogeneous solution to the Poisson equation) can be expressed in terms of the flux-surface shaping given by \eqref{eq:constant_state_psi_n+2}. 

In Section \ref{sec:sec_vacuum_MDE}, we study the general MDE and obtain some of its fundamental properties. These properties allow one to construct homogeneous solutions of the MDEs and annihilators for the MDEs. Later, we use the homogeneous solution for $\psi^{(n+2)}$ and the annihilator to enforce the periodicity constraint \eqref{eq:integral_constraints_alpha}. The last important property is that the poloidal harmonics of the solution of the MDE truncate if the forcing term has finite poloidal harmonics. Using these properties we found that at $O(n)$, $\Phi^{(n+2)}$ will have at most $(n+2)$ harmonics. The structure of $\alpha^{(n)}$ is more complicated, and solving the MDE for $\alpha$ \eqref{eq:vac_MDE_alpha_order_n} is far more involved than MDE for $\psi$ given by \eqref{eq:MDE_psi_order_n+2}. Fortunately, we do not have to solve the MDE for $\alpha$.

Finally, in Section \ref{sec:Phi_ell_alpha_per}, we demonstrate how one can obtain $\alpha^{(n)}$ once $\Phi^{(n+2)},\psi^{(n+2)}$ has been constructed. This brings us to the end of the analysis of the structure of vacuum equations in the Mercier-Weitzner formalism. In Section \ref{sec:illustration_circ} and Appendix \ref{app:various_examples}, we discuss analytically tractable and physically interesting limits of 3D vacuum fields that possess nested surfaces. We can obtain explicit solutions order by order using the NAE formalism we have developed thus far.

\section{NAE of force-free fields \label{sec:NAE_FF}}
For force-free systems, the current $\J$ and magnetic field $\B$ are co-linear, i.e.,
\begin{align}
    \J=\lambda \B, \quad \J=\dl\times \B.
    \label{eq:basic_force_free}
\end{align}
Since the divergence of $\J$ is zero, we have 
\begin{align}
   \BD \lambda =0.
   \label{eq:MDE_lambda}
\end{align}
We further assume that the force-free fields possess nested flux surfaces. Therefore, \eqref{eq:MDE_lambda} implies that $\lambda$ must be a function of only $\psi$ and $\alpha$. Physically, the $\alpha$ dependence of $\lambda$ is special. It is only possible if the rotational transform is a constant rational number. Any amount of magnetic shear will lead to 
\begin{align}
    \lambda= \lambda(\psi),
\end{align}
which we assume from now on. 

\subsection{Basic formalism for the force-Free fields}

In the Mercier-Weitzner formalism, force-free fields that possess nested flux surfaces can be seen to be of the form
    \begin{align}
        \B =\dl \Phi +\Kb(\psi,\alpha) \dl \psi, \quad \B =\dl\psi\times \dl \alpha,
        \label{eq:Grad_Boozer_form_Force_free}
    \end{align}
    since the current takes the form
    \begin{align}
         \J = \dl \Kb\times \dl \psi = -\Kb_{,\alpha}\B.
         \label{eq:J_K_alpha_B}
    \end{align}
    Comparing \eqref{eq:basic_force_free} with \eqref{eq:J_K_alpha_B} we find that
    \begin{align}
        \Kb = -\lambda(\psi)\: \alpha.
        \label{eq:Kb_sol}
    \end{align}
    Note that a homogeneous solution $\Kb'_0(\psi)$ could be added to \eqref{eq:Kb_sol}. However, from \eqref{eq:Grad_Boozer_form_Force_free} we find that the $\Kb'_0(\psi)$ term can be absorbed by redefining $\Phi\to \Phi-\Kb_0(\psi)$. Hence, we can set $\Kb_0$ to zero without loss of generality.   
   
   To carry out the NAE, we assume that $\lambda(\psi)$ is sufficiently smooth that we can expand $\lambda$ as
    \begin{align}
        \lambda(\psi)=\lambda_0 + \psi \lambda_2 + \psi^2 \lambda_4+...,
        \label{eq:lambda_NAE}
    \end{align}
    where, $\lambda_i, i=0,2,4..$ are constants. However, the usual near-axis expansion in direct coordinates \eqref{eq:NAE_power_series} needs to be checked carefully. From the condition  $\dl\cdot \B=0$ we have
    \begin{align}
        \Delta \Phi + \dl\cdot (\Kb \dl \psi)=0
        \label{eq:divB_force_free}
    \end{align}
    We need to show that the solution of \eqref{eq:divB_force_free} can be expressed as a regular power series near the axis. In the vacuum limit, we obtained \eqref{eq:vac_Laplacian_order_n}, where the right-hand side did not have any harmonics that would beat with the operator on the left-hand side. However, it is not guaranteed that
    at each $O(\rho^{n})$, the term $\dl\cdot (\Kb \dl \psi)$ will not generate such resonant harmonics. 
    
    To further illustrate this point, let us consider the case when $\lambda=$ constant, i.e., $K=-\lambda \alpha$. The  $\dl\cdot\B=0$ condition now reads
        $$\Delta \Phi -\lambda \dl\cdot( \alpha \dl \psi)=0$$
    Expanding as in \eqref{eq:NAE_power_series}, we get
        \begin{align*}
            \Delta \Phi&= \sum_n \rho^{n}\lbr \del^2_{\omega}+(n+2)^2\rbr\Phi^{(n+2)}+ ...\\
            \dl\cdot( \alpha \dl \psi)&= \sum_n \rho^{n}\sum_{i=0}^{n}\lbr \alpha^{(i)}\psi^{(n+2-i)}(n+2)(n+2-i)+\del_\omega\lbr \alpha^{(i)}\psi_{,\omega}^{(n+2-i)}\rbr\rbr +...
        \end{align*}
         Using the Fourier representations: $\psi^{(n+2-i)}\sim \sum\psi^{(n+2-i)}_m e^{i m u},\alpha^{(i)}\sim \sum\alpha^{(i)}_p e^{i p u}$,
        \begin{align*}
           \dl\cdot( \alpha \dl \psi) \sim \sum_n\rho^{n}\sum_{i=0}^{n+2}\sum_{m=0}^{n+2-i}\sum_{p=0}^{i}\lbr\psi^{(n+2-i)}_m\alpha^{(i)}_p e^{i (m+p)u}c_{ni}+...\rbr
        \end{align*}
        where, $c_{ni}(\ell)$ is some overall constant factor. Clearly, resonance can occur when $m+p=n+2$. If such a $(n+2)$ harmonic is generated then we cannot invert \[(\del^2_{\omega}+(n+2)^2)\Phi^{(n+2)}\sim e^{ i (n+2)u}, \]
    to solve for a periodic $\Phi^{(n+2)}$. As shown in \citep{weitzner2016expansions}, if such a resonance occurs, then the regular power series will fail, and weakly singular terms like $\rho^n(\log{\rho})^m$ will appear in the Mercier expansion.
    
    In the following Section, we prove that such logarithmically singular terms can indeed be avoided under reasonable assumptions. However, the boundary cannot be arbitrarily specified and must be self-consistent with the NAE.  
    
    \subsection{Definition of new variables}

Equating $\B = \dl\Phi -\lambda \alpha \dl \psi = \dl \psi\times \dl \alpha$, we get the following system of equations for the force-free magnetic fields with flux surfaces in the Mercier-Weitzner formalism:
\begin{subequations}
    \begin{align}
        \Phi_{,\rho}-\lambda \alpha \psi_{,\rho}&= \frac{1}{\rho h}\lbr \psi_{,\omega}\alpha_{,\ell}- \psi_{,\ell}\alpha_{,\omega}\rbr \label{eq:force_free_phi_rho}\\
        \Phi_{,\omega}-\lambda \alpha \psi_{,\omega}&= \frac{\rho}{h}\lbr \psi_{,\ell}\alpha_{,\rho}- \psi_{,\rho}\alpha_{,\ell}\rbr\\
        \Phi_{,\ell}-\lambda \alpha \psi_{,\ell}&= \frac{h}{\rho }\lbr \psi_{,\rho}\alpha_{,\omega}- \psi_{,\omega}\alpha_{,\rho}\rbr
        \end{align}
    \label{eq:basic_force_free_system}
    \end{subequations}
  As discussed in the last Section, unlike the vacuum limit, it is not immediately evident that resonances can be avoided in the system \eqref{eq:basic_force_free_system}. 
  
  Our goal is, therefore, to bring the force-free system of equations \eqref{eq:basic_force_free_system} as close as possible to the vacuum system given by \eqref{eq:vacuum_system}.
  We now try to define a new variable $\xi$ that will play the role analogous to $\Phi$ in the vacuum limit. It is useful to define a variable $\Lambda$ such that
  \begin{align}
     \lambda(\psi)= \Lambda'(\psi).
     \label{eq:Lambda_def}
  \end{align}
Now, we define $\xi$ such that
    \begin{align}
    \xi_{,\rho}\equiv \Phi_{,\rho}-\lambda \alpha \psi_{,\rho}= \Phi_{,\rho}-\alpha \Lambda_{,\rho}.
        \label{eq:xi_definition}
    \end{align}
    Integrating with respect to $\rho$ we get
    \begin{align}
        \xi = \Phi + \cI,\quad \cI\equiv  -\int_0^\rho d\varrho\: \Lambda_{,\varrho} \alpha.
        \label{eq:Lambda_def_integral}
    \end{align}
Note that $\xi,\Phi$, and $\alpha$ have similar secular terms linear in the angles. Therefore, imposing the form \eqref{eq:secular_n_periodic_Phi_alpha_K} order by order on $\alpha$ is enough to impose the right secular structure of $\xi$.
    
Since ultimately we hope to be able to expand in power series of $\rho$, the integral $\cI$ in \eqref{eq:Lambda_def_integral}, amounts to simply reshuffling the terms and multiplying the coefficients by fractions. With this definition in mind we now rewrite \eqref{eq:basic_force_free_system} in terms of $\xi$ and its derivatives as
\begin{subequations}
    \begin{align}
        \xi_{,\rho}&= \frac{1}{\rho h}\lbr \psi_{,\omega}\alpha_{,\ell}- \psi_{,\ell}\alpha_{,\omega}\rbr \label{eq:force_free_xi_rho_alpha}\\
        \xi_{,\omega}-\Lambda_{,\omega} \alpha-\cI_{,\omega} &= \frac{\rho}{h}\lbr \psi_{,\ell}\alpha_{,\rho}- \psi_{,\rho}\alpha_{,\ell}\rbr \label{eq:force_free_xi_omega_alpha}\\
        \xi_{,\ell}-\Lambda_{,\ell} \alpha-\cI_{,\ell}&= \frac{h}{\rho }\lbr \psi_{,\rho}\alpha_{,\omega}- \psi_{,\omega}\alpha_{,\rho}\rbr
        \label{eq:force_free_xi_ell_alpha}
    \end{align}
    \label{eq:froce_free_system_xi_alpha}
\end{subequations}
The $\dl\cdot\B=0$ condition in terms of $\xi$ reads
\begin{align}
    \rho^2 \Delta \xi -\frac{1}{h}\del_\omega \lbr h\lbr\Lambda_{,\omega} \alpha+\cI_{,\omega} \rbr \rbr -\frac{1}{h}\del_\ell \lbr \frac{\rho^2}{h}\lbr \Lambda_{,\ell} \alpha+\cI_{,\ell}\rbr\rbr=0.
    \label{eq:Laplacian_xi}
\end{align}
Comparing \eqref{eq:froce_free_system_xi_alpha}, \eqref{eq:Laplacian_xi}  to the vacuum limit given by \eqref{eq:vacuum_system} and \eqref{eq:Laplace_for_Phi}, we find that the two differ because of the $(\Lambda_{,\omega} \alpha+\cI_{,\omega})$ term, which is an $O(1)$ term in the $\rho$ expansion. All other deviations from the vacuum limit are higher order in $\rho$.
Fortunately, a workaround is possible by noting that
\begin{align}
    \del_\rho\lbr \Lambda \alpha_{,\omega}\rbr-  \del_\omega\lbr \Lambda \alpha_{,\rho}\rbr = \lambda \lbr \psi_{,\rho}\alpha_{,\omega}-\alpha_{,\rho}\psi_{,\omega} \rbr
    \label{eq:identity_force_free}
\end{align}
Integrating once again in $\rho$, using \eqref{eq:force_free_xi_ell_alpha} and simplifying, we get
\begin{align}
    -\Lambda_{,\omega} \alpha-\cI_{,\omega} =\cW,\quad \cW\equiv\int_0^\rho d\varrho\: \frac{\varrho}{h}\lambda \lbr \xi_{,\ell}+\Xi\rbr, \quad \Xi\equiv -\lbr \Lambda_{,\ell}\alpha + \cI_{,\ell}\rbr.
\end{align}
With these new definitions, we can write down the force-free equations as
\begin{subequations}
    \begin{align}
        \xi_{,\rho}&= \frac{1}{\rho h}\lbr \psi_{,\omega}\alpha_{,\ell}- \psi_{,\ell}\alpha_{,\omega}\rbr=B_\rho \label{eq:force_free_xi_rho_Xi}\\
        \xi_{,\omega}+\cW &= \frac{\rho}{h}\lbr \psi_{,\ell}\alpha_{,\rho}- \psi_{,\rho}\alpha_{,\ell}\rbr= \rho B_\omega \label{eq:force_free_xi_omega_Xi}\\
        \xi_{,\ell}+\Xi&= \frac{h}{\rho }\lbr \psi_{,\rho}\alpha_{,\omega}- \psi_{,\omega}\alpha_{,\rho}\rbr
        \label{eq:force_free_xi_ell_Xi}= h B_\ell
       \end{align}
    \label{eq:froce_free_system_xi_Xi}
\end{subequations}
where,
\begin{align}
      \lambda=\Lambda'(\psi),\;\;
        \cI\equiv - \int_0^\rho d\varrho\: \Lambda_{,\varrho} \alpha,\;\; \Xi\equiv -(\Lambda_{,\ell}\alpha + \cI_{,\ell}), \;\; \cW\equiv \int_0^\rho d\varrho\: \frac{\varrho}{h}\lambda \lbr \xi_{,\ell}+\Xi\rbr.
        \label{eq:def_cI_Xi_FF}
\end{align}
Finally, the Poisson equation for $\xi$ reads
\begin{align}
\rho^2\Delta \xi +\frac{1}{h}\del_\omega\lbr h \cW \rbr +\frac{1}{h}\del_\ell\lbr \frac{\rho^2}{h}\Xi\rbr=0.
    \label{eq:Force_free_Laplacian_xi_Xi}
\end{align}
In Appendix \ref{app:Force_free_MHD_structure}, we show in detail that just like in the vacuum limit, \eqref{eq:Force_free_Laplacian_xi_Xi} can be solved order by order without encountering any resonances. In short, the basic idea of the proof is to note that while the first term in \eqref{eq:Force_free_Laplacian_xi_Xi} is $O(1)$ and of the form $(\del^2_{\omega}+(n+2)^2)\xi^{(n+2)}$, the second and the third term both are $O(\rho^2)$ smaller than the first term. Hence, the second and third term has at most $n^\text{th}$ harmonics, thereby avoiding any resonance with the first term. Thus, the logarithmic terms are avoided. Once $\xi$ has been obtained to $O(\rho^n)$, we can use \eqref{eq:Lambda_def_integral} to determine $\Phi$ to the same order of accuracy.

\subsection{ NAE equations for Force-Free equilibrium}
\label{sec:NAE_Force_free_eqns}
We shall now discuss the lowest-order NAE equations for the force-free equilibrium using the variables defined in the last Section. Finally, we shall discuss the lowest two orders before discussing the general $O(\rho^n)$ system of equations.

Substituting the near-axis power series expansion into the definitions of $\cI,\Xi$ given in \eqref{eq:def_cI_Xi_FF}, we find that
\begin{align}
    \cI =-\lambda_0 \rho^2 \alphaO \psiTwo +O(\rho^3), \;\; \Xi=\rho^2 \lambda_0 \psiTwo \alphaOcz +O(\rho^3), \;\; \cW=\frac{\lambda_0}{2}\xi^{(0)}_{,\ell}\rho^2 +O(\rho^3).
    \label{eq:cI_Xi_FF}
\end{align}
We are now in a position to investigate the Poisson equation for $\xi$, \eqref{eq:Force_free_Laplacian_xi_Xi}. First, we note from  \eqref{eq:cI_Xi_FF} that the contribution of $\Xi$ to \eqref{eq:force_free_xi_omega_Xi} starts at $O(\rho^4)$. Next, we find from \eqref{eq:cI_Xi_FF} that the Poisson equation \eqref{eq:Force_free_Laplacian_xi_Xi} matches the vacuum case exactly to $O(\rho^{-2}), O(\rho^{-1})$. Thus, $\xi$ and $\Phi$ differ only from the second order onward, i.e.,
\begin{align}
\xi^{(0)}=\Phi_0(\ell), \quad \xi^{(1)}=0, \qquad B_0=\Phi_0'(\ell).
    \label{eq:xi0_xi1_expressions_FF}
\end{align}
Consequently, we note that even though $\cW\sim O(\rho^2)$ its contribution to \eqref{eq:Force_free_Laplacian_xi_Xi} through the $\del\omega(h \cW)$ term is $O(\rho^3)$ since it is a function of $\ell$ to lowest order. Therefore, even the $O(1)$ Poisson equation assumes the same form as in the vacuum case, namely,
   \begin{align}
       \lbr \del^2_{\omega} +4 \rbr\xiTwo +\Phi_0''(\ell)=0,
       \label{eq:Poisson_xi2}
   \end{align}
whose solution is given by
   \begin{align}
       \xiTwo&=-\dfrac{1}{4}\Phi_0''(\ell)+\Phi_0'(z)\lbr\fTwo_{c2}(\ell)\cos{2u}+\fTwo_{s2}(\ell)\sin{2u}\rbr,\\
       u&=\theta +\delta(\ell)=\omega-\int\tau d \ell +\delta(\ell). \nonumber
   \end{align} 

The components of the magnetic field are given by
\begin{align}
    &B_\rho=2\rho \xiTwo + O(\rho^3), \quad B_\omega = \rho \lbr \xiTwo_{,\omega} + \frac{\lambda_0}{2} \Phi_0' \rbr, \quad B_\ell = (1+\kappa \rho \cos\theta)\Phi_0' + O(\rho^2), \label{eq:Bs_in_FF}
\end{align}
which implies that
\begin{align}
    (\BD)_0^{(n)}= \Phi_0' \del_\ell + \lbr \xiTwo_{,\omega} + \frac{\lambda_0}{2} \Phi_0' \rbr\del_\omega +2 n \xiTwo.
    \label{eq:MDO_FF_MHD}
\end{align}
Another implication of \eqref{eq:Bs_in_FF} is that the $O(\rho^2)$ NAE equations are
\begin{subequations}
\begin{align}
      2\psiTwo \alphaOcw =\Phi_0', \quad 2\psiTwo \alphaOcz =-\lbr \xiTwo_{,\omega} +\frac{\lambda_0}{2} \Phi_0'\rbr \label{eq:alpha0_cw_cz_FF} \\
      2\xiTwo =\alphaOcz \psiTwo_{,\omega}-\alphaOcw \psiTwo_{,\ell}. \label{eq:xi2_PB_alpha0_psi2_eqn_FF}
\end{align}
    \label{eq:Zeroth_order_FF}
\end{subequations}

The lowest order flux surface and field line label can be shown to be the one obtained by Mercier \citep{Mercier1964}
   \begin{subequations}
         \begin{align}
       \psi^{(2)}&=\Phi_0'(\ell)(a(\ell)+b(\ell) \cos{2u})\\ a(\ell)&=\dfrac{1}{2} \cosh(\eta(\ell)),\quad b(\ell)=\dfrac{1}{2} \sinh(\eta(\ell))\\
       \fTwo_{s2}(\ell)&= \dfrac{b}{2a}\lbr \delta'-\tau+\frac{\lambda_0}{2}\rbr\quad ,\quad \fTwo_{c2}(\ell)=-\dfrac{a'}{4b}=-\frac{\eta'}{4}\\
       \alpha^{(0)}&= \tan^{-1}\lbr e^{-\eta}\tan u\rbr-\int d\ell\lbr \dfrac{\lbr \delta'-\tau+\lambda_0/2\rbr}{\cosh{\eta}}\rbr
   \end{align}
   \label{eq:FF_psi2_xi2_alpha0}
   \end{subequations}
As in the vacuum case, we choose $2\pi \psi$ as the toroidal flux. A comparison with the vacuum solutions given in Section \ref{eq:vacuum_lowest_orders} shows minimal changes from vacuum solutions. The main difference arises because of the extra $\lambda_0/2$ factor in the expression for $\alphaOcz$ in \eqref{eq:alpha0_cw_cz_FF}. While the structure of the equations stays the same, the forcing functions, $F_\phi,F_\psi, F_\alpha$, get modified to include the effects of force-free currents. 

The on-axis rotational transform for the force-free case is given by
\begin{align}
    \iota_0&= \alpha(\ell,\theta)-\alpha(\ell+L,\theta)+N = \frac{v(L)-(\delta(L)-\delta(0))}{2\pi}+N, \nonumber\\
    v(\ell)&= \int_0^\ell ds \frac{\delta'(s)-\tau(s)+\frac{\lambda_0}{2}}{\cosh{\eta(s)}}.
    \label{eq:Force_free_iota}
\end{align}
As shown by \citep{mercier1974lectures}, pressure does not affect the on-axis rotational transform. Therefore, the on-axis rotational transform given in \eqref{eq:Force_free_iota} also applies to the general MHD case.

We now proceed to describe the structure of the $O(\rho^{(n+2)})$ equations. We note that just like in the vacuum case, equation \eqref{eq:force_free_xi_rho_Xi} gives us an MDE for $\psi^{(n+2)}$. To obtain the MDE for $\alpha^{(n)}$ we proceed exactly as in the vacuum limit by eliminating $\psi^{(n+2)}$ from \eqref{eq:force_free_xi_omega_Xi} and \eqref{eq:force_free_xi_ell_Xi}. Therefore, following the vacuum case, we can write down the steps involved in solving the Force-Free equations
$(\xi^{(n+2)},\psi^{(n+2)},\alpha^{(n)})$. The steps are precisely the ones we discussed in the vacuum limit.
\begin{enumerate}[I.    ]
    \item Poisson equation \eqref{eq:Force_free_Laplacian_xi_Xi} for $\xi^{(n+2)}$
    \item The MDE for $\psi^{(n+2)}$ from \eqref{eq:force_free_xi_rho_Xi}
    \item The $\alpha_{,\omega}$ equation from \eqref{eq:force_free_xi_ell_Xi} to solve for $\alpha^{(n)}$ up to a function $\mathfrak{\overline{a}}^{(n)}(\ell)$
    \item Poloidal $\omega$ average of the MDE for $\alpha$ to determine $\mathfrak{\overline{a}}^{(n)}(\ell)$.
\end{enumerate}
We present a detailed demonstration of the above steps for $n=1$ in Appendix \ref{app:NAE_order_by_order_MHD}.

\section{MHD equilibrium with finite $\beta\sim p'/B^2$ \label{sec:NAE_MHD} }
\subsection{Basic formalism for the MHD fields: Current potentials}

In the Mercier-Weitzner formalism, as mentioned in Section \ref{sec:Mercier_Weitzner}, we have an inhomogeneous MDE of the form
    \begin{align}
     \BD K =p'(\psi)
     \tag{\ref{eq:MDE_K}}
    \end{align}
that must be solved to obtain the current potential $K$. In the force-free limit, we obtained the homogeneous solution $K=\Kb=-\lambda(\psi)\alpha$. In the presence of finite pressure gradients, in addition to $\Kb$, we also need to solve for the inhomogeneous solution, $G$, such that 
\begin{align}
     K=\Kb(\psi,\alpha) +G(\rho,\omega,\ell), \quad \Kb(\psi,\alpha)=-\lambda \alpha, \quad \BD G=p'(\psi).
     \label{eq:split_of_K}
\end{align}
Equation \eqref{eq:split_of_K} implies
\begin{align}
     \J= \dl K\times \dl \psi = \lambda(\psi) \B +\dl G\times \dl \psi .
    \label{eq:MW_B_J_form}
\end{align}
Thus, the quantity $G$ is the pressure-driven part of the current potential. Note that, in general, the function $\lambda$ is not the parallel current ($\jpl$) since
\begin{align}
    \jpl \equiv \frac{\J \cdot \B}{B^2}= \lambda(\psi) -\frac{\B\times \dl\psi\cdot \dl G}{B^2}
    \label{eq:jpl_def}
\end{align}
Only in the force-free limit ($p'=0, G=0$) $\jpl$ equal $\lambda$. Also, note that $G$ is not single-valued. However, the secular parts of $G$ need to be of the same form as $K$ and $\alpha$ in \eqref{eq:secular_n_periodic_Phi_alpha_K}.

Let us first look at the axisymmetric limit to understand better the relation between $\lambda,G$, and the parallel current. The axisymmetric magnetic field and the current in a tokamak are given by \citep{freidberg_idealMHD}
\begin{align}
    \B =F(\Psi)\dl \zeta + \dl \zeta \times \dl \Psi, \quad \J = -F'(\Psi)\B -R^2p'(\Psi)\dl \zeta,
    \label{eq:axisym_B}
\end{align}
where, $(R,\zeta,z)$ denotes the standard cylindrical coordinate system, $F(\Psi)$ is the poloidal current, and $\Psi$ is the poloidal flux function that satisfies $d\Psi/d\psi=\iota(\psi)$ together the Grad-Shafranov equation
\begin{align}
\dlts \Psi + F F' + R^2 p'(\Psi)=0, \quad \dlts = R^2\dl\cdot \lbr \frac{1}{R^2}\dl \rbr .
    \label{eq:GS_axisym}
\end{align}
Furthermore, in axisymmetry, the toroidal angle dependence can only be through secular terms. Thus,
\begin{align}
    G=\Gb(\Psi(R,z))\zeta + \widetilde{G}(R,z), \qquad \Gb \BD \zeta+\BD \widetilde{G}=\iota p'(\Psi) .
    \label{eq:axisym_G}
\end{align}
Using the following axisymmetric relations \citep{freidberg_idealMHD,helander2014theory} 
\begin{align}
   \jpl= -F'(\Psi)-\frac{p'(\Psi)F}{B^2}, \quad \B\times \dl \psi \cdot \dl = F \BD -B^2 \del_\zeta, \quad 
\end{align}
and \eqref{eq:axisym_G}, the pressure driven terms in the equation for $\lambda$, \eqref{eq:jpl_def}, cancel resulting in
\begin{align}
-\lambda(\psi) = F'(\Psi)+\frac{1}{\iota(\psi)}\Gb(\Psi).
    \label{eq:axisym_lambda_Fp_Gb}
\end{align}
Equation \eqref{eq:axisym_lambda_Fp_Gb} shows that in the force-free limit $-\lambda(\psi)= F'(\Psi)$, but for finite-beta, the pressure-driven current also contributes to $\lambda$ through the secular term.

\subsection{NAE of MHD fields}
We now discuss the NAE of MHD fields. We postulate that $\Kb,G$ have the following expansions
\begin{subequations}
    \begin{align}
    \Kb(\psi,\alpha)&=-\alpha (\lambda_0 +\lambda_2 \psi +\lambda_4 \psi^2+\dots),\\
    G(\rho,\omega,\ell)&=\GO(\ell)+\GOne(\omega,\ell)\rho +\GTwo(\omega,\ell) \rho^2+\dots
\end{align}
\label{eq:NAE_Kbar_G_MHD}
\end{subequations}
where each quantity is finite and non-singular. By allowing power-series expansions of $\Kb,G$, we are explicitly excluding current profiles that are singular near the axis. It is beyond the scope of this work to account for the effects of near-axis current singularities and will be left for future publications. 

Carrying out the NAE assuming power series expansion in $\rho$, we get the following coupled PDE system at each order, $n$:
\begin{subequations}
\begin{align}
        (\BD G)^{(n)}&=p^{(n+2)}+\dots\label{eq:naive_MDE_G},\\
        (\BD \psi)^{(n)}&=\dots\label{eq:naive_MDE_psi},\\
        (\Delta \Phi + \dl\cdot (K \dl \psi))^{(n)}&=0
        \label{eq:naive_Laplacian_Phi}.
    \end{align}
    \label{eq:naive_MHD_NAE}
\end{subequations}
We have precisely two MDE for $G$ and $\psi$ and one elliptic equation for $\Phi$. The function $\alpha$ is obtained from $\Phi,\psi$ without solving an MDE for $\alpha$. Therefore, the PDE is of mixed elliptic-hyperbolic type as is well-known \citep{weitzner2014ideal,Garabedian2012computational}. However, as discussed in Section \ref{sec:NAE_FF}, it is not immediately obvious that \eqref{eq:naive_Laplacian_Phi} can be solved for $\Phi$ without encountering logarithmic singularities. In principle, the logarithmic singularities can still arise even if the current profiles are chosen to be smooth near the axis by imposing \eqref{eq:NAE_Kbar_G_MHD}.
  
\subsection{Definitions of new variables}
The change of variables is the same as in the force-free case, with additional terms from $G$. We shall first the covariant and the contravariant forms of $\B$
\begin{align}
    \dl \Phi +(-\lambda \alpha +G )\dl \psi =\dl \psi\times \dl \alpha
    \label{eq:cov_contra_B}.
\end{align}
Then we define the new variable $\xi$, such that
\begin{align}
    \xi_{,\rho}=\Phi_{,\rho}+(-\lambda \alpha +G)\psi_{,\rho}.
    \label{eq:xi_c_rho_form_MHD}
\end{align}
Integrating \eqref{eq:xi_form_MHD} with respect to $\rho$ we get
\begin{align}
    \xi = \Phi +\cI +\cG, \quad \cI \equiv  -\int_0^\rho d\varrho\:\Lambda_{,\varrho} \alpha, \quad \cG=\int_0^\rho d\varrho\:G \psi_{,\varrho}.
    \label{eq:xi_form_MHD}
\end{align}
The $(\rho,\omega,\ell)$ components of \eqref{eq:cov_contra_B} then yields
\begin{subequations}
    \begin{align}
    \xi_{,\rho}&= \frac{1}{\rho h}\lbr \psi_{,\omega}\alpha_{,\ell}- \psi_{,\ell}\alpha_{,\omega}\rbr=B_\rho \label{eq:MHD_B_rho},\\
     \xi_{,\omega}+\cY -\lbr \cI_{,\omega}+\alpha \Lambda_{,\omega}\rbr &= \frac{\rho}{h}\lbr \psi_{,\ell}\alpha_{,\rho}- \psi_{,\rho}\alpha_{,\ell}\rbr= \rho B_\omega 
        \label{eq:MHD_B_omega},\\
     \xi_{,\ell}+\Xi+\cX&= \frac{h}{\rho }\lbr \psi_{,\rho}\alpha_{,\omega}- \psi_{,\omega}\alpha_{,\rho}\rbr
        \label{eq:MHD_B_ell}= h B_\ell.
\end{align}
\label{eq:preMHD_system}
\end{subequations}
While the $B_\rho$ and $B_\ell$ equations above show minimal change when we go to the vacuum limit, the $B_\omega$ equation needs modifications similar to the force-free case. Utilizing the same trick \eqref{eq:identity_force_free} that we used before in the force-free case, we find that 
\begin{align}
    -\lbr \cI_{,\omega}+\alpha \Lambda_{,\omega}\rbr= \int_0^\rho d\varrho\: \frac{\varrho}{h}\lambda \lbr \xi_{,\ell}+\Xi+\cX\rbr.
\end{align}

Therefore, the final set of MHD equations has the form
\begin{subequations}
    \begin{align}
        \xi_{,\rho}&= \frac{1}{\rho h}\lbr \psi_{,\omega}\alpha_{,\ell}- \psi_{,\ell}\alpha_{,\omega}\rbr=B_\rho \label{eq:MHD_xi_rho_Xi},\\
        \xi_{,\omega}+\cY+ \cW &= \frac{\rho}{h}\lbr \psi_{,\ell}\alpha_{,\rho}- \psi_{,\rho}\alpha_{,\ell}\rbr= \rho B_\omega \label{eq:MHD_xi_omega_Xi},\\
        \xi_{,\ell}+\Xi+\cX&= \frac{h}{\rho }\lbr \psi_{,\rho}\alpha_{,\omega}- \psi_{,\omega}\alpha_{,\rho}\rbr
        \label{eq:MHD_xi_ell_Xi}= h B_\ell,\\
        \text{where,} \quad \lambda=\Lambda'(\psi),\quad \cI&\equiv  -\int_0^\rho d\varrho\: \Lambda_{,\varrho} \alpha \:\:, \:\: \Xi\equiv -\lbr \Lambda_{,\ell}\alpha + \cI_{,\ell}\rbr,\:\: \label{eq:def_cG_cY_cX_MHD}\\
        \cG=\int_0^\rho d\varrho\:G \psi_{,\varrho}, \:\: \cY&\equiv G \psi_{,\omega} - \cG_{,\omega},\:\:\cX\equiv G \psi_{,\ell} - \cG_{,\ell},\;\;
        \cW=\int_0^\rho d\varrho\: \frac{\varrho}{h}\lambda \lbr \xi_{,\ell}+\Xi+\cX\rbr\nonumber .
    \end{align}
    \label{eq:MHD_system_xi_Xi}
\end{subequations}
The Poisson equation for $\xi$ is now given by
\begin{align}
\rho^2\Delta \xi +\frac{1}{h}\del_\omega\lbr h \lbr \cY+ \cW\rbr \rbr +\frac{1}{h}\del_\ell\lbr \frac{\rho^2}{h}(\Xi+\cX)\rbr=0.
    \label{eq:MHD_Laplacian_xi_Xi}
\end{align}
As in the force-free case, $\xi$ plays the analogous role of $\Phi$ in the vacuum limit, and it can be shown (see Appendix \ref{app:Force_free_MHD_structure}) that the Poisson equation for $\xi$ is free of any resonances.

\subsection{NAE of MHD equations}
The expansions of the quantities $\cI, \Xi$ that appear in \eqref{eq:MHD_system_xi_Xi} have already been discussed in \eqref{eq:cI_Xi_FF}. From the NAE, \eqref{eq:NAE_Kbar_G_MHD} and \eqref{eq:NAE_power_series_gen} we find that the quantities $\cG,\cY,\cX,\cI,\Xi$, as defined in \eqref{eq:def_cG_cY_cX_MHD}, have the following forms \begin{align}
    \cG&= \rho^2 \GO(\ell) \psiTwo +O(\rho^3), \quad \cY =O(\rho^3),\quad \cX =-\rho^2 {\GO}'(\ell) \psiTwo +O(\rho^3),
    \label{eq:cG_cY_cX_NAE}\\
    \cI&=-\lambda_0 \rho^2 \alphaO \psiTwo +O(\rho^3), \;\; \Xi=\rho^2 \lambda_0 \psiTwo \alphaOcz +O(\rho^3),\;\;\cW=\frac{\lambda_0}{2}\xi^{(0)}_{,\ell}\rho^2 +O(\rho^3).\nonumber
\end{align}
Therefore, the deviation of the MHD system \eqref{eq:MHD_system_xi_Xi} from the vacuum system \eqref{eq:vacuum_system} is zero to $O(\rho^{-2}),O(\rho^{-1})$, just like the force-free equations \eqref{eq:froce_free_system_xi_Xi}. Moreover, \eqref{eq:cG_cY_cX_NAE} also implies that the expressions for the lowest order components of $\B$ are the same as in the force-free case \eqref{eq:Bs_in_FF}. As a result, MHD and force-free both have the same form of MDO given by \eqref{eq:Bs_in_FF}. Consequently, the basic framework is analogous to the vacuum system and the effects of currents and pressure show up mostly through the forcing terms $F_\phi,F_\psi,F_\alpha$.

The only additional step in MHD is Step 0, which involves solving the MDE for $G$ order by order. Step 0 follows the same procedure that we have discussed for the solution of the generic MDE in Section \ref{sec:sec_vacuum_MDE}. To $O(\rho^0)$, the MDE for $G$ is given by   
\begin{align}
     G^{(0)}_{,\ell}\Phi_0'(\ell)=\pTwo,  \label{eq:MDE_G0}
\end{align}
which can be easily solved to obtain the following 
    \begin{align}
        G^{(0)}=p^{(2)}\int \dfrac{d\ell}{\Phi_0'(\ell)}, \quad K^{(0)}= G^{(0)}-\lambda_0\alphaO.
    \end{align}
The details of the general first-order solutions are given in Appendix \ref{app:NAE_order_by_order_MHD}. Due to the similarities between the force-free and the vacuum problem, we do not explicitly show Steps I-IV.

\section{Construction of analytical solutions in the Mercier-Weitzner formalism}
\label{sec:illustrations}
We have completed the development of the Mercier-Weitzner formalism of the direct NAEs. In this Section, we shall provide explicit analytical solutions to vacuum, force-free, and MHD fields that can be obtained within the Mercier-Weitzner formalism.

\subsection{Examples of vacuum fields}
\label{sec:illustration_circ}
We now discuss the next to leading order, $n=1$, solutions to illustrate some of the key features of the NAE for the vacuum case. Our goal is also to clarify some of the steps in the Mercier-Weitzner formalism discussed earlier. We begin with the straightforward case where the lowest order flux-surface $\psiTwo$ has a circular cross-section. The general case (with rotating ellipse, triangularity etc) is described in Appendix \ref{app:general_first_order_vacuum}. 

We consider the case where $a(\ell)=1/2,b(\ell)=0$ in \eqref{eq:a_b_definitions}, which corresponds to a circular cross-section to the lowest order. For this case, the NAE equations to $O(1)$ \eqref{eq:Phi2_alpha0_psi2_system} imply that
\begin{align}
    \psiTwo=\frac{1}{2}\Phi_0', \quad \alphaO= \omega =\theta-\int \tau d\ell , \quad \PhiTwo= -\frac{\Phi_0''}{4}.
    \label{eq:psi2_alpha0_phi2_circ}
\end{align}
To $O(\rho)$ the relevant equations for the variables $(\PhiThree,\alphaOne,\psiThree)$ takes the form
\begin{subequations}
    \begin{align}
        -(\del^2_\omega+3^2)\PhiThree = \cos\theta \lbr \kappa' \Phi_0'+\frac{5}{2}\kappa \Phi_0''\rbr + \sin\theta\:  \kappa \tau \Phi_0'
        \label{eq:Phi3_eqn_circ},\\
        \lbr \Phi_0'\del_\ell -\frac{3}{2}\Phi_0''\rbr \psiThree + \Phi_0'\lbr 3 \PhiThree+\Phi_0''\kappa \cos \theta \rbr =0
        \label{eq:MDE_psi3_circ},\\
        - \Phi_0'\alphaOne_{,\omega}=\lbr 3\psiThree-\Phi_0' \kappa \cos \theta\rbr \label{eq:alpha1_omega_eqn_circ}.
    \end{align}
    \label{eq:Order_rho_1_system_circ}
\end{subequations}
Equation \eqref{eq:Phi3_eqn_circ} is the Poisson equation for $\PhiThree$ whose solution can be readily seen to be
\begin{align}
    \PhiThree= -\frac{1}{8}\cos u \lbr \kappa'\Phi_0'+\frac{5}{2}\kappa \Phi_0'' \rbr -\frac{\Phi_0'}{8}\kappa \tau \sin u -\frac{\lbr \Phi_0'\rbr^{3/2}}{9} \frac{\del}{\del \ell}\lbr \QThree \cos{3u}+\PThree \sin{3u}\rbr.
    \label{eq:Phi3_soln_circ}
\end{align}
Here, $u=\theta$, and we have used \eqref{eq:choice_free_function}, the Soloviev form of the free function, in representing $\PhiThree$.

Next, we solve equation \eqref{eq:MDE_psi3_circ}, which is the MDE for $\psiThree$. The first step is to eliminate the linear term in $\psiThree$ through a change of variables,
\begin{align}
    \psiThree= \lbr \Phi_0' \rbr^{3/2}\YThree, \quad \YThree_{,\ell}+\frac{1}{\lbr \Phi_0' \rbr^{3/2}}\lbr 3\PhiThree+\Phi_0'' \kappa \cos\theta\rbr=0.
    \label{eq:S3_eqn_circ}
\end{align}
Substituting the expression for $\PhiThree$ \eqref{eq:Phi3_soln_circ} into \eqref{eq:S3_eqn_circ} we find
\begin{align}
    \frac{\del}{\del \ell}\lbr \YThree -\frac{3}{8\sqrt{\Phi_0'}}\kappa \cos{u} -\frac{1}{3}\lbr \QThree \cos{3u}+\PThree \sin{3u}\rbr\rbr+ F^{(3)}_{\psi c1}\cos u =0 \label{eq:S3_MDE_circ},
\end{align}
with
\begin{align}
     F^{(3)}_{\psi c1}=-\frac{1}{{\lbr \Phi_0' \rbr^{3/2}}} \frac{\kappa}{8}\Phi_0''.
\end{align}
The solution of \eqref{eq:MDE_psi3_circ} is, therefore, of the form
\begin{align}
    \psiThree =\frac{3}{8}\Phi_0'\kappa \cos u+\lbr\Phi_0'\rbr^{3/2}\lbr \frac{1}{3}\lbr \QThree \cos{3u}+\PThree \sin{3u}\rbr+\lbr \sigma^{(3)}_{c1}\cos u + \sigma^{(3)}_{s1}\sin u \rbr\rbr,
    \label{eq:psi3sol_circ}
\end{align}
where, $ \sigma^{(3)}_{c1}, \sigma^{(3)}_{s1}$ satisfy
\begin{align}
    {\sigma^{(3)}_{c1}}'-\tau \sigma^{(3)}_{s1}+F^{(3)}_{\psi c1}=0, \quad 
     {\sigma^{(3)}_{s1}}'+\tau \sigma^{(3)}_{c1}=0.
     \label{eq:sigma3c1_s1_eqn}
\end{align}
Note that due to our choice of the free function in \eqref{eq:Phi3_soln_circ}, the third harmonics of $\psiThree$ is obtained explicitly in \eqref{eq:psi3sol_circ}. We can now combine \eqref{eq:sigma3c1_s1_eqn} into a single complex equation
\begin{align}
   &{ \cZ^{(3)}_\psi}'-i \Omega_0 \tau { \cZ^{(3)}_\psi} + \cF^{(3)}_\psi=0,\\
   & \cZ^{(3)}=\sigma^{(3)}_{c1} + i \sigma^{(3)}_{s1}, \quad \cF_\psi^{(3)}=F^{(3)}_{\psi c1} + i F^{(3)}_{\psi s1},\quad  \Omega_0=-\tau.\nonumber
\end{align}
Note that we can add a homogeneous solution to $\psiThree$ of the form \begin{align}
    \psiThree_H=\lbr \Phi_0' \rbr^{3/2}\YThree_H
\end{align}
We now solve \eqref{eq:alpha1_omega_eqn_circ} to obtain $\alphaOne$ up to a function  $\mathfrak{a}^{(1)}(\ell)$. Substituting the solution of $\psiThree$ and integrating 
\eqref{eq:alpha1_omega_eqn_circ} with respect to $\omega$, we obtain
\begin{align}
    \alphaOne &= -\frac{\kappa}{8} \sin u -\frac{\lbr \Phi_0'\rbr^{1/2}}{3}\lbr \QThree \sin{3u}-\PThree \cos{3u}\rbr\\ &\quad +\mathfrak{a}^{(1)}(\ell)-3\lbr \Phi_0'\rbr^{1/2}\lbr \sigma^{(3)}_{c1}\sin u - \sigma^{(3)}_{s1}\cos u \rbr.\nonumber
\end{align}
To determine $\mathfrak{a}^{(1)}(\ell)$ we need the MDE for $\alphaOne$,
\begin{align}
       -\frac{1}{2}\Phi_0'' \alphaOne+ \Phi_0'\alphaOne_{,\ell} +\PhiThree_{,\omega}=0.
        \label{eq:MDE_alpha_eqn_circ}
\end{align}
Since $\alphaOne$ and $\PhiThree$ only have odd harmonics, the poloidal average of \eqref{eq:MDE_alpha_eqn_circ} is zero. Hence, both  $\YThree_H$ and $\mathfrak{a}^{(1)}(\ell)$ are zero. 

When the on-axis magnetic field, $\Phi_0'=B_0$, is a constant, the NAE results simplify even further. For $\Phi_0''=0$, we have $\sigma^{(3)}_{c1}=\sigma^{(3)}_{s1}=0$, and we obtain the following explicit forms for the lower-order quantities:
\begin{subequations}
   \begin{align}
    \PhiTwo&=0,\quad \psiTwo=\frac{1}{2}B_0, \quad \alphaO= \omega =\theta-\int \tau d\ell , \quad u=\theta
    \label{eq:psi2_alpha0_phi2_circ_B0},\\
    \PhiThree&=-B_0\frac{\del}{\del \ell}\lbr \frac{\kappa}{8}\cos{u}+\frac{\sqrt{B_0}}{9}  \lbr \QThree \cos{3u}+\PThree \sin{3u}\rbr \rbr\label{eq:Phi3_circ_B0},\\
    \psiThree&=B_0\lbr\frac{3}{8}\kappa \cos u +\frac{\sqrt{B_0}}{3}  \lbr \QThree \cos{3u}+\PThree \sin{3u}\rbr\rbr\label{eq:psi3_circ_B0},\\
    \alphaOne&=-\frac{\kappa}{8}\sin{u} -\frac{\sqrt{B_0}}{3}\lbr \QThree \sin{3u}-\PThree \cos{3u}\rbr.
\end{align}
\label{eq:lower_order_circ_B0}
\end{subequations}

\subsection{Examples of force-free and MHD fields}
\label{sec:examples_FF_MHD}

\begin{figure}
    \centering
    \includegraphics[width=0.45\textwidth]{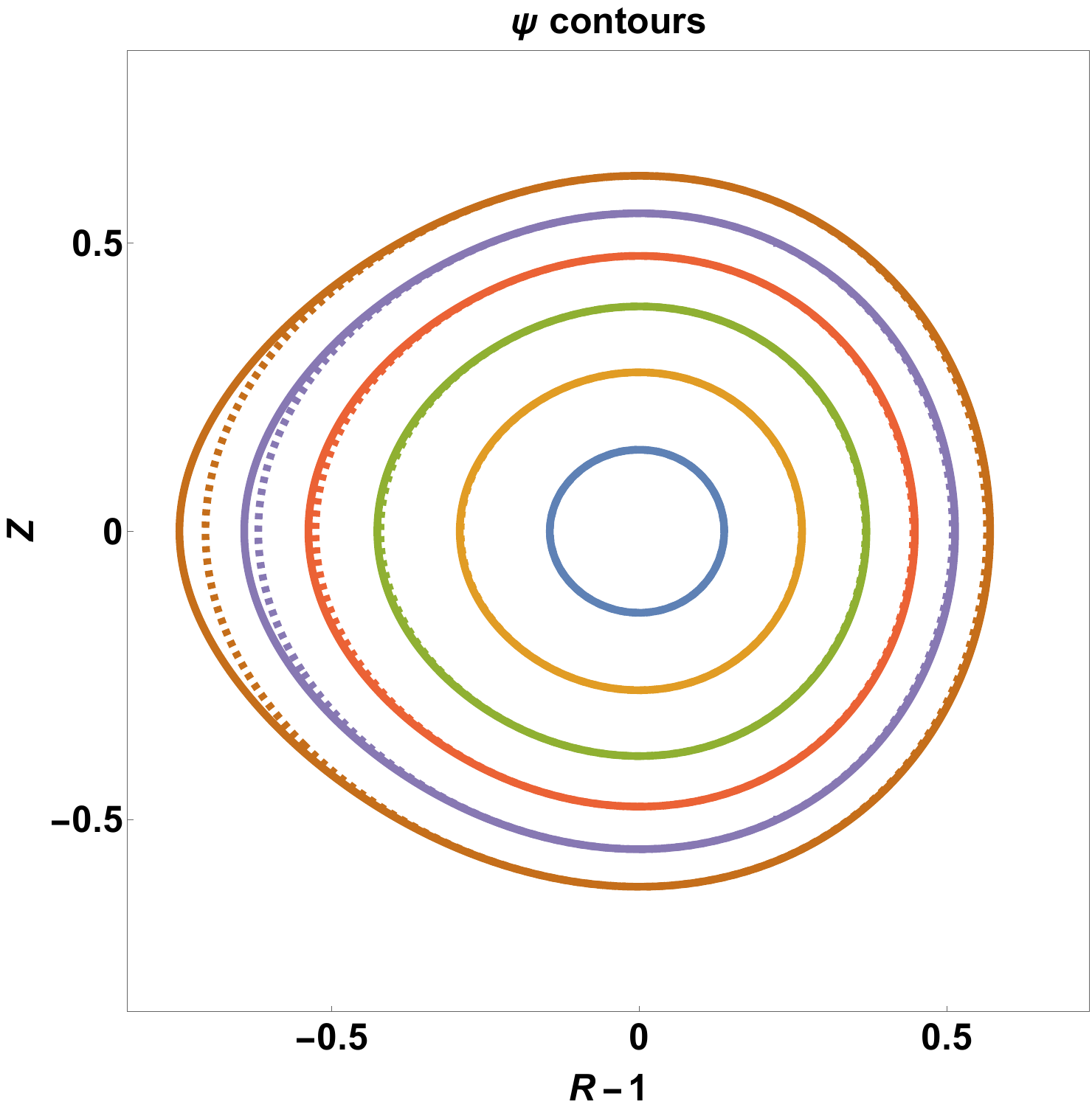}
    \includegraphics[width=0.45\textwidth]{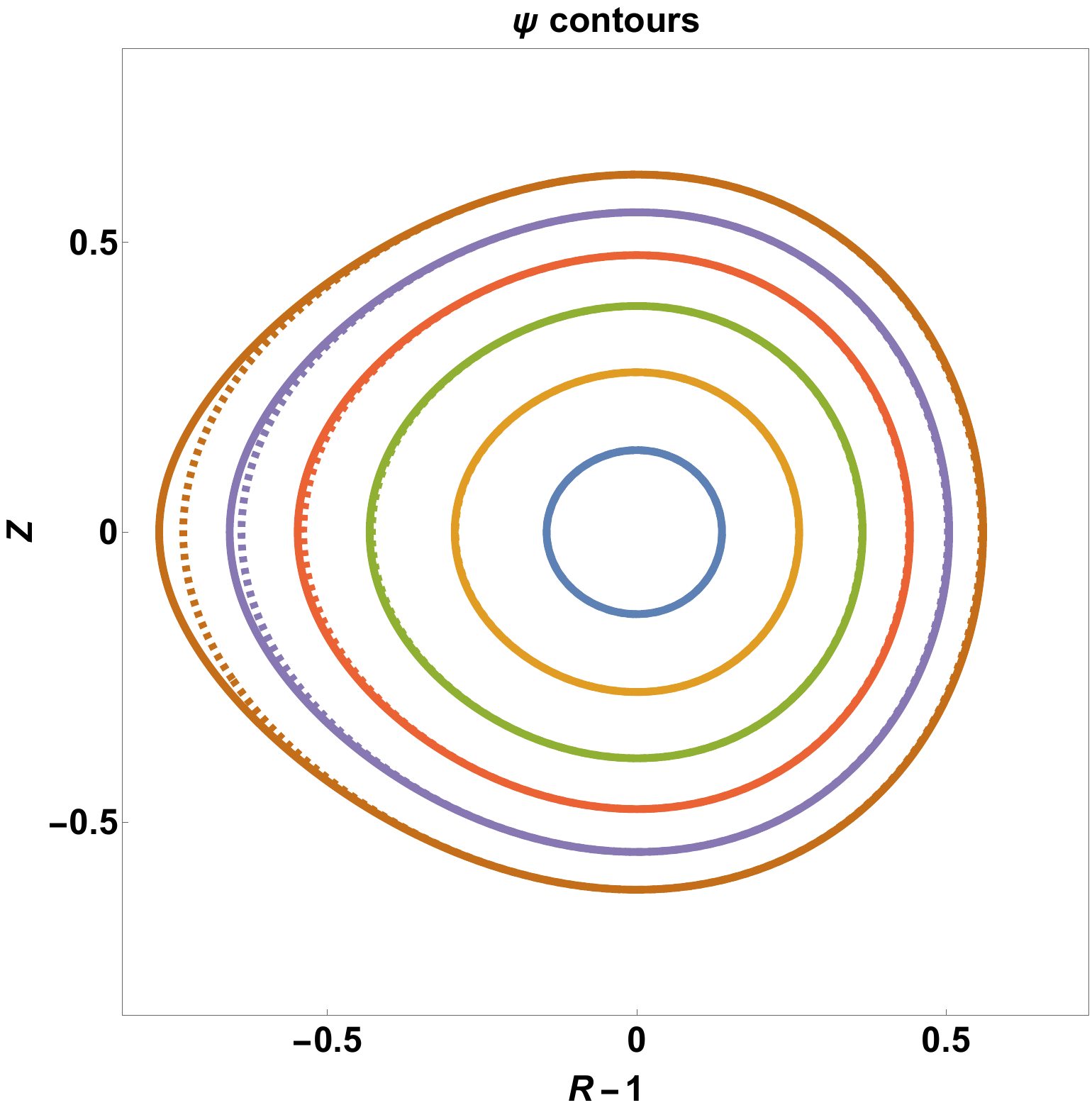}
    \caption{\textbf{Comparison of exact axisymmetric force-free (left) and MHD (right) equilibrium} at 5\% plasma $\beta$ obtained for Soloviev profiles to the NAE. The effect of $\beta$ is barely visible. The NAE matches exactly to the 1-size model up to $O(\rho^4)$. Note that the NAE deviates significantly from the exact solution only when the aspect ratio is sizable ($\approx 1/2$). }
    \label{fig:1size}
\end{figure}

We shall now compare the NAE of force-free and MHD equilibrium with the ``one-size fits all" Soloviev profiles \citep{cerfon2010_1_size}. The Soloviev profiles make the Grad-Shafranov equation \eqref{eq:GS_axisym} linear and, therefore, analytically tractable. In the following, we shall restrict ourselves to first-order NAE, i.e., the accuracy of $\psi$ to $O(\rho^4)$.

To the required order of accuracy, the one-size model is given by \citep{cerfon2010_1_size}
\begin{align}
    \Psi(x,y)=\frac{x^4}{8}+A\lbr \frac{x^2}{2}\log{x} -\frac{x^4}{8}\rbr +c_1 +c_2\lbr x^2\rbr +c_3 \lbr y^2-x^2\log{x}\rbr.
    \label{eq:1size_psi_p}
\end{align}
where, $A$ is a constant and $x,y$ represent the axisymmetric cylindrical $R,z$ coordinates, normalized by the major radius $R_0$. The normalized pressure and current profiles satisfy
\begin{align}
    -\frac{1}{B_0^2}\dfrac{dp}{d\Psi}=1-A, \quad -\frac{1}{B_0^2}F(\Psi)F'(\Psi)=A,
    \label{eq:Soloviev_profile}
\end{align}
Equivalently, the Soloviev profiles are defined by
\begin{align}
   \frac{p(\Psi)}{B_0^2}=\frac{p_0}{B_0^2}-(1-A)\Psi, \quad \frac{F(\Psi)}{B_0}=\sqrt{\lbr \frac{F_0}{B_0}\rbr^2-2 A \Psi}, \label{eq:int_Sol_profile}
\end{align}
where, $p_0,F_0$ are constants. The $A=1$ limit corresponds to force-free equilibrium. The poloidal flux function $\Psi$, given by \eqref{eq:1size_psi_p}, satisfies the following Grad-Shafranov equation 
\begin{align}
    x \del_x (x^{-1}\Psi_{,x})+\Psi_{,yy}=(1-A)x^2+A.
\end{align}

We choose a circular cross-section to the lowest order for simplicity while allowing for arbitrary shaping at higher orders. The details of the NAE are provided in Appendix \ref{app:various_examples} along with other analytical examples. To $O(\rho^4)$, the NAE matches the exact 1-size results for both force-free and MHD cases. Figure \ref{fig:1size} compares the exact 1-size solution to the NAE in the case of force-free ($\beta$=0\%) and MHD at $\beta=5\%$. We see that the deviations of NAE from the exact are more on the inboard side, where the surfaces have a more significant shaping.

\section{Numerical Implementation \label{sec:numerical}}

In this section, we present some numerical examples of the second-order near-axis equilibria constructed numerically using the formalism developed in the previous sections. We verify aspects of this implementation by generating a plasma boundary at a finite aspect ratio, using it to construct a global equilibrium using the VMEC code \citep{hirshman1983steepest}, and comparing some of its properties to the near-axis equilibrium. We detail this procedure in what follows.
\par
The examples presented here are constructed with an axis with constant torsion and a nearly circular cross-section, which we chose for simplicity. The details on the axis are provided in Appendix \ref{app:circ_const_tau}. The choice of nearly circular cross-sections, discussed in detail in Appendix \ref{app:circ_const_tau}, leads to the toroidal flux function, 
\begin{equation}
    \psi = \rho^2\left(\frac{1}{2}+\rho \psi_3\right),
\label{eq:psi_at_this_order}
\end{equation}
where
\begin{equation}
    \psi_3 = \frac{3 \kappa}{8}\cos u + \sigma^{(3)}_{c1}\cos u + \sigma^{(3)}_{s1}\sin u,
\end{equation}
with $\sigma^{(3)}_{c1}$ and $\sigma^{(3)}_{s1}$ given by Eq. (\ref{eq:psi2_alpha0_phi2_circ_MHD_tau0}).
\par
In addition, we choose $\Phi_0=l$ so that the magnetic field is constant on the axis, $B_0=1$, a feature easily checked in the finite aspect ratio equilibrium constructed.
Given the emphasis in this paper on the near-axis construction for finite current and pressure, we choose finite values for these: $\lambda_0=1.2$ and $p_2=-0.1$. 
\par
This information is sufficient to describe the near-axis equilibrium and to construct a flux surface at a finite aspect ratio. Once such a surface has been constructed, find a global equilibrium solution with \texttt{VMEC}. The position vector $\mathbf r$ for a surface at constant $\psi$ is given by
\begin{equation}
    \mathbf r = \mathbf r_0 + \rho(\cos \theta \mathbf n + \sin \theta \mathbf b),
\label{eq:pos_vector_surface}
\end{equation}
\begin{figure}
    \centering
    \includegraphics[width=.31\textwidth]{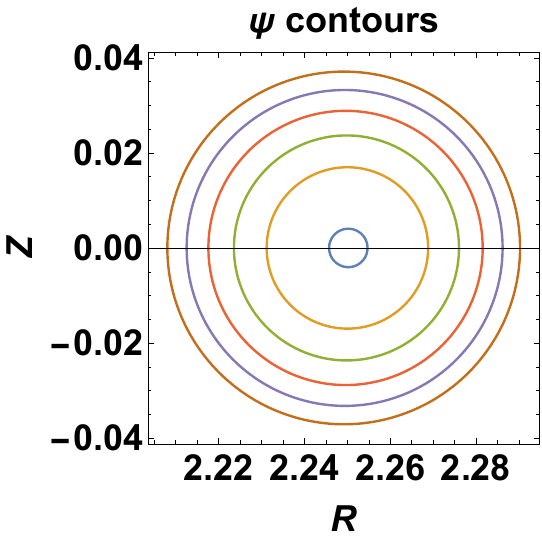}
    \includegraphics[width=.31\textwidth]{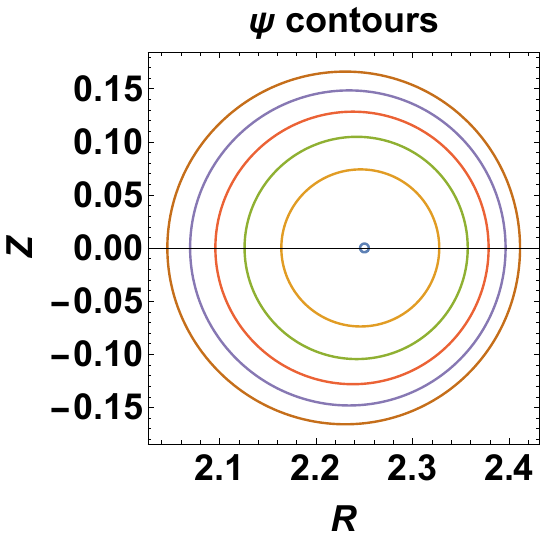}
    \caption{\textbf{Cross-sections of constant $\psi$ for near-axis construction.} Surfaces of constant $\psi$ at $\phi=l=0$ following Eq.~(\ref{eq:pos_vector_surface}) in the range from $\psi=0$, the magnetic axis, to (left) $\psi=0.001$ and (right) $\psi=0.02$. 
    }
    \label{fig:RZ_plots}
\end{figure}
where $\rho$ is found by inverting Eq. (\ref{eq:psi_at_this_order}), namely $\rho = \sqrt{2 \psi}- 2 \psi_3 \psi$, at given values of $\psi$. Surfaces of constant $\psi$ at $\phi=l=0$ are shown in Fig. \ref{fig:RZ_plots} from the magnetic axis to $\psi=0.001$ (left - high-aspect ratio $\epsilon=1/50$) and from the magnetic axis to $\psi=0.02$ (middle - moderate aspect ratio $\epsilon=1/11$). Note the noticeable difference in the Shafranov shift of these configurations. This difference is consistent with their effective plasma $\beta$ on axis: for the high-aspect-ratio $\beta=0.12\%$, and the moderate aspect ratio case, $\beta=2.69\%$.
\par
To proceed further with the comparison, we need to compute the Fourier coefficients of the finite aspect ratio flux surface. That is, the Fourier coefficients of $R(\theta, \phi_c)$ and $Z(\theta, \phi_c)$ in cylindrical coordinates $(R, Z, \phi_c)$ for the plasma surface. Note that this cylindrical coordinate system is not the Mercier coordinate system, the natural coordinate system for near-axis construction. In particular, the toroidal angle in cylindrical coordinates is not the same as the toroidal angle associated with the length $\ell$ along the magnetic axis. Thus, if we need to use the cylindrical angle off-axis $\phi_c(\psi, \theta, \phi)$ as our coordinate for the surface, we must invert its definition, 
\begin{equation}
    \tan \phi_c = \frac{\mathbf r_y(\psi,\theta,\phi)}{\mathbf r_x(\psi,\theta,\phi)}.
\label{eq:tan_phi_off_axis}
\end{equation}
to compute the toroidal angle on-axis $\phi$. This can be achieved using a numerical root solver based on Newton's method.
%
Once this inversion is achieved, we calculate the Fourier components $(RBC,ZBS)$ of $R(\theta,\phi_c)=\sum_{m,n}RBC_{m,n}\cos(m \theta - n\phi_c)$ and $Z(\theta,\phi_c)=\sum_{m,n}ZBS{m,n}\sin(m \theta -n \phi_c)$ for $0\le m \le 10$ and $0-10\le N \le 10$. The surface defined this way is now in a form to be used as an input to the equilibrium solve in \texttt{VMEC}, see Figure~\ref{fig:vmec_3d}.
\par
\begin{figure}
    \centering
    \includegraphics[width=0.8\textwidth]{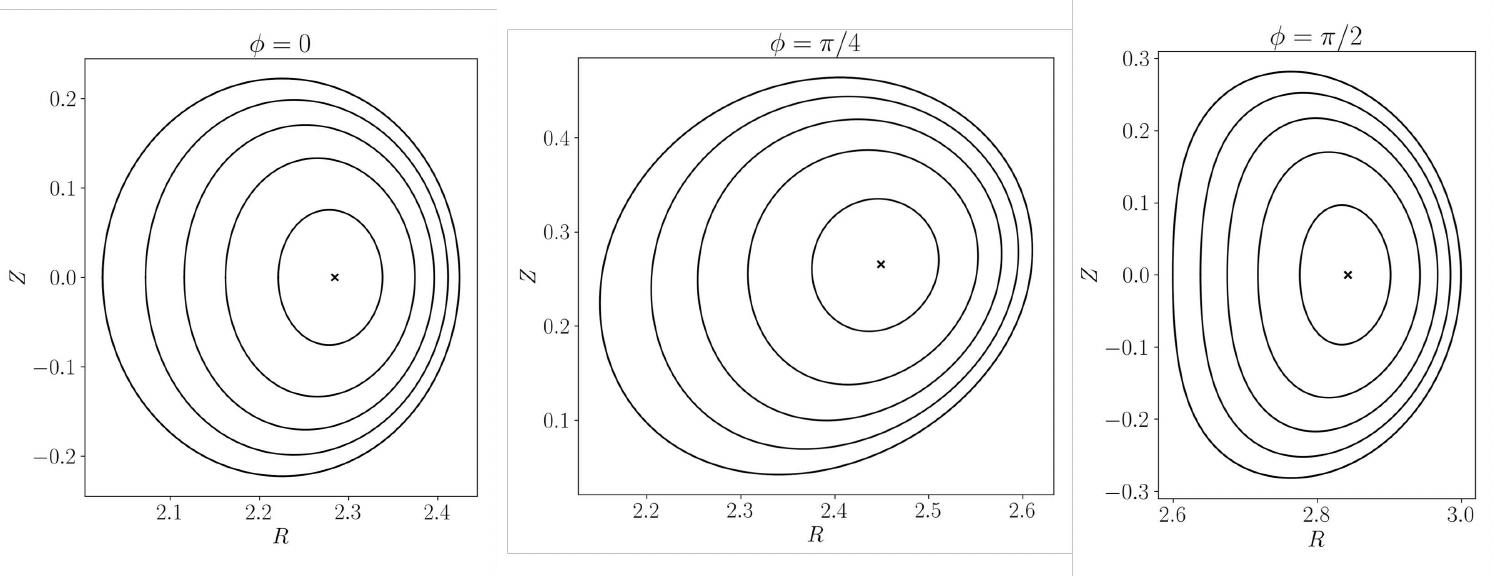}
    \caption{\textbf{Cross-sections of the global equilibrium for $\psi=0.02$.} Example of cross-sections for the global \texttt{VMEC} equilibrium solution for $\psi=0.02$. Cross-sections at $\phi_c=0,~\pi/4$, and $\pi/2$ are shown at values of normalized toroidal flux from $0.1--1.0$. The cross sections show the non-trivial shaping of the example field used in this section.}
    \label{fig:vmec_surfaces}
\end{figure}

\begin{figure}
    \centering
    \includegraphics[width=.48\textwidth]{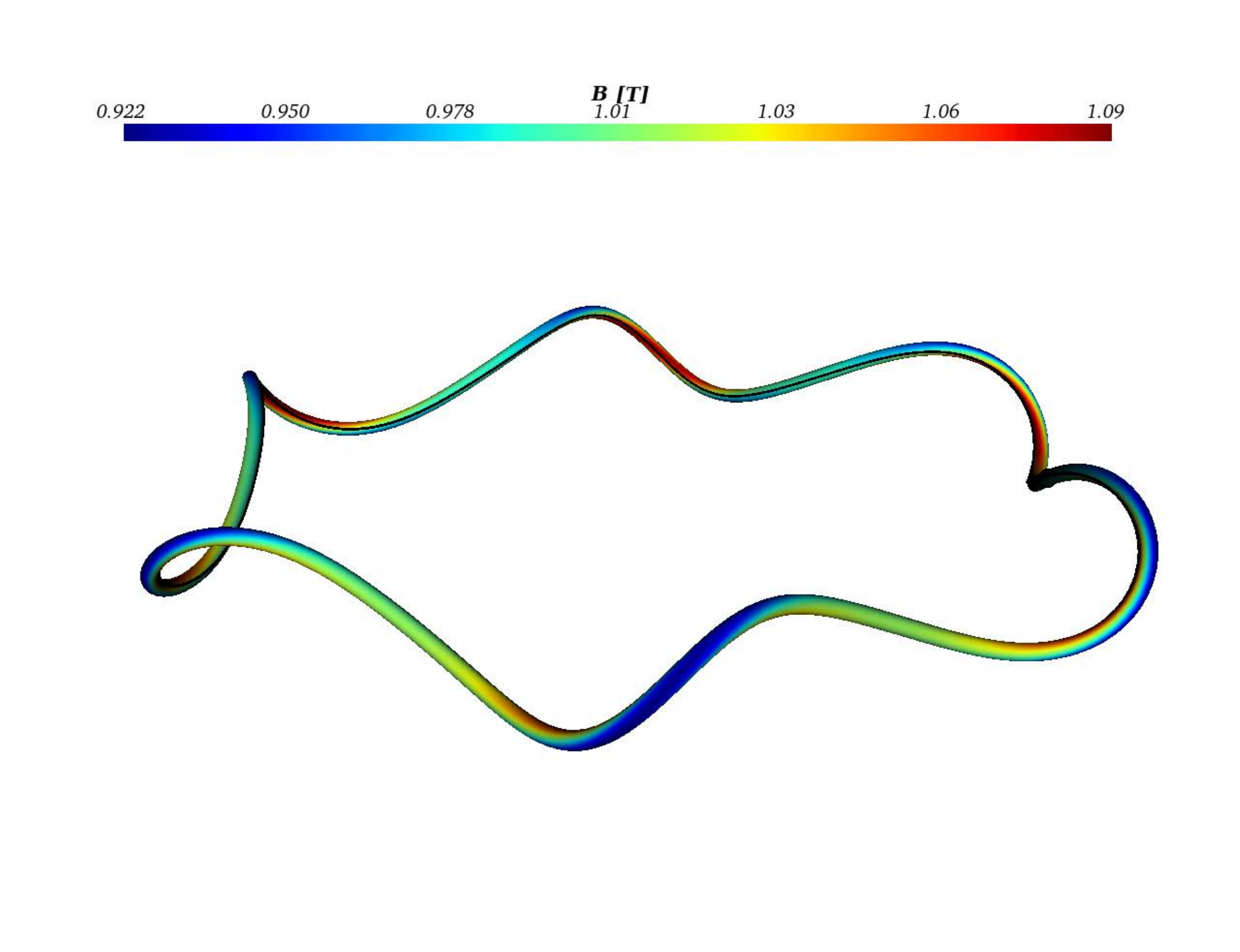}
    \includegraphics[width=.48\textwidth]{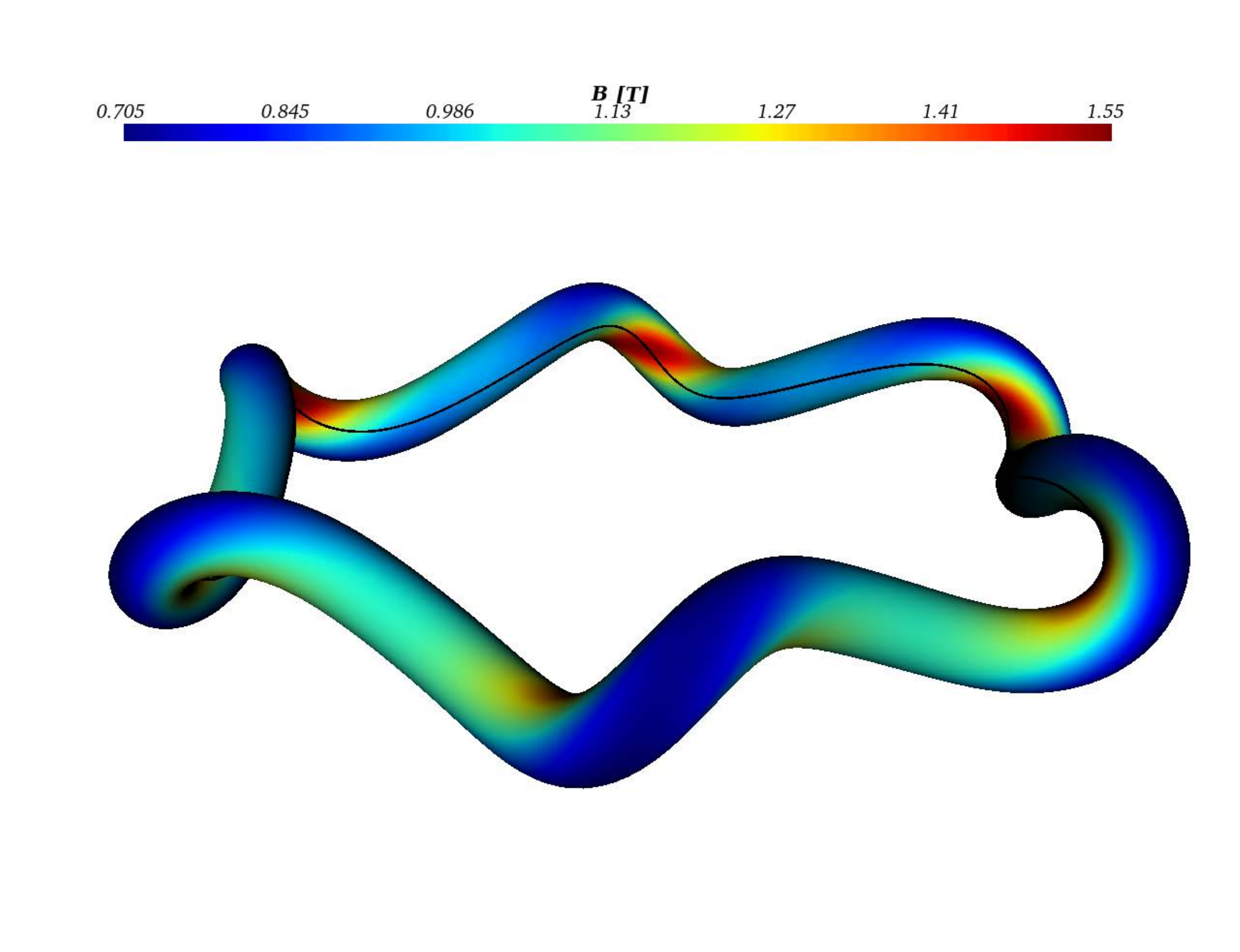},
    \caption{\textbf{Three-dimensional representation of the constructed equilibria.} Plots of the boundary of the equilibria constructed from the near-axis solution for: (left) $\psi=0.001$ and (right) $\psi=0.02$. Some cross-sections for the latter are shown in Figure~\ref{fig:vmec_surfaces}. The color map represents the magnitude of the magnetic field on the surface.}
    \label{fig:vmec_3d}
\end{figure}

Comparing Fig.~\ref{fig:vmec_surfaces} to Fig.~\ref{fig:RZ_plots}, we find that although the flux surfaces are nearly circular in the Frenet-Serret frame of the magnetic axis, they have significant shaping in the lab-frame, most notably in elongation. This difference arises because of the non-circular geometry of the axis.
\begin{figure}
    \centering
    \includegraphics[width=.8\textwidth]{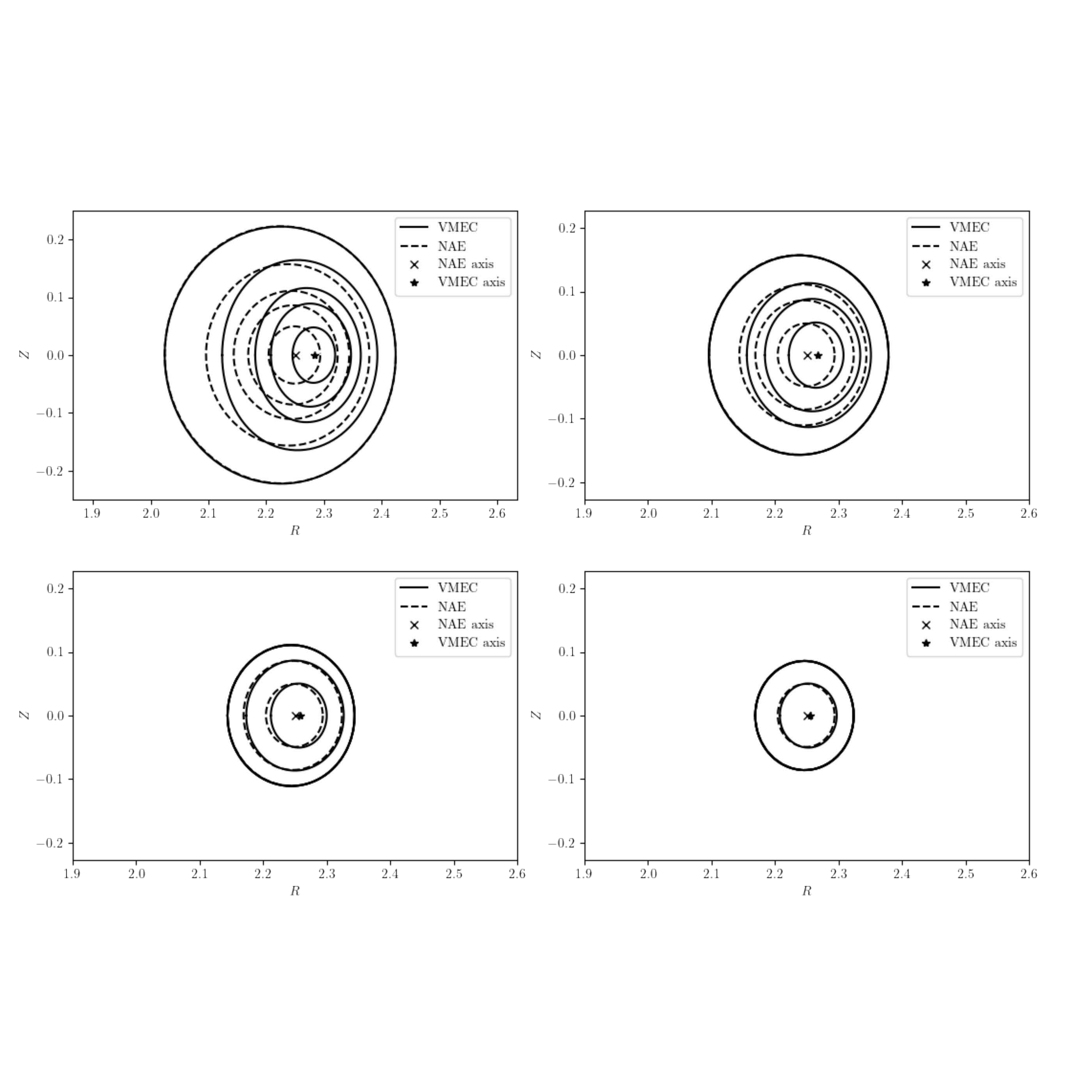}
    \caption{\textbf{Comparison of cross-sections from NAE and \texttt{VMEC} calculations.} The plots show a comparison between the cross-sections at $\phi_c=0$ between the global equilibria computed with \texttt{VMEC} for $\psi=0.02,~0.01,~0.005,~0.003$ (top left to bottom right) and the near axis solution (broken lines). The comparison of the NAE solution to the finite aspect ratio \texttt{VMEC} gets better with increasing aspect ratio. The flux surfaces shown correspond to $\psi=0.02,~0.01,~0.005,~0.003,~0.001$ and the magnetic axis.}
    \label{fig:compare_vmec_surfaces}
\end{figure}

The surfaces from the near-axis construction and the finite aspect ratio equilibrium are compared for several aspect ratios in Fig. \ref{fig:compare_vmec_surfaces}. There are clear differences in the shapes of both equilibria, but these tend to decrease in the limit of increasing aspect ratio. For a more accurate matching, as discussed in \citep{landreman_Sengupta2019_2nd_order}, the error incurred in setting the minor radius $\sqrt{\psi}$ to a finite value, say $a$ when constructing an equilibrium makes a one-to-one comparison between the solutions complicated. A double expansion in both small parameters, $\sqrt{\psi},a$ must be carried out to account for finite aspect ratio. Or construct the finite aspect ratio equilibrium in an alternative way \citep{panici2022_DESC_VMEC_near_axis,panici_rodriguez_NAE2023near}. This is particularly important for comparing higher-order NAE features, such as the magnitude of the Shafranov shift. A more accurate and complete numerical description is left for future work. 
\par
\begin{figure}
    \centering
    \includegraphics[width=.9\textwidth]{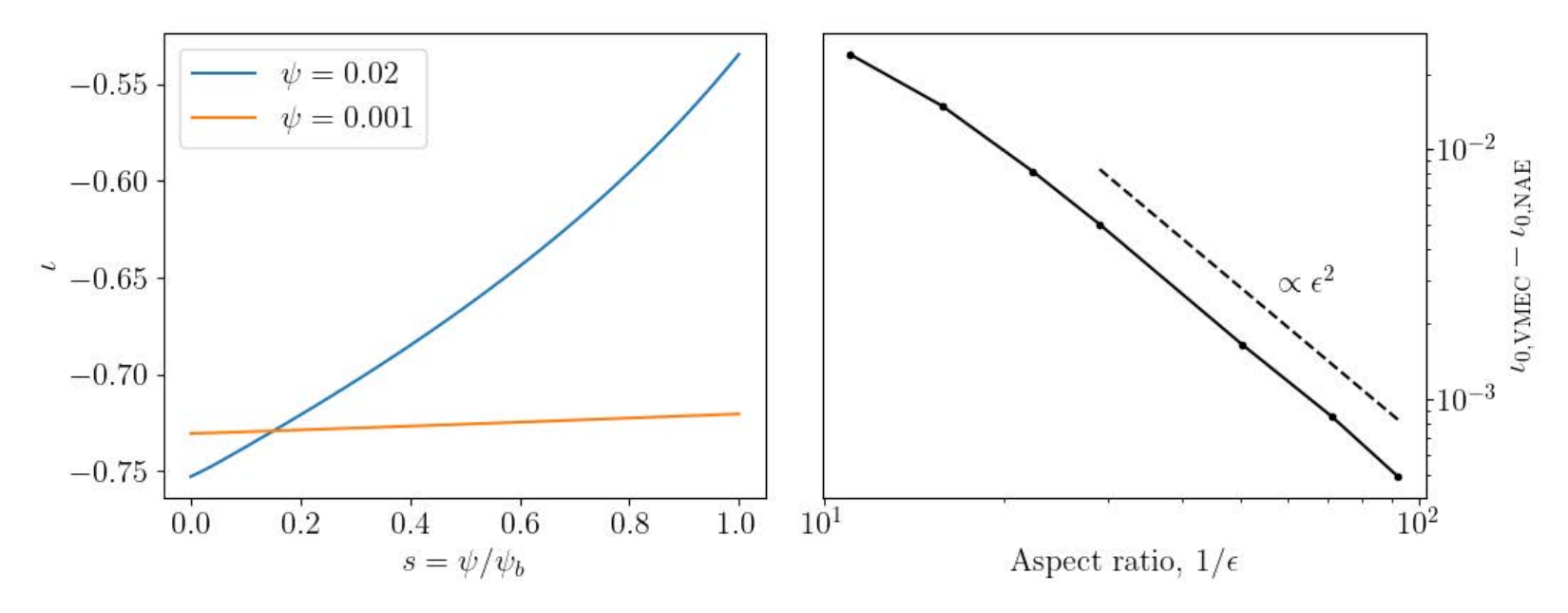}
    \caption{\textbf{Rotational transform profile of global equilibria and difference to the NAE.} (Left) Rotational transform $\iota$ as a function of the normalised toroidal flux $s=\psi/\psi_b$ computed by \texttt{VMEC} for the equilibria with $\psi=0.001$ and $0.02$. The lower aspect ratio case shows higher magnetic shear. (Right) Difference between the on-axis rotational transform between the finite aspect ratio \texttt{VMEC} equilibria and the near axis value, as a function of the aspect ratio. The dashed line shows a scaling $\epsilon^2$.  }
    \label{fig:iota_scaling}
\end{figure}

The rotational transform provides another important feature of the equilibria to study. Using Mercier's formula for the rotational transform on-axis $\iota_0$ \citep{Mercier1964}, the near-axis construction expects a value
\begin{equation}
    \iota_0=\left(\frac{\lambda}{2}-\tau\right)\frac{L}{2\pi}-N=\left(\frac{1.2}{2}+1.643\right)\frac{6\pi}{2\pi}-6=0.729.
\end{equation}
Figure~\ref{fig:iota_scaling} shows the difference between the NAE rotational transform and the one obtained from the finite aspect ratio equilibria. The difference tends to zero quadratically in the aspect ratio, suggesting that the construction of the equilibrium is correct to first order. The differences in rotational transform may be compared in magnitude to the total global shear of the equilibria. 


\begin{figure}
    \centering
    \includegraphics[width=.48\textwidth]{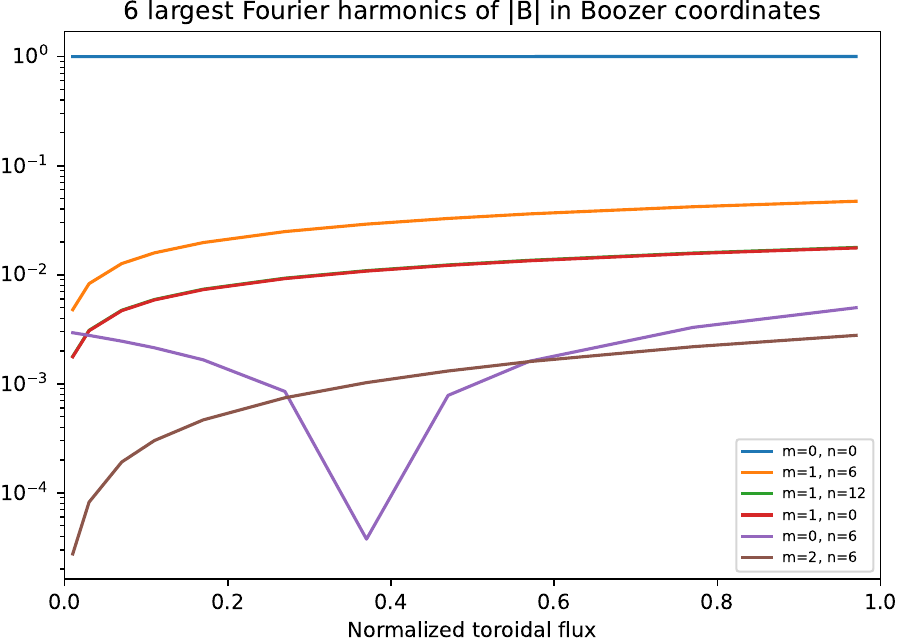}
    \includegraphics[width=.48\textwidth]{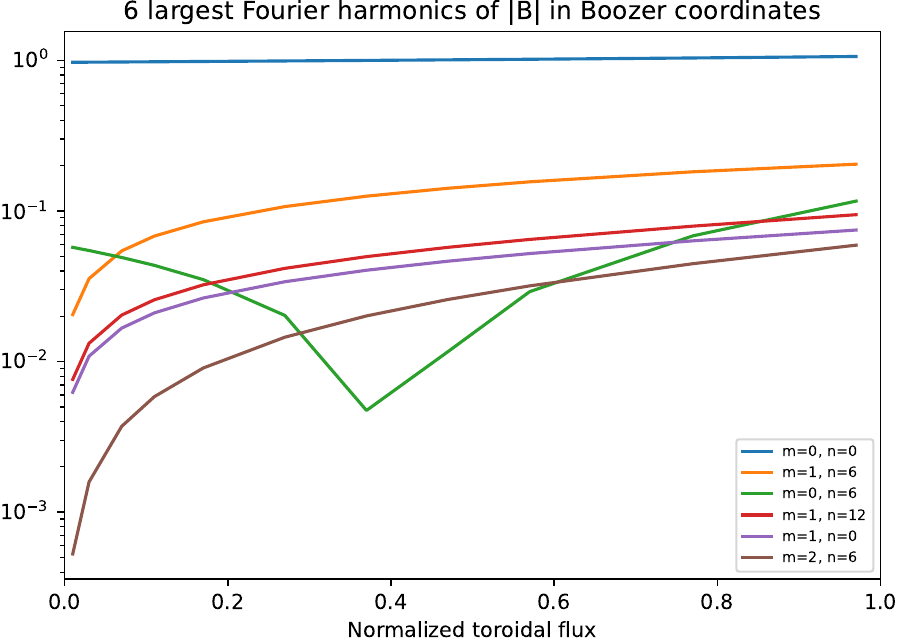}
    \caption{\textbf{Boozer spectrum of $|\mathbf{B}|$ for global equilibria.} The plots show the Boozer spectra of $|\mathbf{B}|$ as a function of the normalised toroidal flux $s=\psi/\psi_b$, for the configurations constructed at $\psi=0.001$ (left) and $\psi=0.02$ (right). The spectra were computed using the \texttt{BOOZXFORM} code.}
    \label{fig:boozer_spectrum}
\end{figure}

Finally, let us recall that the near-axis construction was made for a constant magnetic field on the axis. Thus, studying the spectrum of Fourier modes of the magnetic field strength $B$ should also be informative. Figure~\ref{fig:boozer_spectrum} shows the spectrum computed using the \texttt{BOOZ\_XFORM} code \citep{Sanchez2000a}.
We find that, indeed, on the axis, the magnetic field is approximately constant. A mirror mode growing with decreasing aspect ratio is apparent, in line with previous findings \citep{Jorge_sengupta_landreman_2020_NF,landreman_Sengupta2019_2nd_order}, and a more sophisticated way of generating finite-aspect-ratio equilibria would be necessary \citep{landreman_Sengupta2019_2nd_order} to suppress them.

\section{Discussion \label{sec:discussion}}

In this work, we have presented an asymptotic theory of MHD equilibrium in a stellarator using the near-axis expansion scheme. Our approach generalizes previous work in this area by allowing MHD fields of arbitrary plasma beta. We have shown that the analyticity property of the vacuum fields, i.e., the existence of regular power-series expansions in the vicinity of the magnetic axis, can be retained if one allows for current profiles with regular power-series expansions near the axis. We have studied the vacuum limit carefully as a model for the force-free and MHD cases. We have shown that near the axis, we can transform to variables that allow us to describe the force-free and MHD systems analogous to the vacuum system. 

We have demonstrated that one can carry out the NAE formally to all orders by following the same steps. The zeroth step consists of solving a magnetic differential equation for the pressure-driven part of the current. We have shown how using Birkhoff normal form considerably simplifies this calculation. The first step is to solve a Poisson equation for a scalar representing the magnetic scalar potential in the vacuum limit. The second step is to solve another MDE to obtain the flux function up to a homogeneous solution of the MDE. We fix this homogeneous solution by requiring the periodicity of physical quantities. Finally, with the help of the flux function and the solution to the Poisson equation, we can obtain the field line label through a simple integral over the poloidal angle up to a function of only the toroidal variable. We use the poloidally averaged MDE for the field line label to fix this unknown. The procedure described above is repeated in each order. We have given the complete details of the first two orders of the NAE for vacuum fields in Appendix \ref{app:general_first_order_vacuum} followed by force-free and MHD fields in Appendix \ref{app:NAE_order_by_order_MHD}.

We have illustrated the finite beta near-axis procedure obtained through the Mercier-Weitzer formalism for various problems. We have treated the case of vacuum fields with a nearly circular cross-section in detail in Section \ref{sec:illustration_circ}. The force-free and finite beta analogs of this example are given in Appendix \ref{sec:illustration_circ_MHD}. Furthermore, the special cases of axisymmetry with Solov'ev profiles and magnetic axis with constant torsion are discussed in Section \ref{sec:examples_FF_MHD}. The analytical details of these examples are to be found in Appendices \ref{app:circ_Soloviev} and \ref{app:circ_const_tau}. 

The discussion of physical quantities, such as the magnetic well and the magnetic shear, which plays a crucial role in MHD equilibrium and stability, is out of the scope of the present work. Analytical expression of the near-axis magnetic shear has already been obtained in inverse coordinates \citep{rodriguez2022weakl_QS}, while the near-axis magnetic well has been studied in \citep{landreman2020magnetic_well, Kim_p_Jorge_Dorland_2021_Well_Mercier}. In subsequent work, we shall calculate the magnetic well and the shear in direct coordinates using the formalism developed in this work.%

The calculations presented here, while lengthy, move forward the NAE stellarator optimization program significantly by presenting reduced analytical models that can be automated in fast numerical optimization suites for stellarator design and for testing numerical codes.  


\noindent{\bf Acknowledgements}
W.S. would like to thank H. Weitzner, H. Mynick, S. Hudson, P. Helander, G.G. Plunck, W. Dorland, A. Cerfon, F.P. Diaz, E.J.Paul, N.Nikulsin, A. Wright and G.Roberg-Clark for helpful discussions and feedback.

This research was supported by a grant from the Simons Foundation/SFARI (560651, AB), and the Department of Energy Award No. DE-SC0024548. 

\section{Appendices}
\appendix

\section{Choice of toroidal and poloidal flux label}
\label{app:flux_choice}
 We shall now show that by choosing $\psi$ to be the toroidal flux, we can set $f(\psi)$ to unity in \eqref{eq:sec_n_per_alpha}. Let $\psi$ be the net toroidal flux,
\begin{align}
   2\pi \psi=\int d\psi\oint d\theta \BD\phi\:\sqrt{g} =\oint \B\cdot \bm{dS}\quad  \text{(toroidal flux)},
\end{align}
where
\begin{align}
    dV&=\sqrt{g}d\psi\: d\theta\: d\phi, \quad  \phi=\ell/L, \nonumber\\
    \sqrt{g}&=\frac{1}{\dl\psi \times \dl \theta \cdot \dl \phi}=\frac{(2\pi)^{-1}}{\dl\psi \times \dl \theta \cdot \dl (\ell/L )}.
\end{align}
We can obtain the following expression for $\B$ from \eqref{eq:Basic_Clebsch} and \eqref{eq:sec_n_per_alpha}
\begin{align}
\B=\dl\psi\times \dl \alpha=f(\psi)\lbr \dl\psi\times \dl \theta -2\pi\iota(\psi)\dl\psi\times \dl (\ell/L) \rbr +\dl \psi\times \dl \widetilde\alpha.
    \label{form_of_B}
\end{align}
The last term in \eqref{form_of_B} vanishes on averaging by construction and does not contribute to any net flux. From the expression of the toroidal flux,
\begin{align}
   2\pi \psi=\int d\psi\oint d\theta\: \BD\phi\:\sqrt{g} =2\pi\int d\psi\: f(\psi),
\end{align}
it follows that the poloidal component of the secular part of $\alpha$ is simply $\theta$, i.e., 
\begin{align}
    f(\psi)=1.
    \label{eq:f_is_1}
\end{align}

Similarly, we can show that the poloidal flux is given by
\begin{align}
    2\pi \psi_p = \int d\psi\oint d\phi\: \BD\theta\:\sqrt{g}= 2\pi\int \iota(\psi)d\psi,
\end{align}
which yields the common interpretation of the rotational transform as
\begin{align}
    \iota(\psi)=\frac{d\psi_p}{d\psi}.
\end{align}

\section{Vacuum and MHD Characteristics}
\label{app:MHD_characteristics}
To understand the relationship between the consistency conditions \eqref{eq:Laplace_for_Phi}, \eqref{eq:vacuum_characteristic_psi} and \eqref{eq:vacuum_characteristic_alpha}, we approach the vacuum system with nested flux surfaces from the point of view of characteristics. Equations \eqref{eq:vacuum_characteristics} define the characteristics of the vacuum limit of MHD. As shown in Appendix \ref{app:MHD_characteristics}, the characteristic surfaces, given by $\Omega(\bm{r})$ with $\bm{k}=\dl \Omega$, in the vacuum limit satisfy
\begin{align}
    (\bm{k}\cdot \B)k^2=0, \quad k^2 = \bm{k}\cdot \bm{k}.
    \label{eq:vacuum_disp}
\end{align}
From equation \eqref{eq:vacuum_disp}, we find that for vacuum fields with nested flux surfaces, there are both elliptic characteristics ( $k^2=0$) mixed with one hyperbolic characteristic ($\bm{k}\cdot\B=0$). If the vacuum system of equations satisfies the Laplace equation \eqref{eq:Laplace_for_Phi} and either of the \eqref{eq:vacuum_characteristics}, then the other consistency condition is satisfied automatically. 

We shall derive \eqref{eq:vacuum_disp} for the vacuum limit of the ideal MHD. The full MHD limit was discussed in \citep{weitzner2014ideal}. Let \begin{align}
    \B_0 =\dl \Phi_0=\dl\psi_0 \times \dl \alpha_0,
    \label{eq:vacuum_with_surfaces}
\end{align}
represent a vacuum magnetic fields that possess nested flux surfaces. We shall look for solutions close to \eqref{eq:vacuum_with_surfaces} such that
\begin{align}
    \B=\B_0+\delta \B_0,\quad \Phi=\Phi_0+\delta\Phi,\quad \psi=\psi_0+\delta\psi, \quad \alpha=\alpha_0+\delta \alpha.
\end{align}
It follows that
\begin{align}
    \dl\psi_0 \cdot \dl \delta \Phi= -\B_0\cdot \dl \delta \psi, \quad \dl\alpha_0 \cdot \dl \delta \Phi= -\B_0\cdot \dl \delta \alpha \nonumber\\
    \B_0\cdot \dl\delta \Phi= \B_0 \cdot \lbr \dl\psi_0 \times \dl\delta \alpha+\dl\delta\psi \times \dl \alpha_0\rbr
\end{align}
Going to Fourier space (denoted by hats), we find 
\begin{align}
    k^2 \delta \hat{\phi}=0, \quad (\B_0\cdot \bm{k})k^2 \delta \hat{\psi}=0,\quad (\B_0\cdot \bm{k}) k^2 \delta \hat{\alpha}=0,
\end{align}
which shows that the dispersion relation is given by $(\B_0\cdot \bm{k}) k^2=0$.

In \ref{sec:Alt_Basic_vacuum_eqn}, we describe an alternative set of equations where we solve the Laplace equation ($k^2=0$) and the MDE for $\psi$ $(\bm{k}\cdot \B=0)$. The quantity $\alpha$ is obtained from these quantities. Since we do not solve the MDE for $\alpha$, the characteristics satisfy \eqref{eq:vacuum_disp}.

\section{Normal forms and MDEs}
\label{app:normal_form}

We shall now derive the normal form identities \eqref{eq:BDn+2_normal_form_basis}. Before we do so, we need some basic relations that can be easily obtained from the following normal form definitions \eqref{eq:norm_normal_form} and \eqref{eq:complex_zN_def}:
\begin{subequations}
    \begin{align}
    \cos u = e^{-\eta/2}\sqrt{\frac{2\psiTwo}{\Phi_0'}}\cos\Theta, \quad  \sin u = e^{\eta/2}\sqrt{\frac{2\psiTwo}{\Phi_0'}}\sin\Theta\\
    e^{-\eta}\cos^2{\Theta}+ e^{\eta}\sin^2{\Theta}= \lbr \frac{2\psiTwo}{\Phi_0'}\rbr^{-1}\\
    \sin{2u}=\lbr\frac{2\psiTwo}{\Phi_0'}\rbr\sin{2\Theta},\quad \cos{2u}=\lbr e^{-\eta}\cos^2{\Theta}- e^{\eta}\sin^2{\Theta}\rbr\lbr \frac{2\psiTwo}{\Phi_0'}\rbr 
\end{align}
 \label{eq:basic_normal_form_reln}
\end{subequations}
Starting from the $\PhiTwo$ expression \eqref{eq:phi2_expression} and 
\begin{align}
    \PhiTwo_{,\omega}= \Phi_0'\lbr \frac{\eta'}{2}\sin{2u}+u' \tanh{\eta}\cos{2u}\rbr
    \label{eq:Phi2omega_expression_vacuum}
\end{align}
and using the identities \eqref{eq:basic_normal_form_reln}, we obtain another useful identity
\begin{align}
    u'\Phi_0' +\PhiTwo_{,\omega} = \Phi_0' \lbr \frac{2\psiTwo}{\Phi_0'}\rbr \lbr \frac{\eta'}{2}\sin{2\Theta}+\Omega_0 \rbr, \qquad  \Omega_0 = \frac{\delta'-\tau}{\cosh \eta}.
    \label{eq:nom_form_convenient_identity}
\end{align}
We shall now calculate $(\BD)^{(n+2)}_0 z_\cN$, where $z_\cN$ is given by  
\begin{align}
    z_\cN=\lbr \frac{2\psiTwo}{\Phi_0'}\rbr^{1/2}e^{i\Theta}.
    \label{eq:z_cN_def}
\end{align}
It is convenient to split the MDO $(\BD)^{(n+2)}_0$, as follows:

  \begin{align}
    (\BD)^{(n+2)}_0= (\BD)^{(0)}_0 + 2(n+2)\PhiTwo, \qquad (\BD)^{(0)}_0=\Phi_0'\del_\ell+\PhiTwo_{,\omega}\del_\omega
    \label{eq:BD_split}
\end{align}

From the definition of $z_\cN$ \eqref{eq:z_cN_def} we find that
\begin{align}
    (\BD)^{(n+2)}_0 z_\cN = z_\cN \lbr  i(\BD)^{(0)}_0 \Theta +  \frac{1}{2}\lbr\frac{2\psiTwo}{\Phi_0'}\rbr^{-1}(\BD)^{(0)}_0 \frac{2\psiTwo}{\Phi_0'} +2(n+2)\PhiTwo\rbr.
    \label{eq:BDn+2zcN}
\end{align}
Let us calculate each of the terms in \eqref{eq:BDn+2zcN}. From the relation between $\Theta$ and $u$ \eqref{eq:basic_normal_form_reln} it follows that
\begin{subequations}
    \begin{align}
    \Theta_{,\omega}=\lbr \frac{2\psiTwo}{\Phi_0'}\rbr^{-1}, \qquad  \Theta_{,\ell}= u'\Theta_{,\omega}-\frac{\eta'}{2}\sin{2\Theta}\\
    (\BD)^{(0)}_0 \Theta =\lbr \Phi_0'u'+\PhiTwo_{,\omega}\rbr\lbr\frac{2\psiTwo}{\Phi_0'}\rbr^{-1}-\Phi_0' \frac{\eta'}{2}\sin{2\Theta}
\end{align}
\end{subequations}
From the identity \eqref{eq:nom_form_convenient_identity}, we find that
\begin{align}
    (\BD)^{(0)}_0 \Theta =\Phi_0' \Omega_0.
    \label{eq:norm_form_lovely_identity}
\end{align}
From the MDE for $\psiTwo$ \eqref{eq:lin_eqn_psi2} we get
\begin{subequations}
    \begin{align}
    (\BD)^{(0)}_0 \psiTwo = -4 \PhiTwo \psiTwo, \quad (\BD)^{(0)}_0 \Phi_0'=\Phi_0'\Phi_0''\\
    (\BD)^{(0)}_0\lbr \frac{2\psiTwo}{\Phi_0'}\rbr = -8\frac{\psiTwo}{\Phi_0'}\lbr \PhiTwo+\frac{1}{4}\Phi_0'' \rbr
\end{align}
\end{subequations}

Thus,
\begin{subequations}
    \begin{align}
    (\BD)^{(n+2)}_0 z_\cN &= z_\cN \Phi_0' \lbr i\Omega_0 -\frac{2}{\Phi_0'} \lbr \PhiTwo+\frac{1}{4}\Phi_0'' \rbr +2(n+2)\frac{\PhiTwo}{\Phi_0'}\rbr \\
    (\BD)^{(n+2)}_0 \overline{z}_\cN &= \overline{z}_\cN \Phi_0' \lbr -i\Omega_0 -\frac{2}{\Phi_0'} \lbr \PhiTwo+\frac{1}{4}\Phi_0'' \rbr +2(n+2)\frac{\PhiTwo}{\Phi_0'}\rbr\\
    (\BD)^{(0)}_0 \lbr \overline{z}^m_\cN z^{(n+2-m)}_\cN\rbr &= m \overline{z}^{m-1}{z}^{(n+2-m)}_\cN(\BD)^{(0)}_0 \overline{z}_\cN\\
     &\quad +(n+2-m)\overline{z}^m_\cN {z}^{(n+1-m)}_\cN (\BD)^{(0)}_0 {z}_\cN \nonumber
\end{align}
\label{eq:BDn+2_zN_products_expr}
\end{subequations}
Now, using \eqref{eq:BDn+2_zN_products_expr} it follows that
\begin{align}
    (\BD)^{(n+2)}_{0} \lbr \overline{z}^m_\cN {z}^{(n+2-m)}_\cN\rbr &= (\BD)^{(0)}_0\lbr \overline{z}^m_\cN {z}^{(n+2-m)}_\cN\rbr + 2(n+2)\PhiTwo\lbr \overline{z}^m_\cN {z}^{(n+2-m)}_\cN\rbr\nonumber\\
    &=\Phi_0' \overline{z}^m_\cN {z}^{(n+2-m)}_\cN \lbr i\Omega_0 \lbr n+2-2m \rbr -\frac{n+2}{2}\frac{\Phi_0''}{\Phi_0'} \rbr,
\end{align}
which proves \eqref{eq:BDnorm_form_id_1}. Identity \eqref{eq:BDnorm_form_id_2} follows after straightforward algebra from \eqref{eq:BDnorm_form_id_1} and the expression for $(\BD)^{(n+2)}_0$ given in \eqref{eq:BD_split}. 

From \eqref{eq:BDn+2_zN_products_expr} we find that in general $ (\BD)^{(1)}_0$ acting on $z_\cN$ for $n+2\neq -1$, mixes the different poloidal harmonics because of the $\PhiTwo$ term. However, for $n+2=1$ we have
\begin{align}
     (\BD)^{(1)}_0 z_\cN &= z_\cN \Phi_0' \lbr i\Omega_0 -\frac{\Phi_0''}{2\Phi_0'} \rbr \\
    (\BD)^{(1)}_0 \overline{z}_\cN &= \overline{z}_\cN \Phi_0' \lbr -i\Omega_0 -\frac{\Phi_0''}{2\Phi_0'} \rbr,
\end{align}
which shows that the action of $ (\BD)^{(1)}_0$ on $z_\cN,\overline{z}_\cN$ is simplify a multiplication with a $\omega$ independent function. Furthermore, \eqref{eq:BDnorm_form_id_2} implies that
\begin{align}
    (\BD)^{(1)}_0 \lbr \sqrt{\Phi_0'} z_N \overline{Z}(\ell)\rbr = \lbr \Phi_0' \rbr^{3/2} \lbr \overline{Z}'(\ell) + i \Omega_0 \overline{Z}(\ell)\rbr z_\cN, 
    \label{eq:Id_useful_for_G1}
\end{align}
together with its complex-conjugate equation.

We now point out the changes necessary to generalize the above normal form results to force-free and MHD fields. Firstly, $\PhiTwo$ in \eqref{eq:Phi2omega_expression_vacuum} and elsewhere need to be replaced by $\xiTwo$. The force-free and MHD form of the MDO \eqref{eq:MDO_FF_MHD} has to be used. Next, the identity that replaces \eqref{eq:nom_form_convenient_identity} is
\begin{align}
    \lbr u' +\frac{\lambda_0}{2} \rbr\Phi_0' +\xiTwo_{,\omega} = \Phi_0' \lbr \frac{2\psiTwo}{\Phi_0'}\rbr \lbr \frac{\eta'}{2}\sin{2\Theta}+\Omega_0 \rbr, \quad  \Omega_0 = \frac{\delta'-\tau+\frac{\lambda_0}{2}}{\cosh \eta}.
    \label{eq:nom_form_convenient_identity_MHD}
\end{align}
The rest of the derivation goes through with the replacements of $\PhiTwo$ by $\xiTwo$ and $\Omega_0$ by its new definition given in \eqref{eq:nom_form_convenient_identity_MHD}.

\section{Structure of the vacuum MDEs}
\label{app:MDE_structure}
In the vacuum limit, the equation $\BD\chi=0$ for a function $\chi$ takes the form
\begin{align}
    \lbr \Phi_{,\rho}\del_\rho +\frac{1}{\rho^2}\Phi_{,\omega}\del_\omega +\frac{1}{h^2}\Phi_{,\ell}\del_\ell\rbr \chi=0
    \label{eq:generic_MDE_chi},
\end{align}
where
\begin{align}
    \frac{1}{h^2}=(1-\kappa \rho \cos\theta)^{-2}=\sum_{m=0}^\infty (m+1)(\kappa \cos \theta)^m \rho^m.
    \label{eq:1/h^2}
\end{align}

\subsection{MDE for $\psi$}
\label{app:sec_MDE_Psi}
We first consider the case $\chi=\psi$. Using the near-axis expansions \eqref{eq:NAE_power_series} and \eqref{eq:1/h^2} we find
\begin{align}
    \BD \psi &= \sum_n \rho^{n+2} \sum_{i=0}^{n}\left( \Phi_{,\omega}^{(i+2)}\psi_{,\omega}^{(n+2-i)}+(i+2)(n+2-i)\Phi^{(i+2)}\psi^{(n+2-i)}\right.
    \label{eq:BDpsi_general_structure}\\
    &\qquad\qquad\qquad\qquad  +\left. \sum_{m=0}^{n-i} \Phi_{,\ell}^{(i)}\psi_{,\ell}^{(n+2-i-m)}(\kappa \cos\theta)^m (m+1) \right) \nonumber.
\end{align}
Separating the $i=m=0$ component of \eqref{eq:BDpsi_general_structure} from the rest, we obtain
\begin{align}
    (\BD)^{(n+2)}_0 \psi^{(n+2)} + F^{n+2}_\psi=0.
    \label{eq:MDE_psi_nplu2}
\end{align}
The first term follows from the  $i=m=0$ component, and $F^{n+2}_\psi$ is the collective of the rest of the terms.

As shown in details in \citep{Jorge_Sengupta_Landreman2020NAE}, if for all $n$, $\Phi^{(n+2)},\psi^{(n+2)}$ are of the form \eqref{eq:Phi_nplus2_form}, then products of terms $\Phi^{(i)}\psi^{(n-i)}$ are also periodic in $u$ with harmonics ranging from zero or one (depending on whether $n$ is even or not ) to a maximum of $n+2$. Perhaps, an easy way to demonstrate this is to use complex variables instead of sinusoidal functions.

We shall now assume that for each order $q\leq n+2$, we are led to a forcing term in \eqref{eq:MDE_psi_nplu2}, which has harmonics up to $q$. Thus, $F^{(n+2)}_\psi$ has $n+2$ harmonics and the solution for $\psi^{(n+2)}$ is given by \eqref{eq:F_chi_n_definition} which is doubly-periodic and has at most $n+2$ harmonics in the poloidal angle.

\subsection{Analysis of the magnetic differential equation for $\alpha$}
\label{sec:sec_vacuum_MDE_alpha}
We now bring the MDE for $\alpha$ \eqref{eq:vac_MDE_alpha_order_n} to the standard form of MDE \eqref{eq:MDE_standard_form} by using \eqref{eq:alpha0cw},\eqref{eq:alpha0cz} and \eqref{eq:Phi2_PB_alpha0_psi2_eqn} to replace the angular derivatives of $\alphaO$ and $\psiTwo$ in \eqref{eq:vac_MDE_alpha_order_n} by derivatives of $\PhiO$ and $\PhiTwo$. The resultant form of the MDE is
 \begin{align}
     \lbr2\psiTwo\rbr \lbr \Phi_0'(\ell)\del_\ell + \PhiTwo_{,\omega} \del_\omega+2 n \PhiTwo \rbr\alpha^{(n)}+F_\alpha^{(n+2)}=0,
    \label{eq:MDE_alpha_order_n_standard_form}
 \end{align}
where
\begin{align}
    F_\alpha^{(n+2)}=\Phi^{(n+2)}_{,\omega}+\text{ terms from lower order}.
    \label{eq:F_alpha_n+2_form}
\end{align}
The homogeneous solution of  \eqref{eq:MDE_alpha_order_n_standard_form} is given by
\begin{align}
    {\alpha_H}^{(n)}=\lbr{2\psiTwo}\rbr^\frac{n}{2}{\mathfrak{a}^{(n)}_H}\lbr \alphaO\rbr.
\end{align}
However, unlike $\psi^{(n+2)}$, $\alpha^{(n)}$ does not satisfy the analyticity condition \eqref{eq:analyiticity_psi}. 

Since $\psiTwo$ is non-vanishing, we can  define a quantity $\cA^{(n+2)}$ such that
\begin{align}
    \alpha^{(n)} = \frac{\cA^{(3n)}}{\lbr 2\psiTwo \rbr^n}.
\end{align}
Using the identity \eqref{eq:lin_eqn_psi2} we can show that \eqref{eq:MDE_alpha_order_n_standard_form} implies that
\begin{align}
    \lbr \Phi_0'(\ell)\del_\ell + \PhiTwo_{,\omega} \del_\omega+6 n \PhiTwo \rbr\cA^{(3n)}+F_\cA^{(3n)}=0, \quad F_\cA^{(3n)}=(2\psiTwo)^{(n-1)}F_\alpha^{(n+2)}
    \label{eq:MDE_cA_order_np2_standard_form}
\end{align}

It can be shown inductively (details given in Section \ref{app:sec_MDE_alpha} ) that $ F_\cA^{(3n)}$, at each order $n$, has at most $(3n)$ harmonics in $u$. Furthermore, just like $\Phi^{(n+2)}, F_\cA^{(3n)}$ has even (odd) harmonics for even (odd) $n$. We also note that \eqref{eq:MDO_lowest} has even parity with respect to both $\omega$ and $\ell$ and has only second (even) harmonics as coefficients. Since $\psiTwo$ also has even harmonics, \eqref{eq:MDE_alpha_order_n_standard_form} implies that $\cA^{(3n)},\alpha^{(n)}$ will have odd (even) harmonics if $F_\cA^{(3n)}$ is odd (even). It follows that the solution $\cA^{(3n)}$ must be of the same form as $\chi^{(3n)}$ in \eqref{eq:chi_n_definition} since \eqref{eq:MDE_cA_order_np2_standard_form} is the same as \eqref{eq:MDE_standard_form} with $n$ replaced by $3n$.   
 
There is an alternative approach to solving the MDE that is particularly useful if the lowest order $\alphaO$ is rational. We first note that the MDE in the form
        \begin{align*}
            \lbr{\alphaOcw}\del_{\ell}-{\alphaOell}\del_{\omega}\rbr \lbr\dfrac{\alpha^{(n)}}{\lbr 2\psiTwo\rbr^{n/2}}\rbr = -{\alphaOcw}\frac{\lbr \Phi^{(n+2)}_{,\omega}+.. \rbr+..}{2 \lbr 2\psiTwo\rbr^{n/2+1}}
        \end{align*}
         can be rewritten as  
        \begin{align*}
            \at{{\del_\ell}}{{\alphaO}}\lbr\dfrac{\alpha^{(n)}}{\lbr 2\psiTwo\rbr^{n/2}}\rbr =-\frac{\lbr \Phi^{(n+2)}_{,\omega}+.. \rbr+..}{2 \lbr 2\psiTwo\rbr^{n/2+1}}
        \end{align*}
         Integrating along a constant $\alphaO$ line, we get
        \begin{align*}
            \lbr\dfrac{\alpha^{(n)}}{\lbr 2\psiTwo\rbr^{n/2}}\rbr
            ={\mathfrak{a}}_H^{(n)}\lbr\alphaO\rbr-\int_{\alphaO}d\ell\:\frac{\lbr \Phi^{(n+2)}_{,\omega}+.. \rbr+..}{2 \lbr 2\psiTwo\rbr^{n/2+1}} 
        \end{align*}
If the lowest-order field lines are closed, ${\mathfrak{a}}_H^{(n)}\lbr\alphaO\rbr$ is an arbitrary function periodic in $\alphaO$, which is related to the $n^{\text{th}}$ derivative of the rotational transform. 

In summary, we find that at $O(n)$, $\alpha^{(n)}$ is of the form $\cA^{(3n)}/\lbr 2\psiTwo\rbr^n$, where determining $\cA^{(3n)}$ involves solving $3n$ coupled ODEs. Clearly, it is easier to solve the MDE for $\psi^{(n+2)}$ and obtain $\alpha^{(n)}$ from it.

\subsubsection{Details related to the MDE for $\alpha$}
\label{app:sec_MDE_alpha}

The analysis for $\alpha^{(n)}$ proceeds in a similar manner. We can show that
\begin{align}
    \BD \alpha &= \sum_n \rho^{n} \sum_{i=0}^{n}\left( \Phi_{,\omega}^{(i+2)}\alpha_{,\omega}^{(n-i)}+(i+2)(n-i)\Phi^{(i+2)}\alpha^{(n-i)}+\right.
    \label{eq:BDalpha_general_structure}\\
    &\quad\quad \quad\quad\quad \quad  +\left. \sum_{m=0}^{n-i} \Phi_{,\ell}^{(i)}\alpha_{,\ell}^{(n-i-m)}(\kappa \cos\theta)^m (m+1) \right) \nonumber.
\end{align}
Once again separating the $i=m=0$ piece from the rest of the terms in \eqref{eq:BDalpha_general_structure} multiplying by $2\psiTwo$ and using \eqref{eq:lin_eqn_psi2}, we get
\begin{align}
    (2\psiTwo)(\BD)^{(n)}_0 \alpha^{(n)} + F^{n+2}_\alpha=0.
    \label{eq:MDE_alpha_n_again}
\end{align}
As before, $F^{n+2}_\alpha$ denotes the $i\neq 0, m\neq 0$ terms of \eqref{eq:BDalpha_general_structure}. We note that $F^{n+2}_\alpha$ has derivatives of various $\alpha^{(i)}$ and is not a priory periodic.

Equivalently, in terms of
\begin{align}
    \cA^{(3n)}=\lbr 2\psiTwo \rbr^{n}\alpha^{(n)},\quad F^{(3n)}_\cA=\lbr 2\psiTwo \rbr^{n-1}F^{n+2}_\alpha
    \label{eq:cA_F_cA_def}
\end{align}
the MDE \eqref{eq:MDE_alpha_n_again} can be written as
\begin{align}
    (\BD)^{(3n)}_0 \cA^{(3n)} + F^{(3n)}_\cA=0.
    \label{eq:MDE_cA_n_again}
\end{align}
Provided $F^{(3n)}_\cA$ is a sum of sinusoidal functions of $u$ with a maximum frequency of $3n$, the solution of  $\cA^{(3n)}$ is of the same form. Note that this requires that $F^{(3n)}_\cA$ does not have any denominators such as powers of $(2\psiTwo)$. 

Substituting \eqref{eq:cA_F_cA_def} into \eqref{eq:BDalpha_general_structure}, we find that
\begin{align}
    F^{(3n)}_\cA &= \lbr 2\psiTwo\rbr^{n}\sum_{i=1}^{n}\left( \lbr \Phi_{,\omega}^{(i+2)}\del_\omega  +(i+2)(n-i)\Phi^{(i+2)}\rbr
    \lbr \frac{\cA^{3(n-i)}}{\lbr 2\psiTwo\rbr^{(n-1)}}\rbr+..\right)
    \label{eq:F3n}\\
    &=\lbr \lbr 2\psiTwo\rbr \lbr \Phi_{,\omega}^{(3)}\del_\omega +3(n-1) \PhiThree\rbr\cA^{3(n-1)} -2(n-1) \psiTwo_{,\omega}\PhiThree_{,\omega}\cA^{3(n-1)}\rbr+...\nonumber
\end{align}
We now assume that $\cA^{3(n-j)}, j=1,..n$ have no factors of $2\psiTwo$ in the denominator and are sums of harmonics up to $3(n-j)$. This is certainly true for $n=1$. 

Firstly, we observe that $F^{(3n)}_\cA$ in \eqref{eq:F3n} is a sum of harmonics provided our assumption regarding  $\cA^{3(n-j)}$ hold. Next, we observe that the highest harmonic comes from the term $\cA^{3(n-1)}$ followed by $\cA^{3(n-2)}$, etc. Since $\psiTwo,\PhiThree$ has up to $2^{\text{nd}},3^{\text{rd}}$ harmonics, and  $\cA^{3(n-1)}$ has up to $3(n-1)$ by assumption, a straightforward analysis suggests that $F^{(3n)}_\cA$ can have up to $3n+2$ harmonics. However, a detailed calculation shows that 
\begin{align}
   T_{3(n-1)}= \lbr \lbr 2\psiTwo\rbr \lbr \Phi_{,\omega}^{(3)}\del_\omega +3(n-1) \PhiThree\rbr\cA^{3(n-1)} -2(n-1) \psiTwo_{,\omega}\PhiThree_{,\omega}\cA^{3(n-1)}\rbr
\end{align}
has only up to $3n$ harmonics. To verify this consider the highest frequencies from $\psiTwo,\PhiThree,\cA^{3(n-1)}$ ,which are 2,3 and $3(n-1)$, beating together. The contribution to $ T_{3(n-1)}$ vanishes since
\begin{align}
   T_{3(n-1)}=2b |\PhiThree| |\cA^{3(n-1)}| e^{3i n}e^{2i u }\lbr (6(n-1))-2(n-1)\cdot3\rbr =0.
\end{align}
Therefore, both $F_\cA^{3(n)}$ and the solution to \eqref{eq:MDE_cA_n_again} $\cA^{3(n)}$  \eqref{eq:MDE_cA_n_again} have up to $3n$ harmonics, which is what we wanted to prove. 

We now make an important observation regarding the parity of the harmonics in $\cA^{3(n)}$. We use \eqref{eq:F3n} and the fact that $\PhiThree$ and $\psiTwo$ have odd and even harmonics. We note that if $n$ is even (odd) and $\cA^{3(n-1)}$ has only odd (even) modes then $F^{(3n)}_\cA$ has even (odd) modes, which implies that $\cA^{3(n)}$ has also even (odd) harmonics in $u$. Since $\cA^{(3)}$ has odd harmonics, $\cA^{(3n)}$ will have only even (odd) harmonics if $n$ is 
even (odd).

\section{Structure of force-free and MHD NAE equations}
\label{app:Force_free_MHD_structure}
Owing to the similarity of the vacuum and the force-free and MHD NAE systems, much of the mathematical subtleties can be addressed within the vacuum framework. Two questions remain: whether the logarithmic singularities can occur and whether the periodicity requirements \eqref{sec:periodicity_n_structure_soln} are violated. We can show that both of these problems are avoided within this model. 

Firstly, one can retrace the steps of the vacuum case \citep{Jorge_Sengupta_Landreman2020NAE} but using $\xi$ instead of $\Phi$ to show that the resonances never occur. The critical step here is realizing that the deviation of $\xi$ from $\Phi$ is at least second order. Thus, the terms $\xi^{(n+2)}$ are always separated from $\xi^{(n)}$ and lower order quantities by two poloidal harmonics, avoiding the resonance.

To justify our assertion, let us look carefully at the Poisson equations. Beginning with the force-free case, we can rewrite \eqref{eq:Force_free_Laplacian_xi_Xi} as
\begin{align}
\rho^2 \Delta \xi + \cT_{\Lambda} +\cT_\Xi=0
   \label{eq:Poisson_xi_FF}
\end{align}
where,
\begin{align}
   \cT_{\Lambda}= \frac{1}{h}\del_\omega \lbr h \cT_{\xi} \rbr, \qquad 
  \cT_{\xi}=
    \lbr \int_0^\rho d\varrho\: \frac{\varrho}{h}\lambda \xi_{,\ell}\rbr \nonumber\\ 
   \cT_{\Xi}= \frac{1}{h}\del_\omega\lbr h \lbr \int_0^\rho d\varrho\: \frac{\varrho}{h}\lambda \Xi\rbr \rbr +\frac{1}{h}\del_\ell\lbr \frac{\rho^2}{h}\Xi\rbr.
    \label{eq:Tlambda_Txi_def}
\end{align}
We can show that $\cT_\Xi = O(\rho^4)$ using
\begin{align}
    \cI =\cI^{(2)} \rho^2 +O(\rho^3), \quad \Xi=\Xi^{(2)}\rho^2 +O(\rho^3), \quad h=1 + O(\rho).
    \label{eq:cI_Xi_h_FF}
\end{align}
It also follows from $\xi^{(0)}=\Phi_0(\ell), \xi^{(1)}=0$ that
\begin{align}
    \cT_\xi = \lambda_0 \Phi_0'\frac{\rho^2 }{2} +O(\rho^3), \qquad \cT_{\Lambda}= O(\rho^3).
\end{align}
In the case of MHD, we can replace \eqref{eq:Poisson_xi_FF} by
\begin{align}
\rho^2 \Delta \xi + \cT_{\Lambda} +\cT_\Xi + \cT_\cY + \cT_\cX=0
   \label{eq:Poisson_xi_MHD},
\end{align}
where the additional terms $\cT_\cY, \cT_\cX$ are given by
\begin{align}
   \cT_{\cY}= \frac{1}{h}\del_\omega \lbr h \cY \rbr \quad
   \cT_{\cX}= \frac{1}{h}\del_\omega\lbr h \lbr \int_0^\rho d\varrho\: \frac{\varrho}{h}\lambda \cX\rbr \rbr +\frac{1}{h}\del_\ell\lbr \frac{\rho^2}{h}\cX\rbr.
    \label{eq:TcY_TcX_def}
\end{align}
From the expansions \eqref{eq:cG_cY_cX_NAE}, it follows that
\begin{align}
    \cT_\cY = O(\rho^3), \qquad \cT_\cX = O(\rho^4).
\end{align}
Now, the first term in both \eqref{eq:Poisson_xi_FF} and \eqref{eq:Poisson_xi_MHD} is the only term that persists in the vacuum limit. From equation (3.15) of \citep{Jorge_Sengupta_Landreman2020NAE}, we find that 
\begin{align}
   \rho ^2 \Delta \xi = \sum_{n=0}\rho^n \lbr  \lbr \del^2_\omega + n^2 \rbr \xi^{(n)} + \text{terms with $\xi^{(n-1)}$ or lower }\rbr
\end{align}
Since the current driven terms $\cT_\Lambda, \cT_\Xi, \cT_\cX, \cT_\cY$ are all at least $O(\rho^3)$, to $O(\rho^n)$ only $\xi^{(n-3)}$ or lower can come from them. Therefore, for each $n$, the only terms in \eqref{eq:Poisson_xi_FF} to $O(\rho^n)$  that has a $\xi^{(n)}$ term comes from the vacuum term. Thus, we can solve the driven harmonic oscillator equation
\begin{align}
    \lbr \del_\omega^2 + n^2\rbr \xi^{(n)} = \cF_\xi^{(n)}
\end{align}
order by order without any resonance from the forcing term. Note that fundamental to this argument is the assumption that the current distributions are smooth and can be expanded in power series \eqref{eq:NAE_Kbar_G_MHD}. Thus, provided this assumption holds, a regular power series solution of the fields holds too. In turn, the plasma currents obtained from the smooth fields will satisfy the regularity condition \eqref{eq:NAE_Kbar_G_MHD}.

To answer the second question, we note that the transformation from $\Phi$ to $\xi$ is through a linear integral operator on $\alpha$. Since the integral is in $\rho$, the secular terms are not affected. The essential similarity with the vacuum case as far as dealing with the MDE for $\psi$ and the $\alpha_{,\omega}$ equation means that we can impose the periodicity constraints \eqref{eq:integral_constraints_alpha} order by order in the same way as in the vacuum case. Thus, the single-valued nature of $\B,\J$ can be maintained order by order.

\section{NAE vacuum: general first-order system (an order beyond the rotating ellipse solution) }
\label{app:general_first_order_vacuum}
To $O(\rho)$ the NAE equations are
\begin{subequations}
    \begin{align}
        -(\del^2_\omega+3^2)\PhiThree &= \cos\theta \lbr \kappa' \Phi'_0+2\kappa \Phi_0'' -2\kappa \PhiTwo\rbr + \sin\theta \lbr \kappa \tau \Phi_0'+\kappa \PhiTwo_{,\omega} \rbr
        \label{eq:Phi3_eqn}\\
        \lbr \BD\rbr^{(3)}_0 \psiThree + \FThree_\psi&=0, \quad \FThree_\psi= \lbr \PhiThree_{,\omega}\psiTwo_{,\omega}+ 6\PhiThree\psiTwo\rbr+2\kappa \cos \theta \Phi_0'\psiTwo_{,\ell}\label{eq:MDE_psi3_short}\\
        \alphaOne \psiTwo_{,\omega}- 2 \psiTwo\alphaOne_{,\omega}&=\lbr 3\psiThree-2\psiTwo \kappa \cos \theta\rbr \alphaOcw \label{eq:alpha1_omega_eqn}
    \end{align}
    \label{eq:first_order_system_vacuum}
\end{subequations}
Finally, we have the MDE for $\alphaOne$:
 \begin{align}
 2\psiTwo\lbr \Phi_0' \del_\ell +\PhiTwo_{,\omega}\del_\omega +2 \PhiTwo \rbr\alphaOne+ \FOne_\alpha=0, \quad
     \FOne_\alpha= \Phi_0' \lbr \PhiThree_{,\omega}-2 \kappa \cos\theta\: \PhiTwo_{,\omega}   \rbr. \label{eq:MDE_alpha1_eqn}
 \end{align}
In the following, we explicitly solve the general first-order system and provide all the details of the steps involved. We extensively use the lower-order equations and identities \eqref{eq:Phi2_alpha0_psi2_system},\eqref{eq:phi2_expression} and \eqref{eq:psi2_expression}.    
 
\subsection{Step I: Calculation of $\PhiThree$} 
Equation \eqref{eq:Phi3_eqn} is the Poisson equation for $\PhiThree$ whose solution we now discuss. The solution of \eqref{eq:Phi3_eqn} reads
\begin{align}
    \PhiThree&= \Phi_0'\lbr \fThree_{c1}\cos u +\fThree_{s1}\sin u+ \fThree_{c3}\cos 3u +\fThree_{s3}\sin 3u\rbr \label{eq:Phi3_expression}\\
    \fThree_{c1}&=-\frac{\kappa}{8}\cos \delta \lbr \frac{5}{2} \frac{\Phi''_0}{\Phi_0'}+\frac{\kappa'}{\kappa}- 2 \fTwo_{c2}\rbr+\frac{\kappa}{8}\lbr 2\fTwo_{s2}+\tau\rbr \sin \delta 
    \nonumber\\
    \fThree_{s1}&=
    -\frac{\kappa}{8}\sin \delta \lbr \frac{5}{2} \frac{\Phi''_0}{\Phi_0'}+\frac{\kappa'}{\kappa}+2 \fTwo_{c2}\rbr+\frac{\kappa}{8}\lbr 2\fTwo_{s2}-\tau\rbr \cos \delta,
    \nonumber
\end{align}
where, $\fThree_{c3}(\ell),\fThree_{s3}(\ell)$ are the free-functions. As before, we assume that the free-functions are of the Soloviev-like form
\begin{subequations}
    \begin{align}
  \fThree_{c3}(\ell)\cos{3u}+\fThree_{s3}(\ell)\sin{3u} =-\frac{\sqrt{\Phi_0'}}{9}\frac{\del}{\del \ell}\lbr \QThree \cos3u + \PThree \sin3u \rbr.\label{eq:fthree_free_P3_Q3}
\end{align}
\label{eq:fthree_free_det_Soloviev}
\end{subequations}
\subsection{Step II Calculation of $\psiThree$ using normal forms } 
\label{sec:app_psi3_normal_form}
Now let us turn to the MDE for $\psiThree$ given by \eqref{eq:MDE_psi3_short}. Using the definition \eqref{eq:MDO_lowest}, we can rewrite \eqref{eq:MDE_psi3_short} as
\begin{align}
     \lbr \Phi_0' \del_\ell +\PhiTwo_{,\omega}\del_\omega+6\PhiTwo \rbr \psiThree + \FThree_{\psi}=0
     \label{eq:MDE_psi3_eqn}.
\end{align}
After straightforward algebra the forcing term $\FThree_\psi$ given in \eqref{eq:MDE_psi3_short} reduces to
\begin{subequations}
   \begin{align}
     \frac{ \FThree_\psi}{{\Phi_0'}^2}&\equiv \GammaThree_{\psi c1}\cos u +  \GammaThree_{\psi s1}\sin u+ \GammaThree_{\psi c3}\cos3 u +  \GammaThree_{\psi s3}\sin 3u,\\
   \GammaThree_{\psi c1}&=6 a \fThree_{c1} + 2b \lbr 2\fThree_{c1}+3 \fThree_{c3}\rbr-2b \kappa u'\sin\delta+\frac{\kappa \cos \delta}{\Phi_0'}\lbr (2a+b) \Phi_0'\rbr',\\
   \GammaThree_{\psi s1}&=6 a \fThree_{s1} + 2b \lbr -2\fThree_{s1}+3 \fThree_{s3}\rbr-2b \kappa u'\cos\delta +\frac{\kappa \sin \delta}{\Phi_0'}\lbr (2a-b) \Phi_0'\rbr',\\
    \GammaThree_{\psi c3}&=2 b \fThree_{c1}+6 a \fThree_{c3}+2bu'\kappa \sin \delta +\lbr b \Phi_0'\rbr'\frac{\kappa}{\Phi_0'} \cos \delta,\\
   \GammaThree_{\psi s3}&=2 b \fThree_{s1}+6 a \fThree_{s3}-2bu'\kappa \cos \delta +\lbr b \Phi_0'\rbr'\frac{\kappa}{\Phi_0'} \sin \delta.
\end{align} 
 \label{eq:forcing_F3_psi}
\end{subequations}

As suggested by the forcing term \eqref{eq:forcing_F3_psi}, we now look for a solution of the MDE \eqref{eq:MDE_psi3_eqn} of the form
\begin{align}
    \psiThree= \lbr \Phi_0'\rbr^{3/2}\lbr \YThree_{c3} \cos 3u +  \YThree_{s3} \sin 3u + \YThree_{c1}\cos u + \YThree_{s1}\sin u \rbr.
    \label{eq:form_of_psi3}
\end{align}
Substituting \eqref{eq:form_of_psi3} in \eqref{eq:MDE_psi3_eqn}, we obtain the following coupled ODEs 
\begin{subequations}
\begin{align}
{\YThree_{c3}}'+3 u' {\YThree_{s3}}+2\lbr \YThree_{c1}\fTwocTwo-\YThree_{s1}\fTwosTwo \rbr +  \frac{\GammaThree_{\psi c3}}{\sqrt{\Phi_0'}}&=0,\\
{\YThree_{s3}}'-3 u' {\YThree_{c3}}+2\lbr \YThree_{s1}\fTwocTwo+\YThree_{c1}\fTwosTwo \rbr + \frac{\GammaThree_{\psi s3}}{\sqrt{\Phi_0'}}&=0,\\
{\YThree_{c1}}'+ 4\fTwocTwo {\YThree_{c1}}+\lbr u'+4 \fTwosTwo \rbr {\YThree_{s1}} +6\lbr \YThree_{c3}\fTwocTwo+\YThree_{s3}\fTwosTwo \rbr + \frac{\GammaThree_{\psi c1}}{\sqrt{\Phi_0'}}&=0,\\
{\YThree_{s1}}'  -4\fTwocTwo {\YThree_{s1}}-\lbr u'-4 \fTwosTwo \rbr {\YThree_{c1}}+6\lbr \YThree_{s3}\fTwocTwo-\YThree_{c3}\fTwosTwo \rbr + \frac{\GammaThree_{\psi s1}}{\sqrt{\Phi_0'}}&=0.
\end{align}
 \label{eq:Coupled_Y3_ODEs}
\end{subequations}

To decouple \eqref{eq:Coupled_Y3_ODEs} we will use the third order complex normal form. As discussed in Section \ref{sec:sec_Step2_normal_form}, at $O(n)$, instead of a Fourier series of $\psi^{(n+2)}$ with $(n+2)$ poloidal harmonics, one could also look for solutions that are polynomial of order $(n+2)$ in the complex normal form coordinates in the form \eqref{eq:psin+2_normal_form}. The major advantage of the complex normal form \eqref{eq:psin+2_normal_form} is that it diagonalizes the MDO, thereby significantly simplifying the analysis. The complex normal form for $\psiThree$ is given by
\begin{align}
    \psiThree = \frac{1}{8}\lbr \Phi_0'\rbr^{3/2} \lbr \cZb_{\psi1}(\ell) z^2_\cN\zb_\cN +\cZ_{\psi1}(\ell) {\zb}^2_\cN z_\cN - \cZb_{\psi3}(\ell) z_\cN^3 - \cZ_{\psi3}(\ell) {\zb}^3_\cN  \rbr.
\label{eq:complex_normal_form_psi3}
\end{align}
We note here that the choice of numerical factors such as $\pm 1/8$ multiplying $\cZ_{\psi1},\cZ_{\psi3}$ are arbitrary. We have made the choices such that our results are as close as possible to \citep{Solovev1970}. Equating the two equivalent representations of $\psiThree$ \eqref{eq:form_of_psi3} (Fourier), and \eqref{eq:complex_normal_form_psi3} (complex normal form), using the definition of the complex normal form,
\begin{align}
    z_\cN \equiv x_\cN + i y_\cN, \quad x_\cN = e^{\eta/2}\cos{u}, \quad y_\cN = e^{\eta/2}\cos{u},
    \label{eq:defn_complex_normal_form}
\end{align}
and the useful identities (with $\vep=\tanh{\eta}$)
\begin{align}
    e^{+\eta}+e^{-\eta}=2\cosh{\eta}=\frac{2}{\sqrt{1-\vep^2}},\quad \frac{3 e^{-\eta}+e^{+\eta}}{e^{+\eta}+e^{-\eta}}=2-\vep \quad \frac{3 e^{\eta}+e^{-\eta}}{e^{+\eta}+e^{-\eta}}=2+\vep\nonumber\\
    \fTwo_{s2} =\frac{u'\varepsilon}{2},\quad \lbr  2 \fTwo_{s2} +u'\rbr = e^\eta \Omega_0, \quad  \lbr -2 \fTwo_{s2} +u' \rbr= e^{-\eta}\Omega_0 , \quad \Omega_0 = \frac{u'}{\cosh \eta}.
   \label{eq:Lambda0_identities}
\end{align}
we find that
\begin{align}
     \YThree_{c1} &=\frac{e^{+\eta/2}}{8\sqrt{1-\vep^2}}\Re{\lbr (2+\vep)\cZ_{\psi1}-3 \vep \cZ_{\psi3}\rbr},\\
     \YThree_{s1}
     &=\frac{e^{-\eta/2}}{8\sqrt{1-\vep^2}}\Im{\lbr (2-\vep)\cZ_{\psi1}-3 \vep \cZ_{\psi3}\rbr},\\
    \YThree_{c3} &=\frac{e^{+\eta/2}}{8\sqrt{1-\vep^2}}\Re{\lbr -(2-\vep)\cZ_{\psi3}+ \vep \cZ_{\psi1}\rbr},\\
    \YThree_{s3}  &=\frac{e^{-\eta/2}}{8\sqrt{1-\vep^2}}\Im{\lbr -(2+\vep)\cZ_{\psi3}+ \vep \cZ_{\psi1}\rbr}.
    \label{eq:Y3_soln}
\end{align}
The coupled ODEs \eqref{eq:Coupled_Y3_ODEs} are now replaced by two decoupled complex ODEs for $\cZ_{\psi1},\cZ_{\psi3}$  
\begin{align}
    {\cZ'_{\psi1}}- i \Omega_0 \cZ_{\psi1} &+  \cF_{\psi1}=0, \qquad \cF_{\psi1}=\cF_{1/2} +  3\cF_{3/2} \nonumber\\
     {\cZ'_{\psi3}}- 3i \Omega_0\cZ_{\psi3} &+ \cF_{\psi3}=0, \qquad  \cF_{\psi3}= \overline{\cF}_{1/2} -  \overline{\cF}_{3/2},
    \label{eq:ZOne_Z3_eqn}
\end{align}
where,
\begin{subequations}
    \begin{align}
     \sqrt{\Phi_0'}  \cF_{1/2}&=-e^{\eta/2}(3\GammaThree_{\psi c3}-\GammaThree_{\psi c1}) + i e^{-\eta/2}(3\GammaThree_{\psi s3}+\GammaThree_{\psi s1}),\\
    \sqrt{\Phi_0'} \cF_{3/2}&= +e^{-3\eta/2}(\GammaThree_{\psi c3}+\GammaThree_{\psi c1}) - i e^{3\eta/2}(\GammaThree_{\psi s3}-\GammaThree_{\psi s1}).
    \end{align}
     \label{eq:complex_forcing_psi3}
\end{subequations}

we can bring $ \cF_{\psi1}, \cF_{\psi3}$ to the following form:
\begin{align}
   \cF_{\psi1} &= \frac{2\lbr\Phi_0' \rbr^{-1/2}}{\sqrt{1-\vep^2}}\lbr e^{-\eta/2}\lbr -3\vep \GammaThree_{\psi c3}+\GammaThree_{\psi c1}(2-\vep)\rbr+i e^{\eta/2}\lbr -3\vep \GammaThree_{\psi s3}+\GammaThree_{\psi s1}(2+\vep)\rbr\rbr, \nonumber\\
    \cF_{\psi3} &= \frac{2\lbr\Phi_0' \rbr^{-1/2}}{\sqrt{1-\vep^2}}\lbr e^{-\eta/2}\lbr +\vep \GammaThree_{\psi c1}-\GammaThree_{\psi c3}(2+\vep)\rbr-i e^{\eta/2}\lbr \vep \GammaThree_{\psi s1}+\GammaThree_{\psi s3}(2-\vep)\rbr\rbr.
\end{align}
The solution of \eqref{eq:ZOne_Z3_eqn} subject to periodic boundary condition in $\ell$ is straightforward to obtain and has been discussed in \citep{Mercier1964,Solovev1970,Jorge_Sengupta_Landreman2020NAE}. We shall omit the derivation. 

Our results given in \eqref{eq:Y3_soln}  matches with \citep{Solovev1970} ( see Section Section 6.2 on p.60 of \citep{Solovev1970}) when accounting for the slight modification of the form of the free functions \eqref{eq:choice_free_function}. To show that our diagonalization approach based on complex normal form is indeed the same as the \citep{Solovev1970} approach, we take a closer look at the diagonalization approach in \citep{Solovev1970}, which uses the so-called ``rotating coordinates" \eqref{eq:Rotating_coords}, $(X_\cN, Y_\cN)$, with $\psiThree$ given by the following third-order homogeneous polynomial in $(X_\cN,Y_\cN)$
\begin{align}
\psiThree \rho^3= \lbr \Phi_0'\rbr^{3/2}\lbr \zeta_1(\ell) X_\cN Y_\cN^2+\zeta_3(\ell) X_\cN^2 Y_\cN+\zeta_2(\ell) X_\cN^3  + \zeta_4(\ell) Y_\cN^3\rbr,
    \label{eq:psi3_in_rotating_coords}
\end{align}
which they obtained assuming analyticity. Using \eqref{eq:XThree_i_def}, \eqref{eq:Rotating_coords}, and standard trigonometric identities involving third harmonics, we can show that \eqref{eq:psi3_in_rotating_coords} is identical to \eqref{eq:form_of_psi3} provided

\begin{align}
    \YThree_{c1}-3 \YThree_{c3} &= e^{-\eta/2}\zeta_1, \qquad \YThree_{c1}+ \YThree_{c3}= e^{+3\eta/2}\zeta_2, \nonumber\\ 
    \YThree_{s1}+3 \YThree_{s3} &= e^{+\eta/2}\zeta_3,\qquad
    \YThree_{s1}- \YThree_{s3}= e^{-3\eta/2}\zeta_4.
    \label{eq:XThree_i_def}
\end{align}

The equations for $\zetaThree_i$ then take the form 
\begin{subequations}
    \begin{align}
        \zeta_2' +\Omega_0 \zeta_3 &=-e^{-3\eta/2}(\GammaThree_{\psi c3}+\GammaThree_{\psi c1})/{\sqrt{\Phi_0'}},\\
        \zeta_4' -\Omega_0 \zeta_1 &=+e^{+3\eta/2}(\GammaThree_{\psi s3}-\GammaThree_{\psi s1})/{\sqrt{\Phi_0'}},\\
        \zeta_1' -2\Omega_0 \zeta_3 +3\Omega_0 \zeta_4  &=+e^{+\eta/2}(3\GammaThree_{\psi c3}-\GammaThree_{\psi c1})/\sqrt{\Phi_0'},\\
        \zeta_3' +2\Omega_0 \zeta_1 -3\Omega_0 \zeta_2  &=-e^{-\eta/2}(3\GammaThree_{\psi s3}+\GammaThree_{\psi s1})/\sqrt{\Phi_0'}.
    \end{align}
     \label{eq:Coupled_zetai_ODEs}
\end{subequations}
Compared to \eqref{eq:Coupled_Y3_ODEs}, \eqref{eq:Coupled_zetai_ODEs} is a step forward in terms of simplification. In particular, all the $\fTwocTwo,\eta'$ terms drop out. Next, \citep{Solovev1970} define the complex variables,
\begin{align}
    \cZ_{\psi1}&= (\zeta_1 + i \zeta_3) +3(\zeta_2+i \zeta_4),\\
    \cZ_{\psi3}&= (\zeta_1 - i \zeta_3) -(\zeta_2-i \zeta_4),
\end{align}
which from \eqref{eq:Coupled_zetai_ODEs} can be seen to satisfy the decoupled complex ODEs \eqref{eq:ZOne_Z3_eqn}. Finally, the variables are given by
\begin{align}
\zeta_1 &=\frac{1}{4}\Re{\lbr \cZ_{\psi 1}+3\cZ_{\psi 3} \rbr }, \qquad \zeta_2 =\frac{1}{4}\Re{ \lbr \cZ_{\psi 1}-\cZ_{\psi 3} \rbr}\nonumber,\\
\zeta_3 &=\frac{1}{4}\Im{ \lbr \cZ_{\psi 1}+3\cZ_{\psi 3}\rbr}, \qquad \zeta_4 =\frac{1}{4}\Im{ \lbr \cZ_{\psi 1}-\cZ_{\psi 3} \rbr}.
   \label{eq:normal_form_variables} 
\end{align}
Besides extending the pioneering approach of \citep{Solovev1970} to MHD fields, our method also systematizes their approach so that we can avoid the intermediate variables $\zeta_i, (i=1,2,3,4)$. This is particularly needed for going to arbitrary orders in $n$, where the coefficients that connect the variables $\zeta_i$ to $\cZ_{\psi i}$ are not known apriori. 

\subsection{Step III \& IV : Calculation of $\alphaOne$} 
We shall now solve \eqref{eq:alpha1_omega_eqn} for $\alphaOne$. Rewriting \eqref{eq:alpha1_omega_eqn} as
\begin{align}
    \del_\omega \lbr \frac{\alphaOne}{\sqrt{2\psiTwo}}\rbr = \cTThree_\alpha, 
    \quad \cTThree_\alpha=-\frac{3\Phi_0'}{\lbr 2\psiTwo \rbr^{5/2}}\lbr \psiThree-\frac{2}{3}\kappa \cos\theta\: \psiTwo\rbr,
    \label{eq:alphaOne_omega_rewrite}
\end{align}
and upon using \eqref{eq:form_of_psi3}, the expression for $\cTThree_\alpha$ can be further simplified to
\begin{align}
    \cTThree_\alpha &=-\frac{3}{(2a)^{5/2}}\frac{\left(  \cTThree_{\alpha c3}(\ell)\cos{3u}+\cTThree_{\alpha s3}(\ell)\sin{3u}+\cTThree_{\alpha c1}(\ell)\cos{u}+\cTThree_{\alpha s1}(\ell)\sin{u} \right)}{\lbr 1+(b/a) \cos 2 u\rbr^{5/2}}\nonumber\\
    \cTThree_{\alpha c3}&=\lbr \YThree_{c3}-\frac{b \kappa}{3\sqrt{\Phi_0'}}\cos \delta\rbr, \qquad \qquad \quad   \cTThree_{\alpha s3}=\lbr \YThree_{s3}-\frac{b \kappa}{3\sqrt{\Phi_0'}}\sin \delta\rbr\\
    \cTThree_{\alpha c1}&=\lbr \YThree_{c1}-\frac{(2a+b) \kappa}{3\sqrt{\Phi_0'}}\cos \delta\rbr, \qquad     \cTThree_{\alpha s1}=\lbr \YThree_{s1}-\frac{(2a-b)\kappa}{3\sqrt{\Phi_0'}}\sin \delta\rbr\nonumber.
\end{align}
Using the identity (with $b/a=\varepsilon=\tanh{\eta}$)
\begin{align}
    &\int du\: \frac{\left(  \cTThree_{\alpha c3}\cos{3u}+\cTThree_{\alpha s3}\sin{3u}+\cTThree_{\alpha c1}\cos{u}+\cTThree_{\alpha s1}\sin{u} \right)}{\lbr 1+\varepsilon \cos 2 u\rbr^{5/2}}\\ &=\frac{-(1+\varepsilon)^2\lbr A^{(1)}_{c1}\cos{u}+A^{(1)}_{c3}\cos{3u}\rbr +(1-\varepsilon)^2\lbr A^{(1)}_{s1}\sin{u}+A^{(1)}_{s3}\sin{3u}\rbr}{3\lbr 1-\varepsilon^2\rbr^2\lbr 1+\varepsilon \cos{2u} \rbr^{3/2}}\nonumber\\
    & A^{(1)}_{c1}= 3\lbr \cTThree_{\alpha s1}-\varepsilon \cTThree_{\alpha s3}\rbr,  \qquad A^{(1)}_{c3}= \varepsilon \cTThree_{\alpha s1}+(1-2\varepsilon) \cTThree_{\alpha s3} \nonumber\\ 
    & A^{(1)}_{s1}= 3\lbr \cTThree_{\alpha c1}-\varepsilon \cTThree_{\alpha c3}\rbr, \qquad A^{(1)}_{s3}= \varepsilon \cTThree_{\alpha c1}+(1+2\varepsilon) \cTThree_{\alpha c3}\nonumber,
\end{align}
we can integrate equation  \eqref{eq:alpha1_omega_eqn} to obtain the following expression for $\alphaOne$:
\begin{subequations}
    \begin{align}
    \alphaOne&=\frac{\cAOne}{2\psiTwo} +  {\overline{a}}^{(1)}{\sqrt{2\psiTwo}},\\
    \cAOne &= (\Phi_0')^{3/2}\lbr \cAOne_{c1}\cos u + \cAOne_{s1}\sin u+\cAOne_{c3}\cos 3u + \cAOne_{s3}\sin 3u \rbr,\\
    \cAOne_{c1}&=+\frac{A^{(1)}_{c1}}{2a\lbr 1-\vep \rbr^2} ,\quad  \cAOne_{c3}=+\frac{A^{(1)}_{c3}}{2a\lbr 1-\vep \rbr^2},\\
    \cAOne_{s1}&=-\frac{A^{(1)}_{s1}}{2a\lbr 1+\vep \rbr^2},\quad  \cAOne_{s3}=-\frac{A^{(1)}_{s3}}{2a\lbr 1+\vep \rbr^2}.
\end{align}
\end{subequations}

To determine ${\overline{a}}^{(1)}$ we need the MDE \eqref{eq:MDE_alpha1_eqn} averaged over $\omega$. The forcing term $\FOne_\alpha$ in \eqref{eq:MDE_alpha1_eqn} can be written down explicitly in the form
\begin{subequations}
    \begin{align}
   \frac{ \FOne_\alpha}{{\Phi_0'}^2}&= \GammaOne_{\alpha c1}\cos u +  \GammaOne_{\alpha s1}\sin u+ \GammaOne_{\alpha c3}\cos3 u +  \GammaOne_{\alpha s3}\sin 3u,
   \label{eq:forcing_F1}\\
   \GammaOne_{\alpha c1}&=  +\fThree_{s1}-2\kappa\lbr \cos \delta \fTwosTwo -\sin \delta \fTwocTwo\rbr,\\
   \GammaOne_{\alpha s1}&= -\fThree_{c1}+2\kappa\lbr \cos \delta \fTwocTwo +\sin \delta \fTwosTwo\rbr,\\
   \GammaOne_{\alpha c3}&= +3 \fThree_{s3}-2\kappa\lbr \cos \delta \fTwosTwo +\sin \delta \fTwocTwo\rbr,\\
   \GammaOne_{\alpha s3}&= 
   -3\fThree_{c3}+2\kappa\lbr \cos \delta \fTwocTwo -\sin \delta \fTwosTwo\rbr.
\end{align}
\end{subequations}
Since only odd harmonics are involved in $\FOne_\alpha$, the poloidal average of \eqref{eq:MDE_alpha1_eqn} is zero, which implies ${\overline{a}}^{(1)}=0$.

Before we end the discussion on the first-order solutions, we note that it is also possible to express \eqref{eq:MDE_alpha1_eqn} in the same form as \eqref{eq:MDE_psi3_eqn}, i.e.,
\begin{align}
     \lbr \Phi_0' \del_\ell +\PhiTwo_{,\omega}\del_\omega+6\PhiTwo \rbr \cAOne + \FOne_\alpha =0, \quad \alphaOne=\frac{\cAOne}{2\psiTwo}
     \label{eq:MDE_for_cA1}
\end{align}

Therefore, we could have also solved the MDE for $\alphaOne$ instead of $\psiThree$.

\section{NAE force-free and MHD: First-order }
\label{app:NAE_order_by_order_MHD}
We now provide the details of the first order NAE for finite beta MHD. The force-free limit can be obtained by setting the pressure-driven current potentials, $\GO,\GOne$, to zero.

To the relevant order the functions $\cY,\cX,\Xi,\cW$ are
\begin{align}
    \cY &=\rho^3 \cY^{(3)} ,\;\; \cX = \rho^2 \cX^{(2)} + \rho^3 \cX^{(3)},\;\; \Xi = \rho^2\Xi^{(2)} +\rho^3\Xi^{(3)},\;\; \cW=\rho^2 \cW^{(2)}(\ell)+\rho^3 \cW^{(3)}  ,\nonumber\\
    \cY^{(3)}&=\frac{1}{3}\lbr \GOne \psiTwo_{,\omega}-2\psiTwo \GOne_{,\omega}\rbr,\;\; \cW^{(2)}=\frac{\lambda_0}{2}\Phi_0', \quad \cW^{(3)}=\frac{\lambda_0}{3}\Phi_0'\kappa \cos\theta ,\nonumber\\
    \cX^{(2)}&=-{G^{(0)} }'\psiTwo,\;\; \cX^{(3)}=-{G^{(0)}}'\psi^{(3)} +  \lbr \GOne \psiTwo_{,\ell}-2\psiTwo \GOne_{,\ell}\rbr ,\label{eq:cYtoXi} \\
    \Xi^{(2)}&=\lambda_0 \alpha^{(0)}_{,\ell} \psiTwo, \;\; \Xi^{(3)}= \lambda_0 \lbr \alphaOcz \psiThree +\frac{1}{3}\lbr \alphaOne_{,\ell}\psiTwo -2 \alphaOne \psiTwo_{,\ell} \rbr \rbr.
   \nonumber
\end{align}

The first-order NAE MHD system is given by 
\begin{subequations}
    \begin{align}    
         \lbr \BD\rbr^{(1)}_0 \GOne + \FOne_G&=0 \label{eq:MDE_G1_MHD},\\
        (\del^2_\omega+3^2)\xiThree +\FThree_{\xi} &= 0,
        \label{eq:xi3_eqn_MHD}\\
        \lbr \BD\rbr^{(1)}_0 \psiThree + \FThree_\psi&=0\label{eq:MDE_psi3_short_MHD},\\
        \alphaOne \psiTwo_{,\omega}- 2 \psiTwo\alphaOne_{,\omega}&=\lbr 3\psiThree-2\psiTwo \kappa \cos \theta\rbr \alphaOcw \label{eq:alpha1_omega_eqn_MHD},\\
        \lbr 2\psiTwo\rbr \lbr \BD\rbr^{(1)}_0 \alphaOne + \FOne_\alpha&=0 \label{eq:MDE_alpha1_MHD}.
    \end{align}
    \label{eq:Order_rho_1_system_MHD}
\end{subequations}
Here, the magnetic differential operator is of the form \eqref{eq:MDO_FF_MHD}, i.e.,
\begin{align}
    (\BD)^{(n)}_0=\lbr \Phi_0'\del_\ell + \lbr \xiTwo_{,\omega}+\frac{\lambda_0}{2}\Phi_0' \rbr \del_\omega + 2n\; \xiTwo \rbr, 
\end{align}
and the forcing terms $\FOne_G,\FThree_\xi,\FThree_\psi,\FOne_\alpha$ are given by
\begin{subequations}
    \begin{align}
\FOne_G &= 2\kappa \cos\theta\; \pTwo \label{eq:F1_G_MHD},\\
\FThree_{\xi}&=\cos\theta \lbr \kappa' \Phi'_0+2\kappa \Phi_0''\rbr +\kappa \tau \Phi_0' \sin\theta  + \kappa\lbr  \xiTwo_{,\omega}\sin\theta -2 \xiTwo \cos\theta\rbr \label{eq:F3_xi_MHD}\\
&\qquad +\frac{\lambda_0}{6}\Phi_0'\kappa \sin\theta -\frac{1}{3}\del_\omega\lbr  2\psiTwo\GOne_{,\omega} -\GOne\psiTwo_{,\omega}\rbr \nonumber,\\
\FThree_\psi&= \lbr \xiThree_{,\omega}\psiTwo_{,\omega}+ 6\xiThree\psiTwo\rbr+2\kappa \cos \theta \;\Phi_0'\psiTwo_{,\ell,}
\label{eq:F3_psi_MHD} \\
&\qquad +\frac{1}{3}\psiTwo_{,\omega}\lbr \lambda_0 \kappa \cos \theta\; \Phi_0' +\GOne \psiTwo_{,\omega}-2\psiTwo\GOne_{,\omega} \rbr \nonumber,\\
\FOne_\alpha&=\Phi_0'\lbr \xiThree_{,\omega}-2\kappa \cos \theta\; \xiTwo_{,\omega} +\frac{1}{3}\lbr -2\lambda_0 \kappa \cos\theta\; \Phi_0' +\GOne \psiTwo_{,\omega}-2\GOne_{,\omega} \psiTwo \rbr \rbr.
\end{align}
\label{eq:MHD_forcing_F_G_xi_psi}
\end{subequations}
We recover the vacuum equations \eqref{eq:first_order_system_vacuum} in the limit $\pTwo\to 0, \GOne\to 0,\lambda_0\to 0, \xi\to \Phi$.

We follow the same steps as outlined in the vacuum case (see Appendix \ref{app:general_first_order_vacuum}) with the addition of Step 0, which involves the solution of the MDE for $\GOne$. 

\subsection{Step 0: Calculation of $\GOne$}
We extensively use the normal form variables and their identities (see Appendix \ref{app:normal_form} ) to solve the MDE for $\GOne$. We first note that $\cos\theta= \cos{(u-\delta(\ell))}$. Using the definition of the normal form variable $z_\cN$, it is straightforward to show that
\begin{align}
    \cos\theta = \frac{1}{2}z_\cN \lbr e^{-\eta/2}\cos \delta-i e^{\eta/2}\sin \delta \rbr + \text{c.c.}
    \label{eq:form_of_forcing_F1_G}
\end{align}
where ``c.c." stands for complex conjugate. Therefore, the forcing term $\FOne_G$ is linear in $z_\cN$. The form of the forcing term \eqref{eq:form_of_forcing_F1_G} and the identity \eqref{eq:Id_useful_for_G1},
\begin{align}
    (\BD)^{(1)}_0 \lbr \sqrt{\Phi_0'} z_N \overline{Z}(\ell)\rbr = \lbr \Phi_0' \rbr^{3/2} \lbr \overline{Z}'(\ell) + i \Omega_0 \overline{Z}(\ell)\rbr z_\cN, 
    \tag{\ref{eq:Id_useful_for_G1}}
\end{align}
suggests that $\GOne$ must be of the form
\begin{align}
\GOne= \frac{1}{2}\sqrt{\Phi_0'}\lbr  z_\cN \overline{\cZ}^{(1)}_G  + \overline{z}_\cN\cZ^{(1)}_G\rbr,
    \label{eq:form_of_G1}
\end{align}
where 
\begin{align}
\cZOne_G(\ell)= \gOne_{c1}(\ell)+ i  \gOne_{s1}(\ell),
\label{eq:Z1_G_def}
\end{align}
is a complex function of $\ell$.
Substituting \eqref{eq:form_of_G1} in the MDE for $\GOne$, and using \eqref{eq:form_of_forcing_F1_G} and \eqref{eq:Id_useful_for_G1}, we get
\begin{align}
    {{\overline{\cZ}}^{(1)}_G}'(\ell) +i \Omega_0 {{\overline{\cZ}}^{(1)}_G} + \cF^{(1)}_G =0, \quad 
    \cF^{(1)}_G = \frac{2 \pTwo \kappa}{\lbr \Phi_0' \rbr^{3/2}}  \lbr e^{-\eta/2}\cos \delta-i e^{\eta/2}\sin \delta \rbr
    \label{eq:ODE_Z1_G}
\end{align}
With the solution of \eqref{eq:ODE_Z1_G} we can reconstruct $\GOne$ in terms of Mercier variables as
\begin{align}
    \GOne&= \GOne_{c1}(\ell)\cos u+  \GOne_{s1}(\ell)\sin u \label{eq:_sol}\\
    \GOne_{c1}&=\sqrt{\Phi_0'}e^{\eta/2} \gOne_{c1}\quad \GOne_{s1}=\sqrt{\Phi_0'}e^{-\eta/2}  \gOne_{s1}.\nonumber
\end{align}

\subsection{Step 1: Calculation of $\xiThree$}
This step is identical to the vacuum case. The only differences are the additional terms in the forcing $F_\xi$ that arise from currents. The solution of \eqref{eq:xi3_eqn_MHD} is
\begin{align}
    \xiThree&= \Phi_0'\lbr \fThree_{c1}\cos u +\fThree_{s1}\sin u+ \fThree_{c3}\cos 3u +\fThree_{s3}\sin 3u\rbr \label{eq:xi3_expression_MHD}\\
    \fThree_{c1}&=-\frac{\kappa}{8}\cos \delta \lbr \frac{5}{2} \frac{\Phi''_0}{\Phi_0'}+\frac{\kappa'}{\kappa}- 2 \fTwo_{c2}\rbr+\frac{\kappa}{8}\lbr 2\fTwo_{s2}+\tau\rbr \sin \delta 
    \nonumber\\
    &\quad + \frac{1}{12}\lbr (b(\ell)-a(\ell))\GOne_{c1}+\frac{\lambda_0}{4}\kappa \sin\delta\rbr \nonumber,\\
    \fThree_{s1}&=
    -\frac{\kappa}{8}\sin \delta \lbr \frac{5}{2} \frac{\Phi''_0}{\Phi_0'}+\frac{\kappa'}{\kappa}+2 \fTwo_{c2}\rbr+\frac{\kappa}{8}\lbr 2\fTwo_{s2}-\tau\rbr \cos \delta
    \nonumber\\
    &\quad - \frac{1}{12}\lbr (b(\ell)+a(\ell))\GOne_{s1}+\frac{\lambda_0}{4}\kappa \cos\delta\rbr, \nonumber
\end{align}
where $\fThree_{c3}(\ell),\fThree_{s3}(\ell)$ are the free functions. The free functions are chosen to be of the Soloviev-like form
\begin{subequations}
    \begin{align}
  \fThree_{c3}(\ell)\cos{3u}+\fThree_{s3}(\ell)\sin{3u} =-\frac{\sqrt{\Phi_0'}}{9}\frac{\del}{\del \ell}\lbr \QThree \cos3u + \PThree \sin3u \rbr.\label{eq:fthree_free_P3_Q3_MHD}
\end{align}
\label{eq:fthree_free_det_Soloviev_MHD}
\end{subequations}

\subsection{Step 2: Calculation of $\psiThree$}
The solution of \eqref{eq:MDE_psi3_short_MHD} can be obtained by proceeding exactly as in the vacuum case discussed in Section \ref{sec:app_psi3_normal_form}. Therefore, we only provide essential details here.

The forcing $\FThree_\psi$ given in \eqref{eq:F3_psi_MHD} reduces to
\begin{subequations}
   \begin{align}
     \frac{ \FThree_\psi}{{\Phi_0'}^2}&\equiv \GammaThree_{\psi c1}\cos u +  \GammaThree_{\psi s1}\sin u+ \GammaThree_{\psi c3}\cos3 u +  \GammaThree_{\psi s3}\sin 3u,\\
   \GammaThree_{\psi c1}&=6 a \fThree_{c1} + 2b \lbr 2\fThree_{c1}+3 \fThree_{c3}\rbr+\frac{\kappa \cos \delta}{\Phi_0'}\lbr (2a+b) \Phi_0'\rbr'\\
   &\qquad -2b \kappa \lbr u' +\frac{\lambda_0}{6}\rbr\sin\delta -\frac{2}{3}b(a-b)\GOne_{c1}\nonumber,\\
   \GammaThree_{\psi s1}&=6 a \fThree_{s1} + 2b \lbr -2\fThree_{s1}+3 \fThree_{s3}\rbr +\frac{\kappa \sin \delta}{\Phi_0'}\lbr (2a-b) \Phi_0'\rbr'\\
   &\qquad -2b \kappa \lbr u' +\frac{\lambda_0}{6}\rbr \cos\delta +\frac{2}{3}b(a+b)\GOne_{s1}\nonumber,\\
    \GammaThree_{\psi c3}&=2 b \fThree_{c1}+6 a \fThree_{c3} +\lbr b \Phi_0'\rbr ' \frac{\kappa}{\Phi_0'} \cos \delta \\
  &\qquad +2b \kappa \lbr u' +\frac{\lambda_0}{6}\rbr \sin\delta +\frac{2}{3}b(a-b)\GOne_{c1}\nonumber,\\
   \GammaThree_{\psi s3}&=2 b \fThree_{s1}+6 a \fThree_{s3} +\lbr b \Phi_0'\rbr '\frac{\kappa}{\Phi_0'} \sin \delta\\
   &\qquad -2b \kappa \lbr u' +\frac{\lambda_0}{6}\rbr \cos\delta +\frac{2}{3}b(a+b)\GOne_{s1}.\nonumber
\end{align} 
 \label{eq:forcing_F3_psi_MHD}
\end{subequations}
The rest of the calculation follows the vacuum case with the replacement
\begin{align}
    u'\to u'+\frac{\lambda_0}{2},
\end{align}
in \eqref{eq:Coupled_Y3_ODEs} and \eqref{eq:Lambda0_identities}.

\subsection{Step III \& IV : Calculation of $\alphaOne$} 
The calculation is once again identical to the vacuum limit. The only difference occurs in the forcing term $\FOne_\alpha$, which for MHD is given by
\begin{subequations}
    \begin{align}
   \frac{ \FOne_\alpha}{{\Phi_0'}^2}&= \GammaOne_{\alpha c1}\cos u +  \GammaOne_{\alpha s1}\sin u+ \GammaOne_{\alpha c3}\cos3 u +  \GammaOne_{\alpha s3}\sin 3u,\\
   \GammaOne_{\alpha c1}&= + \fThree_{s1}-2\kappa\lbr \cos \delta \lbr \frac{\lambda_0}{3}+\fTwosTwo\rbr -\sin \delta \fTwocTwo\rbr -\frac{2}{3}(a+b)\GOne_{s1},\\
   \GammaOne_{\alpha s1}&= -\fThree_{c1}-2\kappa\lbr \sin \delta \lbr \frac{\lambda_0}{3}-\fTwosTwo\rbr -\cos \delta \fTwocTwo\rbr +\frac{2}{3}(a-b)\GOne_{c1},\\
   \GammaOne_{\alpha c3}&= +3 \fThree_{s3}-2\kappa\lbr \cos \delta \fTwosTwo +\sin \delta \fTwocTwo\rbr,\\
   \GammaOne_{\alpha s3}&= 
   -3\fThree_{c3}+2\kappa\lbr \cos \delta \fTwocTwo -\sin \delta \fTwosTwo\rbr.
\end{align}
 \label{eq:forcing_F1_MHD}
\end{subequations}

\section{Various examples of vacuum, force-free, and MHD fields}
\label{app:various_examples}
\subsection{Vacuum fields with helical symmetry}
\label{sec:illustration_helix}
While vacuum fields with axisymmetry are trivial, fields with helical symmetry are not. Here, we consider a vacuum field with a magnetic axis with constant curvature and torsion $\kappa_0,\tau_0$, and hence is a helix with pitch $\tau_0/\kappa_0$. 

The magnetic field is assumed to possess helical symmetry, i.e.,
\begin{align}
    \B =\B(u_h), \quad u_h = \omega -\tau_0 \ell.
\end{align}
As a consequence, $(\psi,\Phi,\alpha)$ must be of the form
\begin{align}
    \psi=\psi(u_h), \quad \Phi=F \theta +G \phi +\widetilde{\Phi}(\psi,u_h), \quad \alpha=\theta -\iota \phi +\widetilde{\alpha}(\psi,u_h).
\end{align}
Since $\omega$ is not defined on the axis, $B$ must be constant on the axis, i.e., $\Phi_0=B_0 \ell$. 

The lowest-order solution is given by
\begin{align}
\PhiTwo=-B_0 \tau_0\frac{\vep}{2} \sin{2 u_h}, \; \psiTwo=a B_0 (1+\vep\cos{2u_h}), \;\alphaO=\tan^{-1}\lbr e^{-\eta} \tan{u_h}\rbr +\frac{\tau_0}{2a}\ell,
    \label{eq:helix_low}
\end{align}
where, $a,b,B_0,\eta$ are all constants. We can obtain all the next order relevant physical quantities by setting $\kappa,\tau,\Phi_0',a,b,\delta$ to constants. However, we go through the steps again to better illustrate the Mercier-Weitzner formalism through this simple analytically tractable example.

To first order, the Poisson equation for $\PhiThree$,
\begin{align}
    \lbr \del_\omega^2 +3^2\rbr\PhiThree=-\lbr 1+\varepsilon\rbr B_0 \kappa_0 \tau_0\: \sin{u_h},
\end{align}
has a solution of the form
\begin{align}
\PhiThree= -\frac{1}{8}\lbr 1+\varepsilon\rbr B_0 \kappa_0 \tau_0\: \sin{u_h}+\PhiThree_{c3}\cos 3u_h+\PhiThree_{s3}\sin 3u_h,
    \label{eq:Phi3_helix}
\end{align}
with $\PhiThree_{c3},\PhiThree_{s3}$ being the free-functions.

The MDE for $\psiThree$ is 
\begin{align}
    \lbr \BD \rbr^{(3)}_0 \psiThree +\cFThree_\psi=0,
    \label{eq:MDE_psi3_helix}
\end{align}
with
\begin{align}
   \cFThree_\psi&=\lbr 6 \PhiThree \psiTwo+ \PhiThree_{,\omega} \psiTwo_{,\omega}\rbr +2 B_0\kappa_0 \cos{u_h} \psiTwo_{,\omega},
\end{align}
which has only first and third harmonics. Therefore, $\psiThree$ must be of the form
\begin{align}
    \psiThree=\psiThree_{c1}\cos u_h+\psiThree_{s1}\sin u_h+\psiThree_{c3}\cos 3u_h+\psiThree_{s3}\sin 3u_h
    \label{eq:psi3_helix}
\end{align}
Substituting \eqref{eq:psi3_helix} into \eqref{eq:MDE_psi3_helix} leads to linear \textit{algebraic} equations with solution
\begin{align}
    \psiThree_{c1}= \frac{1}{4} (3-\varepsilon)\kappa_0 a B_0, \;\;  \psiThree_{s1}=0, \;\;  \psiThree_{c3}= -a\lbr \frac{2\PhiThree_{s3}}{\tau_0}+\frac{\vep}{3} B_0 \kappa_0\rbr,\:\:\psiThree_{s3} =\frac{2 a \PhiThree_{c3}}{\tau_0}.
\end{align}
We have algebraic equations instead of ODEs because of the continuous (helical) symmetry. Furthermore, because of the helical symmetry we can construct a solution that has odd and even parity in $\PhiThree$ and $\psiThree$ by choosing $\PhiThree_{c3}=0$. The expressions for $\PhiThree,\psiThree$ derived above matches with \citep{Solovev1970}, with 
\begin{align}
    \PhiThree_{s3}=-\frac{1}{3}\tau_0 B_0 \QThree.
\end{align}

Proceeding to find $\alphaOne$, from $\PhiThree_{,\ell}$ equation we get
\begin{align}
    \del_\omega \lbr \frac{\alphaOne}{\sqrt{2\psiTwo}}\rbr= \frac{B_0}{\lbr 2\psiTwo\rbr^{5/2}}\lbr 2\kappa_0 \psiTwo \cos u_h -3\psiThree\rbr.
    \label{eq:alpha1_w_helix}
\end{align}
Choosing even parity for $\psiThree$ and integrating with respect to $\omega$, we get
\begin{align}
    \alphaOne= \frac{\cAThree_{s1}\sin u_h+\cAThree_{s3}\sin 3u_h}{1+\varepsilon \cos 2u_h} +\sqrt{2\psiTwo}\mathfrak{a}^{(1)}(\ell)
    \label{eq:alpha1_soln_helix}
\end{align}
with
\begin{align}
    \cAThree_{s1}&= \frac{-\kappa_0 (1-(7-8 \varepsilon ) \varepsilon )+ 8 \QThree\vep}{16 a (\varepsilon +1)^2},\\ \cAThree_{s3}&=\frac{\kappa_0 \varepsilon  (23 \varepsilon +7)-8\QThree(1+2\vep)}{48 a (\varepsilon +1)^2} \nonumber.
\end{align}
Finally, from the averaged MDE for $\alphaOne$, we find that $\mathfrak{a}^{(1)}$ is identically zero.

\subsection{Force-free and MHD fields with nearly circular flux-surfaces}
\label{sec:illustration_circ_MHD}
Similar to the vacuum case, we now consider the circular cross-section case with $a(\ell)=1/2,b(\ell)=0$ in \eqref{eq:FF_psi2_xi2_alpha0}. As with the previous example, we show all the steps involved. The relevant lowest-order quantities are given by
\begin{align}
    \GO=p^{(2)}\int \dfrac{d\ell}{\Phi_0'(\ell)},\quad \xiTwo= -\frac{\Phi_0''}{4},\quad  \psiTwo=\frac{1}{2}\Phi_0', \quad \alphaO= \omega -\frac{\lambda_0}{2} \ell.
    \label{eq:psi2_alpha0_phi2_circ_MHD}
\end{align}
To $O(\rho)$ the equations for $(\GOne,\xiThree,\psiThree,\alphaOne)$ are
\begin{subequations}
    \begin{align}
     \lbr \del_\ell + \frac{\lambda_0}{2}\del_\omega-\frac{1}{2}\frac{\Phi_0''}{\Phi_0'}\rbr \GOne +\frac{2\pTwo}{\Phi_0'}\kappa \cos\theta=0 \label{eq:MDE_G1_circ_MHD}\\
        -(\del^2_\omega+3^2)\xiThree = \cos\theta \lbr \kappa' \Phi_0'+\frac{5}{2}\kappa \Phi_0''\rbr + \sin\theta\:  \kappa \lbr\tau+\frac{\lambda_0}{6} \rbr\Phi_0'-\frac{1}{3}\Phi_0' \GOne_{,\omega\omega}
        \label{eq:Phi3_eqn_circ_MHD}\\
        \lbr \del_\ell + \frac{\lambda_0}{2}\del_\omega -\frac{3}{2}\frac{\Phi_0''}{\Phi_0'}\rbr \psiThree + \lbr 3 \xiThree+\Phi_0''\kappa \cos \theta \rbr =0
        \label{eq:MDE_psi3_circ_MHD}\\
        - \Phi_0'\alphaOne_{,\omega}=\lbr 3\psiThree-\Phi_0' \kappa \cos \theta\rbr \label{eq:alpha1_omega_eqn_circ_MHD}
    \end{align}
    \label{eq:Order_rho_1_system_circ_MHD}
\end{subequations}
Equations \eqref{eq:alpha1_omega_eqn_circ_MHD}  generalize the vacuum limit \eqref{eq:alpha1_omega_eqn_circ} by including current terms $\lambda_0,\GO,\GOne$ but preserve the overall form of the equations.  

Equation \eqref{eq:MDE_G1_circ_MHD} is the MDE for $\GOne$, whose solution can be obtained following a procedure similar to that we used to solve the MDE for $\psi3$ in \eqref{sec:illustration_circ}. Substituting the following form for $\GOne$
\begin{align}
    \GOne= \sqrt{\Phi_0'}\lbr \gOne_{c1}(\ell)\cos{u} +\gOne_{s1}(\ell)\sin{u}\rbr, \quad u=\omega-\int \tau dl  \label{eq:G1_form}
\end{align}
into the MDE \eqref{eq:MDE_G1_circ_MHD}, we get the following coupled ODEs: 
\begin{align}
    {\gOne_{c1}}'+ \Omega_0 \gOne_{s1} +\frac{2\pTwo}{(\Phi_0')^{3/2}}\kappa =0, \quad {\gOne_{s1}}'- \Omega_0 \gOne_{c1}=0, \quad \Omega_0= \frac{\lambda_0}{2}-\tau.
    \label{eq:coupled_g1_ODEs}
\end{align}
The coupled ODEs \eqref{eq:coupled_g1_ODEs} can be combined into the single complex ODE
\begin{align}
    {\cZOne_g}'-i \Omega_0  \cZOne_g + \cFOne_g=0, \quad \cZOne_g= \gOne_{c1} + i \gOne_{s1}, \quad \cFOne_g= \frac{2\pTwo}{(\Phi_0')^{3/2}}\kappa.
\end{align}

With the solution of $\GOne$ in hand, we now turn to 
equation \eqref{eq:Phi3_eqn_circ_MHD}, which is the Poisson equation for $\xiThree$. The solution takes the form
\begin{align}
    \xiThree &= \xiThree_{1H}+\xiThree_{3H}\label{eq:Phi3_soln_circ_MHD},\\
    \xiThree_{1H} &= \fThree_{c1}\cos u +\fThree_{s1}\sin u \nonumber,\\
    \fThree_{c1} &= 
    -\frac{1}{8}\lbr \kappa'\Phi_0'+\frac{5}{2}\kappa \Phi_0'' \rbr -\frac{\sqrt{\Phi_0'}}{24}\gOne_{c1} \quad 
     \fThree_{s1}=-\frac{\Phi_0'}{8}\kappa \lbr \tau +\frac{\lambda_0}{6}\rbr -\frac{\sqrt{\Phi_0'}}{24}\gOne_{s1}, \nonumber\\
    \xiThree_{3H} &= -\frac{\lbr \Phi_0'\rbr^{3/2}}{9} \lbr \del_\ell +\frac{\lambda_0}{2}\del_\omega\rbr \lbr \QThree \cos{3u}+\PThree \sin{3u}\rbr. \nonumber
\end{align}
The reason for choosing the free-function $\xiThree_{3H}$ in this particular Soloviev-like form will become apparent when we solve for $\psiThree$. The choice made earlier in \eqref{eq:fthree_free_det_Soloviev} is equivalent to this choice as can be seen through explicit calculation.  

Proceeding to the MDE for $\psiThree$  \eqref{eq:MDE_psi3_circ_MHD} we proceed exactly as in the vacuum case to get
\begin{align}
    \psiThree= \lbr \Phi_0' \rbr^{3/2}\YThree, \quad \lbr \del_\ell +\frac{\lambda_0}{2}\del_\omega\rbr\YThree+\frac{1}{\lbr \Phi_0' \rbr^{3/2}}\lbr 3\PhiThree+\Phi_0'' \kappa \cos\theta\rbr=0.
    \label{eq:S3_eqn_circ_MHD}
\end{align}
Due to the choice of the free-function in \eqref{eq:Phi3_soln_circ_MHD}, we can clearly separate the first and third harmonics of $\psiThree$.The solution of \eqref{eq:MDE_psi3_circ_MHD} can then be written in the form analogous to the vacuum case
\begin{align}
    \psiThree =\frac{3}{8}\Phi_0'\kappa \cos u+\lbr\Phi_0'\rbr^{3/2}\lbr \frac{1}{3}\lbr \QThree \cos{3u}+\PThree \sin{3u}\rbr+\lbr \sigma^{(3)}_{c1}\cos u + \sigma^{(3)}_{s1}\sin u \rbr\rbr,
    \label{eq:psi3sol_circ_MHD}
\end{align}
where, $ \sigma^{(3)}_{c1}, \sigma^{(3)}_{s1}$ now satisfy
\begin{align}
    {\sigma^{(3)}_{c1}}'+\Omega_0 \sigma^{(3)}_{s1}+F^{(3)}_{\psi c1}=0, \quad 
     {\sigma^{(3)}_{s1}}'-\Omega_0 \sigma^{(3)}_{c1}+F^{(3)}_{\psi s1}=0.
     \label{eq:sigma3c1_s1_eqn_MHD}\\
     F^{(3)}_{\psi c1}=-\frac{1}{8}\lbr \kappa\frac{\Phi_0''}{{(\Phi_0')}^{3/2}} + \gOne_{c1}\rbr, \quad  F^{(3)}_{\psi s1}=-\frac{1}{8}\lbr \frac{2\lambda_0\kappa}{{(\Phi_0')}^{1/2}} + \gOne_{s1}\rbr.\nonumber
\end{align}
We then combine \eqref{eq:sigma3c1_s1_eqn_MHD} into a single complex equation
\begin{align}
   &{ \cZ^{(3)}_\psi}'-i \Omega_0 { \cZ^{(3)}_\psi} + \cF^{(3)}_\psi=0,\\
   & \cZ^{(3)}=\sigma^{(3)}_{c1} + i \sigma^{(3)}_{s1}, \quad \cF_\psi^{(3)}=F^{(3)}_{\psi c1} + i F^{(3)}_{\psi s1}.\nonumber
\end{align}

The $\alpha$ equation \eqref{eq:alpha1_omega_eqn_circ_MHD} is identical to its vacuum analog \eqref{eq:alpha1_omega_eqn_circ}. Thus, we get the same solution,
\begin{align}
    \alphaOne &= -\frac{\kappa}{8} \sin u -\frac{\lbr \Phi_0'\rbr^{1/2}}{3}\lbr \QThree \sin{3u}-\PThree \cos{3u}\rbr\\ &\quad +\mathfrak{a}^{(1)}(\ell)-3\lbr \Phi_0'\rbr^{1/2}\lbr \sigma^{(3)}_{c1}\sin u - \sigma^{(3)}_{s1}\cos u \rbr.\nonumber
\end{align}
The function $\mathfrak{a}^{(1)}(\ell)$ must be determined from the poloidal average of the MDE for $\alphaOne$,
\begin{align}
      \lbr \del_\ell +\frac{\lambda_0}{2}\del_\omega -\frac{\Phi_0''}{2 \Phi_0'} \rbr \alphaOne+\del_\omega\lbr \frac{\xiThree}{\Phi_0'}-\frac{2}{3}\lambda_0 \kappa \sin\theta -\del_\omega \frac{\GOne}{3}\rbr=0.
        \label{eq:MDE_alpha_eqn_circ_MHD}
\end{align}
Since the poloidal average of \eqref{eq:MDE_alpha_eqn_circ_MHD} is identically zero, both  $\YThree_H$ and $\mathfrak{a}^{(1)}(\ell)$ are zero. 

Unlike the vacuum case, the above solutions to $(\GOne,\xiThree,\psiThree,\alphaOne)$ do not simplify much when the on-axis magnetic field, $\Phi_0'=B_0$ is a constant. However, there are two exceptional cases where further simplification is possible: the axisymmetric limit with a planar circular axis and when the magnetic axis has constant torsion. We shall discuss this limit in the following two examples.

\subsection{Axisymmetric Soloviev profiles with nearly circular cross-section}
\label{app:circ_Soloviev}

Assuming axisymmetry, circular cross-section, planar circular axis $\kappa=1/R_0, \tau=0, \omega=\theta$, and constant $\Phi_0'=B_0$, the analysis carried out in Appendix \ref{sec:illustration_circ_MHD} simplify considerably. The lowest order quantities are
\begin{align}
    \GO= \dfrac{\pTwo \ell}{B_0},\quad \xiTwo= 0,\quad  \psiTwo=\frac{B_0}{2}, \quad \alphaO= \omega -\frac{\lambda_0}{2} \ell, \quad \iota_0 =\frac{\lambda_0 B_0}{2}.
    \label{eq:psi2_alpha0_phi2_circ_MHD_axisym}
\end{align}

Using $\del_\ell=0$ on axisymmetric quantities the next order quantities can be shown to be
\begin{align}
    &\GOne= -\frac{4 \pTwo}{\lambda_0 B_0 R_0 }\sin\omega, \quad  \xiThree= \frac{\sin\omega \left(8 \pTwo/\lambda_0-B_0 \lambda_0\right)}{48 R_0}+ \lbr -\QThree\sin{3 \omega} +\PThree \cos{3 \omega}\rbr\nonumber\\
    &\frac{\psiThree}{B_0}= \frac{1}{\lambda_0 B_0 R_0}\cos{\omega} \left(\frac{\pTwo}{ \lambda_0}-\frac{\lambda_0 B_0}{8}\right)-\frac{2}{\lambda_0 B_0}\lbr \PThree \sin{3 \omega}+\QThree\cos{3 \omega}\rbr \label{eq:axisym_circ_B0_MHD}\\
    &\alphaOne=\frac{1}{R_0}\lbr \frac{11}{8}-3\frac{\pTwo/\lambda_0}{\lambda_0 B_0} \rbr \sin\omega -\frac{2}{\lambda_0 B_0}\lbr -\QThree\sin{3 \omega} +\PThree \cos{3 \omega}\rbr. \nonumber
\end{align}
The force-free limit of \eqref{eq:psi2_alpha0_phi2_circ_MHD_axisym} and \eqref{eq:axisym_circ_B0_MHD} is 
\begin{align}
    & \xiTwo= 0,\quad  \psiTwo=\frac{B_0}{2}, \quad \alphaO= \omega -\frac{\lambda_0}{2}\ell, \nonumber\\
    &\xiThree= -\frac{\sin\omega \left(B_0 \lambda_0\right)}{48 R_0}+ \lbr -\QThree\sin{3 \omega} +\PThree \cos{3 \omega}\rbr\nonumber,\\
    &\frac{\psiThree}{B_0}= -\frac{1}{8 R_0}\cos{\omega} -\frac{2}{\lambda_0 B_0}\lbr \PThree \sin{3 \omega}+\QThree\cos{3 \omega}\rbr\label{eq:axisym_circ_B0_FF},\\
    &\alphaOne=\frac{1}{R_0}\lbr \frac{11}{8} \rbr \sin\omega -\frac{2}{\lambda_0 B_0}\lbr -\QThree\sin{3 \omega} +\PThree \cos{3 \omega}\rbr. \nonumber
\end{align}

We first describe the force-free case assuming a circular cross-section and constant on-axis magnetic field strength. This corresponds to $A=1$ in the Soloviev profiles \eqref{eq:int_Sol_profile}. Next, we will set the major radius $R_0=1$.

To $O(\rho^4)$ the poloidal and toroidal $\Psi,\psi$ fluxes are, 
\begin{align} 
    \Psi=\iota_0 \psi +O(\rho^4), \quad \frac{\psi}{B_0}=\lbr \frac{1}{2}\rho^2+ \frac{\psiThree}{B_0}\rho^3 \rbr.
\end{align}
where, $\psiThree$ is given by \eqref{eq:axisym_circ_B0_FF}. 
Expanding $F(\Psi)$ as given in \eqref{eq:int_Sol_profile} we get
\begin{align}
    F(\Psi)= F_0 -\frac{1}{F_0}\Psi+ O(\Psi^2), \quad F'(\Psi)= -\frac{1}{F_0} +O(\Psi)
\end{align}
Now, using the relation between $\lambda(\psi)$ and $F(\psi)$, $-\lambda(\psi)=F'(\Psi)$, we get $\lambda_0=1/F_0$. Since the on-axis axisymmetric $B_0=F_0/R_0$, we obtain
\begin{align}
    \lambda_0 B_0 =1, \qquad \iota_0 =\dfrac{1}{2}.
    \label{eq:iota_is_half}
\end{align}
Therefore, the third order shaping $\psiThree$ is now given by
\begin{align}
    \frac{\psiThree}{B_0}= -\frac{1}{8}\cos{\omega} -2\lbr \PThree \sin{3 \omega}+\QThree\cos{3 \omega}\rbr \label{eq:axisym_circ_B0_FF_simp}
\end{align}
Let us now compare with the one-size model. The exact solution is given by $\Psi(x,y)$ in \eqref{eq:1size_psi_p}, where
\begin{align}
     x=1+\sqrt{2\iota_0 B_0}\rho \cos\theta, \quad y=\sqrt{2\iota_0 B_0}\rho \sin\theta
\end{align}
For convenience, let us set $B_0=1$. Since $\iota_0=1/2$ we have $x-1=\rho\cos\theta, y=\rho \sin \theta$. To match the NAE expression, we now expand the exact solution \eqref{eq:1size_psi_p} in power series of $\rho$. Imposing that $\Psi=0$ on the axis and the cross-section be circular to the lowest order in the $\rho$ expansion, we get
\begin{align}
    c_1=\frac{1}{8},\quad c_2=-\frac{1}{8},\quad  c_3=\frac{1}{4}.
    \label{eq:cis_for_FF}
\end{align}

Further using $\iota_0=1/2$, we get $\Psi=\psi/2$ where,
\begin{align}
    \quad \frac{\psi}{B_0}=\frac{1}{2}\rho^2+ \frac{1}{8}\lbr\cos\theta +\frac{1}{3}\cos{3\theta}  \rbr \rho^3+O(\rho^4).
\end{align}
Clearly, the power series expansion of the exact solution and the NAE match exactly when
\begin{align}
    \theta= \pi+\omega , \quad \PThree=0, \quad \QThree= -\frac{1}{48}.
\end{align}

For the MHD case, we use $A=1+\beta, \beta\neq 0$. For finite values of $\beta$, $\lambda_0,\iota_0$ are unchanged. However, the coefficients get modified to 
\begin{align}
    c_1=\frac{1+\beta}{8},\quad c_2=-\frac{1}{8},\quad c_3=\frac{1}{4},
     \label{eq:cis_for_MHD}
\end{align}
which generalizes \eqref{eq:cis_for_FF}. The toroidal flux is now given by
\begin{align}
 \frac{\psi}{B_0}=\frac{1}{2}\rho^2+\frac{1-4\beta}{8}\lbr \cos\theta +\frac{1}{3}\cos{3\theta}  \rbr \rho^3+O(\rho^4).
\end{align}
The exact match with NAE now occurs when
\begin{align}
    \theta= \pi+\omega , \quad \PThree=0, \quad \QThree= -\frac{1-4\beta}{48}.
\end{align}

\subsection{Axis with constant torsion and nearly circular cross-section}
\label{app:circ_const_tau}
The final example is for an axis that has constant torsion. We also choose the magnetic field magnitude on the axis as a constant, which we normalize to unity. We use the results obtained in \ref{sec:illustration_circ_MHD}, specializing them to an axis with a constant torsion.

Curves with torsion and curvature both constant are helices, which are not closed. Therefore, closed space curves with constant nonzero torsion must have non-constant curvature. In the following, we choose the following closed curve as the magnetic axis:
\begin{align}
    \kappa(\ell) = \kappa_0 + 2 \kappa_1 \cos{2\ell}, \quad \tau(\ell)=\tau_0. \label{eq:kappa_const_tau}\\
     \kappa_0=0.5109, \quad \kappa_1= -0.6895,\quad \tau_0=-1.74809\nonumber
     \label{eq:axis_kappa_tau}.
\end{align}
The periodicity of $\kappa(\ell),\tau(\ell)$ is a necessary but not sufficient condition for the closure of a curve. Periodicity of \eqref{eq:axis_kappa_tau} then implies that the length of the axis, $L$, must be an integer multiple of $\pi$, i.e.,
\begin{align}
    L=n \pi , \quad n=1,2,3,..
\end{align}

The lowest-order quantities are
\begin{align}
    \GO=\pTwo \ell,\quad \xiTwo= 0,\quad  \psiTwo=\frac{1}{2}, \quad \alphaO= \omega -\frac{\lambda_0}{2} \ell.
    \label{eq:psi2_alpha0_phi2_circ_MHD_tau0}
\end{align}
To first order, the solution of $\GOne$ is given by \eqref{eq:G1_form} with $\Phi_0'=1$ and
\begin{align}
    \gOne_{c1}= \frac{8\pTwo\kappa_1}{\Omega_0^2 -4}\sin{2\ell}, \quad \gOne_{s1}= -2\pTwo\lbr \frac{\kappa_0}{\Omega_0}+\frac{2\Omega_0 \kappa_1 }{\Omega_0^2 -4}\cos{2\ell} \rbr.
\end{align}
Note that $\GOne$ thus obtained is automatically periodic; hence, we do not need to separately enforce periodicity by adding a homogeneous solution to the MDE for $\GOne$. 

The periodic solutions of $\xiThree$ and $\psiThree$ are given by \eqref{eq:Phi3_soln_circ_MHD} and \eqref{eq:psi3sol_circ_MHD}  with $\Phi_0'=1$ and
\begin{align}
    \sigmaThree_{c1}&= \frac{1}{4}\frac{\kappa_0}{\Omega_0}\lbr\frac{\pTwo}{\Omega_0}-
    \lambda_0 \rbr+\frac{\kappa_1}{2(\Omega_0^2-4)}\lbr -\lambda_0 \Omega_0 + \frac{\pTwo(\Omega_0^2+4)}{\Omega_0^2-4} \rbr \cos{2\ell} \nonumber\\
     \sigmaThree_{s1}&= \frac{\kappa_1}{\Omega_0^2-4} \lbr 2\pTwo \frac{\Omega_0}{\Omega_0^2-4} -\lambda_0 \rbr \sin{2\ell}.
\end{align}
Finally, $\alphaOne$ is given by
\begin{align}
    \alphaOne&= \alphaOne_{s1}\sin u +\alphaOne_{s(+)}\sin{(u+2\ell)}+\alphaOne_{s(-)}\sin{(u-2\ell)}\\
    \alphaOne_{s1}&=-\frac{3\kappa_0}{4}\lbr \frac{\frac{\pTwo}{\Omega_0}-\lambda_0}{\Omega_0}+\frac{1}{6} \rbr, \quad \alphaOne_{s(\pm)}=-\frac{3\kappa_1}{4}\lbr \frac{\frac{\pTwo}{\Omega_0\pm 2}-\lambda_0}{\Omega_0\pm 2}+\frac{1}{6} \rbr.
\end{align}

\begin{figure}
    \centering
    \includegraphics[width=.38\textwidth]{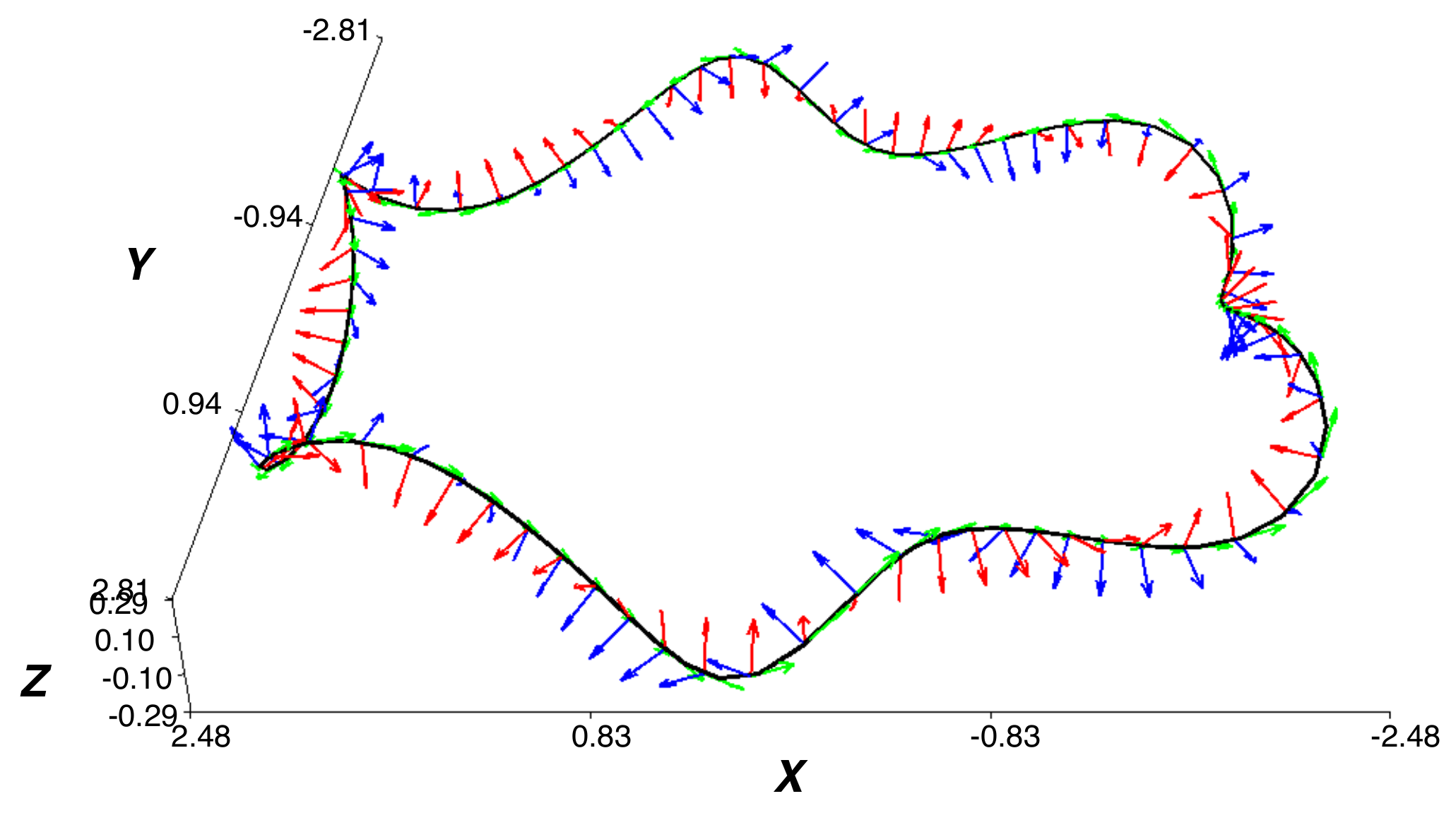}
    \includegraphics[width=.3\textwidth]{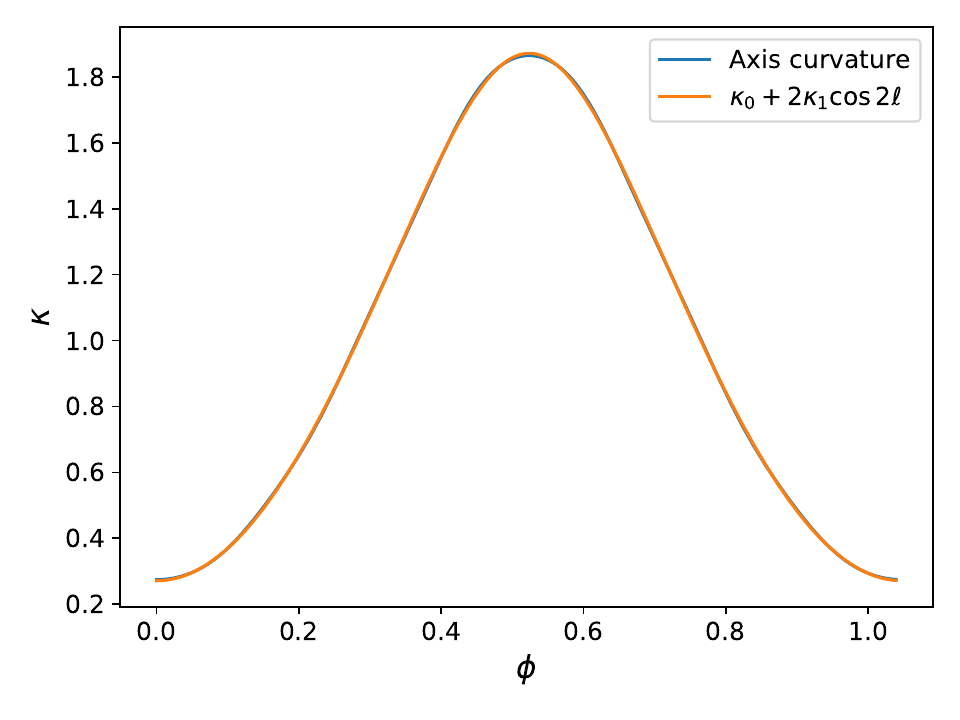}
    \includegraphics[width=.3\textwidth]{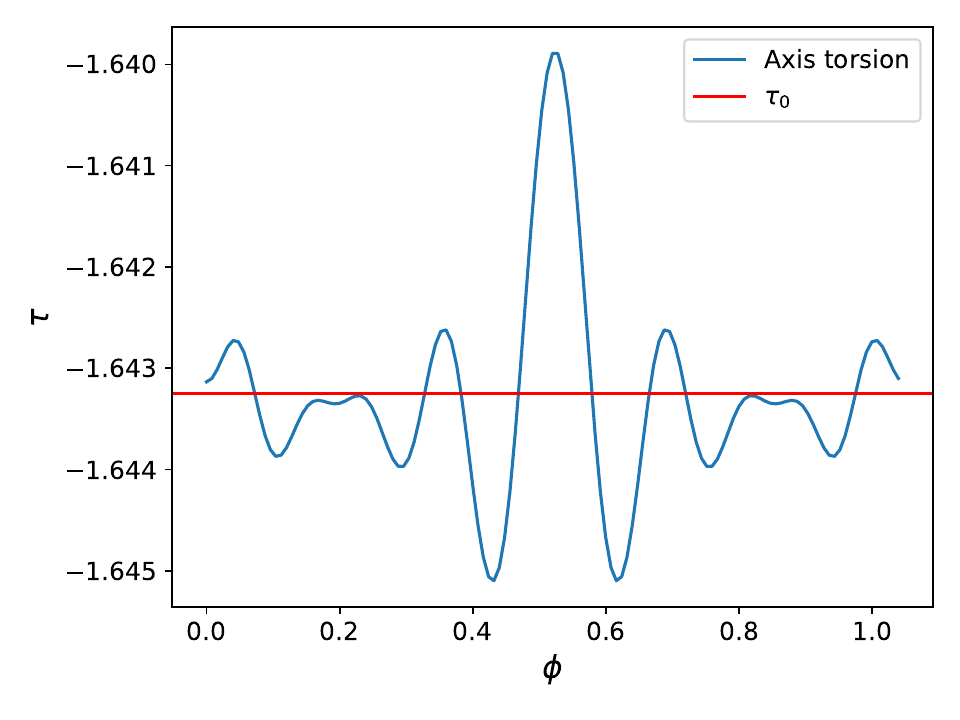}
    \caption{Left: axis curve $\mathbf r_0$ together with its Frenet-Serret frame, normal (red), binormal (blue) and tangent (green); Middle: axis curvature $\kappa$ (blue) and the target curvature (orange); Right: axis torsion (blue) and the target torsion (red).}
    \label{fig:nfp6_axis}
\end{figure}

We now present a numerical implementation of a curve of almost constant torsion. we define the axis shape as a closed curve $\mathbf r_0(\phi)$ parameterized by an angle $\phi$ (cylindrical angle on-axis) such that $\mathbf r_0 = R(\phi) \mathbf e_R + Z(\phi) \mathbf e_Z$ with $\mathbf e_R$ and $\mathbf e_Z$ the radial and vertical unit vectors in cylindrical coordinates, respectively.
In order to ensure stellarator symmetry and field period symmetry, i.e., $R(\phi)=R(-\phi)=R(\phi+2\pi/n_{fp})$ and $Z(\phi)=-Z(-\phi)=Z(\phi+2\pi/n_{fp})$ with $n_{fp}$ a natural number, the functions $(R, Z)$ are written as a $\cos$ and $\sin$ Fourier series with period $2\pi/n_fp$ and coefficients $R_n$ and $Z_n$, respectively.
We then solve the non-linear least-squares optimization problem using the Trust Region Reflective algorithm by minimizing the following cost function.
\begin{equation}
    f=\left[\kappa-(\kappa_0 + 2 \kappa_1 \cos{2\ell})\right]^2+(\tau-\tau_0)^2,
    \label{eq:cost_function_opt_axis}
\end{equation}
and vary the axis coefficients $(R_n, Z_n)$, the components of the curvature $\kappa_0, \kappa_1$ and $\tau_0$.
In Eq. (\ref{eq:cost_function_opt_axis}), $\kappa$ and $\tau$ are the axis curvature and torsion, respectively, and $\ell(\phi)$ is found by solving the equation $|\rho'(\phi)|=\ell'(\phi)$ at each optimization step.
For the case $n_{fp}=6$ used here, we find the following parameters
\begin{align}
    R&=2.465291791214-0.263382920320\cos(6\phi)+0.059015974280\cos(12\phi)\nonumber
    \\&-0.013188484035\cos(18\phi)+0.003189343968\cos(24\phi)-0.000792016256\cos(30\phi)\nonumber
    \\&+0.000193995152\cos(36\phi)-0.000049828064\cos(42\phi)+0.000005385980\cos(48\phi)\nonumber
    \\&+0.000000609205\cos(54\phi)+0.000000009755\cos(60\phi),\\
    Z&=-0.267425343068\sin(6\phi)+0.059377909679\sin(12\phi)-0.013267277590\sin(18\phi)\nonumber
    \\&+0.003222245044\sin(24\phi)-0.000828131438\sin(30\phi)+0.000226082714\sin(36\phi)\nonumber
    \\&-0.000052599459\sin(42\phi)+0.000004090212\sin(48\phi)+0.000000577500\sin(54\phi)\nonumber
    \\&+0.000000012491\sin(60\phi),\\
    \kappa_0&=1.071535923214,~\kappa_1=-0.400217124412,~\tau_0=-1.643251931600,
\end{align}
yielding an objective function $J=5.72\times10^{-5}$ for a grid in $\phi$ with 131 points, a total axis length of $L=6 \phi$ and a normal vector that rotates $N=6$ times around the axis when translating $\mathbf n$ from $\phi=0$ to $2\pi$.
Due to the fact that $N\not=0$, the poloidal angle $\theta$ is replaced by an untwisted poloidal angle $\overline \theta = \theta - 2\pi N \ell/L$ similar to the angle used in \citep{landreman_Sengupta_Plunck2019direct} to generate a finite radius boundary.
The axis shape, its associated Frenet-Serret frame, and its curvature and torsion are shown in Fig. \ref{fig:nfp6_axis}.

\bibliographystyle{jpp}
\bibliography{plasmalit}

\end{document}